# Cyber Catalysis: N$_2$ Dissociation over Ruthenium Catalyst with Strong Metal-Support Interaction

Gerardo Valadez Huerta*,[1], Kaoru Hisama[1], Katsutoshi Sato[2], Katsutoshi Nagaoka[2], Michihisa Koyama*,[1,3]

[1]Research Initiative for Supra Materials, Shinshu University, Nagano 380-8553, Japan.

[2]Department of Chemical Systems Engineering, Nagoya University, Nagoya 464-8601, Japan.

[3]Open Innovation Institute, Kyoto University, Kyoto 606-8501, Japan

*Corresponding authors: valadez@shinshu-u.ac.jp, koyama_michihisa@shinshu-u.ac.jp

**Abstract**

**Catalysis informatics is constantly developing, and significant advances in data mining, molecular simulation, and automation for computational design and high-throughput experimentation have been achieved. However, efforts to reveal the mechanisms of complex supported nanoparticle catalysts in cyberspace have proven to be unsuccessful thus far. This study fills this gap by exploring N$_2$ dissociation on a supported Ru nanoparticle as an example using a universal neural network potential. We calculated 200 catalyst configurations considering the reduction of the support and strong metal-support interaction (SMSI), eventually performing 15,600 calculations for various N$_2$ adsorption states. After successfully validating our results with experimental IR spectral data, we clarified key N$_2$ dissociation pathways behind the high activity of the SMSI surface and disclosed the maximum activity of catalysts reduced at 650 °C. Our method is well applicable to other complex systems, and we believe it represents a key first step toward the digital transformation of investigations on heterogeneous catalysis.**

Catalysis informatics can be classified into three categories[1–3]: i) data mining mainly based on machine learning[2,4,5]; ii) first-principles calculations[6], including the development of kinetic models[7]; iii) and computational automated design[8]. These approaches overlap within cyberspace[4,9,10]. Recent developments in data-driven approaches have enabled the exploration of previously undiscovered catalysts without the need for numerous experiments and computations[11]. Machine-learning approaches can guide high-throughput experiments as counterparts to traditional trial-and-error studies by narrowing the parameter space, resulting in an adaptive experimental design[12]. Obtaining information on catalytic properties from supervised learning based on digital data is another data-driven approach in catalysis[13]. Rational catalyst design using first-principles calculations is ongoing[14,15], and the automated computational high-throughput design of catalysts is primarily practiced in organic chemistry[8]. However, despite the availability of countless catalysis studies and publicly available digital databases, such as Catalysis-Hub[16], significant gaps inhibiting the full exploration of heterogeneous catalyst systems through catalysis informatics remain. This problem persists even in up-to-date discussions[9,17,18] and was confirmed by a machine-generated summary of recent publications for heterogeneous catalysis (see the Supplementary Data) created with the same tool[19] used to obtain machine-generated books[20].

Supervised learning is a data-driven approach used to predict the catalytic properties of heterogeneous catalysts based on descriptors from first-principles data[21]. State-of-the-art first-principles calculations allow single calculations for isolated and supported nanoparticles[22,23] containing up to 2,406 and 1,169 atoms, respectively[22]. This approach can also determine reaction paths for the dissociation of molecules on stepped surfaces[24] or supported clusters[25]. Unfortunately, first-principles calculations, a critical tool for describing electronic and atomistic properties, are not yet widely applicable to the study of real heterogeneous catalysis systems requiring a simplified and ideal model[3]. A computationally inexpensive and widely applicable framework is necessary to examine heterogeneous systems with real solid catalysts, which may have local properties different from those of bulk phases owing to defects[26,27], surface site heterogeneity[28], or a strong metal-support interaction (SMSI)[29–31]. Constant interfacial interactions between various phases may result in surface reconstruction processes[32] influencing catalyst activity[33], which prevents the use of real system models in the first-principles method due to the computational cost[34]. Higher-scaled computations can be performed with potentials such as the ReaxFF[35]. However, the models obtained from such computations must be parameterised for each case, which limits their applicability and transferability[36,37]. Moreover, ReaxFF parameterisation to achieve high accuracy is challenging[38]. Neural network potentials have been used to reproduce first-principles results via machine learning. Calculations using such models are expected to have high computational performance[39–41] and can break through the current theoretical barrier of high reductionist models[42]. Universal neural network potentials (UNNPs) have emerged as a result of the connatural evolution of catalysis informatics[43–45]. The simultaneous calculation of different phases for multiple elements, including all interfacial interactions, is possible using these models. Chemical reactions and charge prediction can also be achieved[43,45]. Thus, these models open the possibility of bringing automated computer design into the discipline of heterogeneous catalysis.

This study transcends the limits of DFT by exploiting the computational possibilities of a UNNP for exploring complex heterogeneous catalyst systems. From the countless material combinations that can be tackled with the UNNP, we chose, as a representative example, the dissociation of $N_2$ on a Ru nanoparticle on a $La_{0.5}Ce_{0.5}O_{1.75-x}$ support for ammonia synthesis[46]. Our modelling approach artificially reproduced the experimentally observed nanoparticle–cationic interfacial interaction resulting from the SMSI with 200 different catalyst configurations considering the reduction degree of the oxide support. We extensively observed $N_2$ adsorption on Ru on-top sites in all catalyst configurations. Calculations of the immense variety of adsorption configurations were unavoidable to obtain a wide-ranging perspective of the microscopic processes. We found catalyst configurations with $N_2$ wavenumbers similar to experimentally measured IR spectra[46]. The activation energy of $N_2$ dissociation for select adsorption sites showed potential adsorption modes leading to the high activity observed in the literature[46].

## Results

**Strategy and Catalyst Model Preparation.** Our approach for modelling and simulating $N_2$ dissociation on supported Ru nanoparticles is depicted in Fig. 1. First, we modelled the catalyst. We prepared $Ru/La_{0.5}Ce_{0.5}O_{1.75-x}$ models consisting of a $Ru_{143}$ half-nanoparticle on a 4 × 4 × 3 $La_{0.5}Ce_{0.5}O_{1.75-x}$(111) slab with a fluorite structure (space group, $Fd\bar{3}m$) (Fig. 1a; Supplementary Fig. 1). We refer to the upper layer of the cationic slab as the cationic surface hereafter. We prepared Ce, La, superlattice (SL), and five solid solution (SS) cationic surfaces. The SL surface was prepared by intercalating the Ce and La cations and SS surfaces via random distribution. The remaining cationic bottom layers were modelled as an SS phase. After assembling the half-nanoparticle and support models, we reduced the slab by removing oxygen atoms from the cationic surface. The onset of an SMSI was observed experimentally for this particular catalyst[46], indicating an increase in nanoparticle–metal oxide interfacial area[29,30]. To model the catalyst structure with an artificial SMSI, we randomly selected cations from the cationic surface and placed them on the lowest nanoparticle layer. The final catalyst configurations were obtained by optimising the models with different degrees of cationic surface reduction and artificial SMSI (Supplementary Fig. 2).

We observed different degrees of SMSI onset, depending on the extent of cationic surface reduction and artificial SMSI (Fig. 1b; Supplementary Note 1). For surfaces with full reduction, that is, all oxygen atoms in the surface layer were removed, surface reconstruction after optimisation resulted in only one or two cations moving toward the upper layers of the Ru nanoparticle without the artificial SMSI. When more cations are initially placed on the nanoparticle at this high reduction degree, they remain at this position or move toward upper layers. By contrast, the cations tend to descend to the support as the reduction degree decreases. We observed the nanoparticle beginning to convulse when the configuration with the highest SMSI and reduction degree simulated in this study was adopted (Supplementary Fig. 3). This type of system deterioration resulting from a high SMSI is a well-known experimentally observed process[29].

Next, we mapped the optimised catalyst configurations to obtain a complete picture of the configurational space. For this purpose, we need to find suitable descriptors. As widely discussed[47,48], the electronic metal-support interaction (EMSI) is more suitable for describing the catalyst support effect than the SMSI. The EMSI describes the charge transfer between a nanoparticle and its support. Therefore, it can be represented by the average charge of Ru. The charge distribution of Ru atoms and, thus, their average charge shift toward more negative values, depending on the degree of cationic surface reduction and SMSI (see blue atoms in Fig. 1c). Thus, we used the average charge $\bar{q}_{Ru,0}$ of the Ru atoms at the surface nanoparticle layer and the value of $x$ in $La_{0.5}Ce_{0.5}O_{1.75-x}$ to spread the $Ru/La_{0.5}Ce_{0.5}O_{1.75-x}$(111) optimised configurations. For simplicity, we only considered the composition of the whole 4 × 4 × 3 slab, which does not correspond to the real system, owing to the size of the $La_{0.5}Ce_{0.5}O_{1.75-x}$ layer in our model. We refer to $x$ as the reduction degree hereafter. The configuration map in Fig. 1d was constructed according to these descriptors. Using different upper cationic slab layer compositions and reduction degrees, as well as artificial modelling of the SMSI, we obtained 200 catalyst configurations.

Finally, we accounted for $N_2$ activation. For each of the 200 catalyst configurations, we placed a $N_2$ molecule with an end-on orientation on each available Ru on-top site by referring to previous experimental and theoretical observations of the adsorption structure[46,49]. This step resulted in 15,600 optimisations. To work with this massive amount of data, we scanned, evaluated, and plotted the properties for each adsorption state using an automated screening approach (Fig. 1e). This step enabled us to automatically create spatial maps for each catalyst configuration by assigning, for example, specific $N_2$ wavenumber values to the Ru site position (compare Fig. 1e and Supplementary Fig. 4).

**Exhaustive Calculation Results of $N_2$ Adsorption.** Histograms showing the distribution of all calculated $N_2$ adsorption states over the N–N bond length are depicted in Fig. 2a. The coverage at 25 °C and 6 kPa is used as the weighting factor for the weighted histogram. Two dense regions with bond length peaks of approximately 1.14 and 1.17 Å were identified. Two adsorption peaks at approximately 2,150 and 1,850 cm$^{-1}$, corresponding to high and low wavenumbers, respectively, were also observed (see bottom diagram in Fig. 2a).

Spatial maps of the $N_2$ wavenumber are illustrated in Fig. 2c. Here, we used the top view of the nanoparticles and polar coordinate systems and kept the size of the markers proportional to the simulated $N_2$ coverage $\theta$ ($T = 25$ °C, $p = 6$ kPa) for better comparison. The adsorbed $N_2$ molecules had a high wavenumber (>2,000 cm$^{-1}$) when $x \approx \bar{q}_{Ru,0} \approx 0$. As the reduction degree $x$ increased and the EMSI became stronger, a lower wavenumber (<2,000 cm$^{-1}$) was observed. Sites with a low wavenumber were located near Ce/La cations wrapped around the nanoparticle. Furthermore, adsorption sites became more active at higher reduction degrees and EMSI, as indicated by the size of the markers. To explain this effect, we constructed spatial maps of the $N_2$ adsorption energy, as shown in Fig. 2d. The adsorption on the edge and vertex sites is stronger than that on the $(10\bar{1}1)$ and $(0001)$ surfaces for configurations with $x \approx \bar{q}_{Ru,0} \approx 0$. As the reduction degree and EMSI increased, the adsorption energy decreased at sites near Ce/La cations. Because of this effect, the coverage increased in these adsorption sites. When the non-weighted and weighted distributions of the histograms in Fig. 2a are compared, the activation of sites with a lower wavenumber become more evident. Moreover, as the number of cations wrapping the nanoparticle increased, the Ru sites became increasingly occupied, thereby preventing the adsorption of $N_2$ molecules (Fig. 2c, $(10\bar{1}1)$ surfaces for the catalyst configuration with $x = 0.27$ and $\bar{q}_{Ru,0} = -0.37$ e). Finally, the wavenumber and $N_2$ molecular charge showed a linear relationship with the $N_2$ bond length (see bottom diagram in Fig. 2a). The N–N bond became weaker owing to electron transfer to the $N_2$ antibonding $\pi$-orbital[50]. While our calculations effectively reproduced this trend, we stress that this statement is simply an interpretation of the results because the UNNP model does not consider electron transfer explicitly.

After performing the exhaustive scan described above, we searched for representative adsorption states. A representative adsorption state should have high coverage and lie in the statistical bulk density within two defined regions. We depict, in Fig. 2b, examples of adsorption states following these criteria (A–I) (Supplementary Fig. 5). $N_2$ molecules adsorbed on the upper layers of the nanoparticle showed a high wavenumber. The wavenumber for adsorption state A, for example, is 2,161 cm$^{-1}$. We can identify two main mechanisms that determine by what extent the wavenumber may decrease. The first effect is applicable to molecules near cations on the nanoparticle (e.g. states B, D, F, G, and H). When a large number of cations are present near the adsorption site, this effect is enriched. However, when these cations are (partially) oxidised, the effect is dampened (compare F and G). The second mechanism is applicable to molecules adsorbed at the interface between the nanoparticle and support, as depicted in adsorption states C and E; in this case, a lower wavenumber is observed even without cations near the adsorption site. However, this finding applies only to supports with a non-zero reduction degree or interfacial regions with oxygen vacancies when $x$ is zero. When both mechanisms are combined, frequencies lower than 1,800 cm$^{-1}$ can be observed, even when the cations attached to the nanoparticle are partially oxidised, as can be observed in adsorption state I.

**Identification of a Real System Catalyst Configuration**. We require validation to identify a catalyst configuration representing a real system. Thus, we calculated the spectrum of the wavenumber for each catalyst configuration by estimating weighted Gaussians for each adsorption state (middle diagram in Fig. 3a) and compared the results with experimental IR spectra. We considered the peak broadening of each vibrational state using the actual detector resolution, Doppler effect, and collisions. The collision contributions consider the gas atmosphere under experimental conditions, surrounding catalyst atoms, and neighbouring adsorbed N$_2$ molecules. The coverage at 25 °C and 6 kPa was used as a weighting factor. The contribution of each effect is shown in the top diagram of Fig. 3a. The spectrum of the wavenumber of the catalyst configuration was obtained by summing the Gaussians of each adsorption configuration (Supplementary Note 2).

We identified regions with higher and lower wavenumbers using different distributions. In general, the shape of a distribution follows a single Gaussian. Lower coverage or higher broadening results in lower peaks; higher broadening also leads to a wider distribution. Because the simulated spectrum represents the sum of the Gaussians, the height of the spectrum increases with the number of N$_2$ molecules with wavenumbers close to a specific value. In our simulated spectra (see, for example, the bottom diagram in Fig. 3a), the broadening of the high wavenumber region can simply be represented by the collisions with the Ru site and adjacent N$_2$ molecules, as the Doppler and gas atmosphere effects are negligible. The asymmetry observed in this case is due to the wide distribution of adsorption states with wavenumbers between 2,000 and 2,100 cm$^{-1}$. The relatively low coverage within this wavenumber interval (see histograms in Fig. 2a) results in Gaussians with low peaks. The high number of neighbouring collisions in the low-frequency region causes a larger broadening of the calculated distribution compared with that observed in the high-frequency region. Moreover, the distribution curve around the high-frequency peak flattens as the distribution in the low-wavenumber region rises. This effect is consistent with the activation of adsorption sites within the low-frequency region discussed in the previous section.

The simulated spectra can be compared with the IR spectra of catalysts reduced at 500 and 650 °C[46]. We can find 14 and 8 out of the 200 simulated spectra that are statistically similar to the experimental spectra measured at 500 and 650 °C, respectively (Supplementary Note 3). The average of these spectra, together with the experimental spectra, are provided in Fig. 3b. The average spectra (with $x = 0.095(21)$ and $\bar{q}_{\text{Ru},0} = -0.133(19)$ e on average) corresponding to the catalyst reduced at 500 °C showed a peak at a low wavenumber of 1,879 cm$^{-1}$. The average spectra ($x = 0.175(72)$ and $\bar{q}_{Ru,0} = -0.233(71)$ e on average) corresponding to the catalyst reduced at 650 °C showed a peak at a low wavenumber of 1,852 cm$^{-1}$. Thus, we can reproduce the measured values of 1,883 and 1,844 cm$^{-1}$, whereas all identified peaks correspond to frequencies with stretching vibrational modes. We further calculated the relative peak intensity by dividing the intensity of peaks at high wavenumbers by the intensity of peaks at low wavenumbers. The intensities were determined relative to the baselines of the spectra given by the thin solid lines in Fig. 3b. The simulated and experimental relative intensities were 7.4 and 6.8, respectively, for a reduction temperature of 500 °C, and 5.9 and 4.1, respectively, for a reduction temperature of 650 °C. While the simulated and experimental relative intensities show a discrepancy when the reduction temperature is 650 °C, the simulated and experimental values obtained for a reduction temperature of 500 °C are in good agreement. Because the shape of the calculated spectra is similar to that observed in experimental observations, we can provide the experimental observations with a physical background. We note here that our simple bottom-up approach shows a certain discrepancy with the measured broadening, which may originate from the overestimation of the broadening mechanisms considered and other effects associated with the measurement proper that are not included in our calculations.

**Analysis of Catalytic Activity**. In experiments, Ru/La$_{0.5}$Ce$_{0.5}$O$_{1.75-x}$ catalysts show maximum activity when reduced at 650 °C (Fig. 3c)[46]. To elucidate the physical origin of this phenomenon, we calculated the dissociation paths (Fig. 3d) for the representative N$_2$ adsorption states A, C, F, G, H, and I (shown in Fig. 2b) at 350 °C. States A, F, G, and H show wavenumbers close to the experimental values. The activation Gibbs energy for the rate-determining step of dissociation on top of the nanoparticle (A) was 1.94 eV. This value is comparable with the reported energy barrier for N$_2$ dissociation on a Ru slab (Supplementary Table 1). The adsorption state at the interfacial layer on the Ru(10$\bar{1}$1) surface (C) showed a lower activation Gibbs energy for dissociation of 0.90 eV, which could be expected from its lower wavenumber of 2,066 cm$^{-1}$. The activation Gibbs energies for the rate-determining steps of adsorption states G and H were 1.09 and 0.98 eV, respectively. Lower values of 0.65 and 0.64 eV were observed for states F and I, respectively. Support reduction is fundamental to achieve a low dissociation barrier because the sites at the interface become active (Fig. 2c) and stabilise the onset of an SMSI. Moreover, wavenumbers lower than 2,000 cm$^{-1}$ can be observed only when the onset of an SMSI occurs (Fig. 2c), resulting in sites corresponding to the states F, G, H, and I. Among these states, states G and H, with a relatively low wavenumber, show higher activation barriers than states F and I.

We further performed structural analysis to clearly differentiate these adsorption states. We counted the number of Ce/La/O atoms $N_{Ce/La}$ and $N_O$ surrounding the N$_2$ molecule. A cut-off radius of 5 Å was sufficiently large to consider different local structures. The cations surrounding the N$_2$ molecule for states F and I were partially oxidised ($N_{Ce/La} >= N_O$ and $N_O > 0$), whereas the cations surrounding the N$_2$ molecule for states G and H were not ($N_{Ce/La} > 0$ and $N_O = 0$). Catalysts with a higher number of adsorption states F and I had higher catalytic activity. Finally, we sought to determine whether our hypothesis can explain the measured catalytic activity of the catalysts (Fig. 3c). We counted the emergences of states F and I for each configuration considering the N$_2$ coverage at 350 °C and 1 MPa (Table 1). On average, we observed 20% more of these states in catalyst configurations corresponding to a reduction temperature of 650 °C compared with those corresponding to a reduction temperature of 500 °C, thus supporting the higher activity of the catalyst system reduced at 650 °C.

**Conclusions**

This study shows the excellent validity of the proposed approach for investigating heterogeneous catalyst systems in cyberspace for catalysts based on supported nanoparticles showing the onset of an SMSI. In the proposed framework, the modelling approach for the catalyst is more compliant with an artificial mapping of configurations than a completely physics-based approach, that is, guided by the chemical potentials of the components and reducing atmosphere. This mapping was necessary to achieve the adequate and systematic preparation of catalyst models used to explore the adsorption and catalytic properties of molecules with which they interact. The exhaustive calculation of adsorbed molecules on the catalyst was unavoidable. The massive size of our models and extensive surface reconstruction necessary during the long optimisation of the supported nanoparticle systems prohibit the use of first-principles calculations. Even if optimising some adsorption sites via this technique is possible, the results obtained cannot be assumed to be meaningful for such a complex heterogeneous system, as discussed throughout the article. Moreover, we successfully validated the numerical results with experimental data and clarified specific experimental observations for these types of catalysts. Notably, we provided experimental observations for the IR spectra of $N_2$ molecules adsorbed on the studied catalyst with a robust physical background, which may be applicable to other similar catalysis systems.

Achieving the number of calculations required to obtain the dissociation reaction paths similar to that employed for the 15,600 calculated adsorption sites in our work will be challenging. However, we can produce an accurate state-of-the-art approximation of such properties for the heterogeneous catalyst system studied here. We believe that if the distribution of cations wrapped around a nanoparticle can be skilfully controlled by modifying the composition of the nanoparticle to avoid the deactivation of strategic adsorption sites, new catalysts with higher catalytic activity could be designed in cyberspace, synthesised, and characterised experimentally for use in real technical systems.

Our approach can be directly used to investigate other material combinations, including other supports and nanoparticles based on binary, ternary, and multinary alloys. Indiscriminate computationally guided high-throughput experimentation for material discovery may now be possible. Moreover, the many unresolved questions arising from the unique properties and behaviours of the innumerable synthesised complex supported nanoparticle catalysts developed thus far can now be solved using the proposed framework. Thus, we believe that this study is a key first step toward the true digital transformation of investigations on heterogeneous catalysis.

**Methods**

**Model Preparation and Calculation.** We configured the 4 × 4 × 3 $La_{0.5}Ce_{0.5}O_{1.75}$ bulk models prior to optimisation using a well-known method for binary metal-oxides with vacancies[51] by accounting for a Warren–Cowley[52] parameter close to zero for the SS cationic phase and a lattice parameter of 7.93 Å corresponding to the DFT value calculated from the Materials Project Database[53]. Energy minimisation was applied to the cell size and atom positions. After optimisation, we added a 30 Å vacuum layer perpendicular to the (111) surface to create a slab and then optimised the atomic positions. We conducted a molecular dynamics (MD) simulation on the canonical ensemble to equilibrate the anions of the slab at 650 °C, corresponding to the reduction temperature reported in the literature[46] (Supplementary Fig. 9). The configuration of the slab trajectory obtained after equilibration was used for further calculations, and the upper cationic layer was reduced by different degrees by randomly removing oxygen anions.

We optimised a $Ru_{238}$ nanoparticle with an HCP structure and extracted a $Ru_{143}$ half-nanoparticle with a large (0001) facet downside. We tested different orientations of the nanoparticle on the slab and chose the most energetically favourable one (Supplementary Methods 1). Different supported nanoparticle configurations were created using this nanoparticle orientation. We then placed random cations from the cationic surface at the lower nanoparticle layer at various degrees and optimised the structure. Individual $N_2$ molecules were optimised and placed at all Ru on-top sites at a distance of 1.6 Å. The N atoms were oriented so that they pointed parallel to the direction from the Ru site to the middle of the nanoparticle (Supplementary Fig. 1). After optimisation, molecules that did not change the Ru site and with a distance to this site of less than 2.3 Å, bond length less than 2.5 Å, and wavenumber higher than 1,200 $cm^{-1}$ were used for further evaluation. These criteria ensured that the molecules to be analysed were end-on adsorbed on Ru on-top sites without repetition.

To approximate configurations for two different dissociated states per adsorption site, we placed N atoms at the neighbouring site of Ru in the $\mathbb{R}^2$ projection of the nanoparticle surface layer and optimised them. We calculated saddle-point configurations using two configurations estimated as the linear interpolation with a factor of 0.5 between the initial state and two approximated final states. Ten saddle points were calculated for each configuration. We minimised a second configuration downhill by extrapolating it 105% away from the initial state toward the saddle point and conducting a final optimisation to find the nearest local minima. We then calculated the minimum energy path and disregarded reaction paths that were repeated or inconsistent. We considered non-consistent paths as those with energies higher than the calculated transition state, energy barriers higher than 4 eV, a non-dissociated final state, or a poorly converged transition state showing more than one imaginary frequency.

**Simulation Methods.** All simulations were conducted using the commercial software UNNP Preferred Potential version 1.0.0 within the Matlantis™ distribution[45]. Although Matlantis™ includes numerous features, we used the freely available ASE[54] tools for all calculations. All all-atom energy optimisations were performed using the FIRE algorithm[55] with a force threshold of 0.001 eV/Å unless otherwise stated. The MD simulation was performed with a timestep of 0.5 fs. The Nosé–Hoover[56] thermostat, as implemented in the FLARE library[57], was used with a damping factor $Q$ corresponding to 100 timesteps[58]. A simulation time of 100 ps was sufficient to achieve equilibration. The Dimer method[59] was used to search for saddle points. We performed downhill optimisation using the Broyden–Fletcher–Goldfarb–Shanno line search method[54]. Only atoms within a cut-off radius of 3.0 Å relative to the N atoms were allowed to move. The same was true for the NEB routine[60], which searched for the minimum energy path with a force threshold of 0.05 eV/Å. The NEB routine was performed by separately calculating segments of three images and adding in-between configurations along the path, resulting in nine images after two iterations, including the initial, transition, and final states. The NEB calculation was repeated for segments with a local maximum by climbing the images[61] with a force threshold of 0.005 eV/Å. Configurations corresponding to the local minimum within the path were further all-atom optimised. A spring constant of 0.1 eV/Å was used for all NEB calculations. The UNNP is validated in Supplementary Methods 2.

**Calculation of thermophysical properties.** We calculated the coverage $\theta$ assuming Langmuir adsorption as $\theta = K_{ads} \cdot p/(1 + K_{ads} \cdot p)$ with the equilibrium constant $K_{ads}$ calculated from $\ln(p^{\ominus} \cdot K_{ads}) = -\Delta G(T,p)/(k_B T)$. Here, $p$ denotes the pressure, $p^{\ominus}$ denotes the standard pressure, $T$ refers to the temperature, and $k_B$ is the Boltzmann constant. The adsorption Gibbs energy $\Delta G(T,p)$ for a single $N_2$ molecule was calculated by assuming the ideal gas limit, and the Gibbs energy for the adsorbed molecule was approximated using the Helmholtz energy from the harmonic limit; here, both models were used as implemented in ASE[54].

To calculate the wavenumber spectrum (Supplementary Fig. 10), we determined the Doppler broadening $\delta\tilde{\nu}_D$ as follows[62]:

$$\delta\tilde{\nu}_D = \frac{2\nu}{c}\left(\frac{2 \cdot k_B T \ln(2)}{m_{N_2}}\right)^{\frac{1}{2}},$$

where $\nu$ denotes the wavenumber, $c$ is the speed of light in vacuum, and $m_{N_2}$ is the mass of the $N_2$ molecule. The broadening due to collisions was given by[62]:

$$\delta\tilde{\nu}_{C,i} = \frac{\pi r^2}{2c}\left(\frac{8 k_B T}{\pi \mu_{N_2-i}}\right)^{\frac{1}{2}} \cdot \rho_i,$$

where the reduced mass $\mu_{A-B} = m_A m_B/(m_A + m_B)$ and $\rho_i$ is the number density of component $i$ corresponding to the ideal gas density for collisions with the $N_2$ atmosphere or the number of atoms/molecules within the volume of the sphere with a radius equal to the collision cross-section $\pi r^2 = 3.4 \text{ nm}^2$ for $N_2$ molecules[62]. The density was weighted using the coverage $\theta$ by accounting for possible adjacent adsorbed $N_2$ molecules. Finally, we considered the detector resolution by applying the Sparrow rule[63] as follows:

$$\delta\tilde{\nu}_{\text{Res}} = \sqrt{2\ln(2)} \cdot \text{FWHM},$$

where the full width at half maximum (FWHM) for each Gaussian is given by the experimental resolution[46] of 4 cm$^{-1}$.

**Data availability**

The data supporting the findings and statements presented in this study can be found in this article and Supplementary Information. Other data sources can be obtained from the authors upon reasonable request.

# Figures

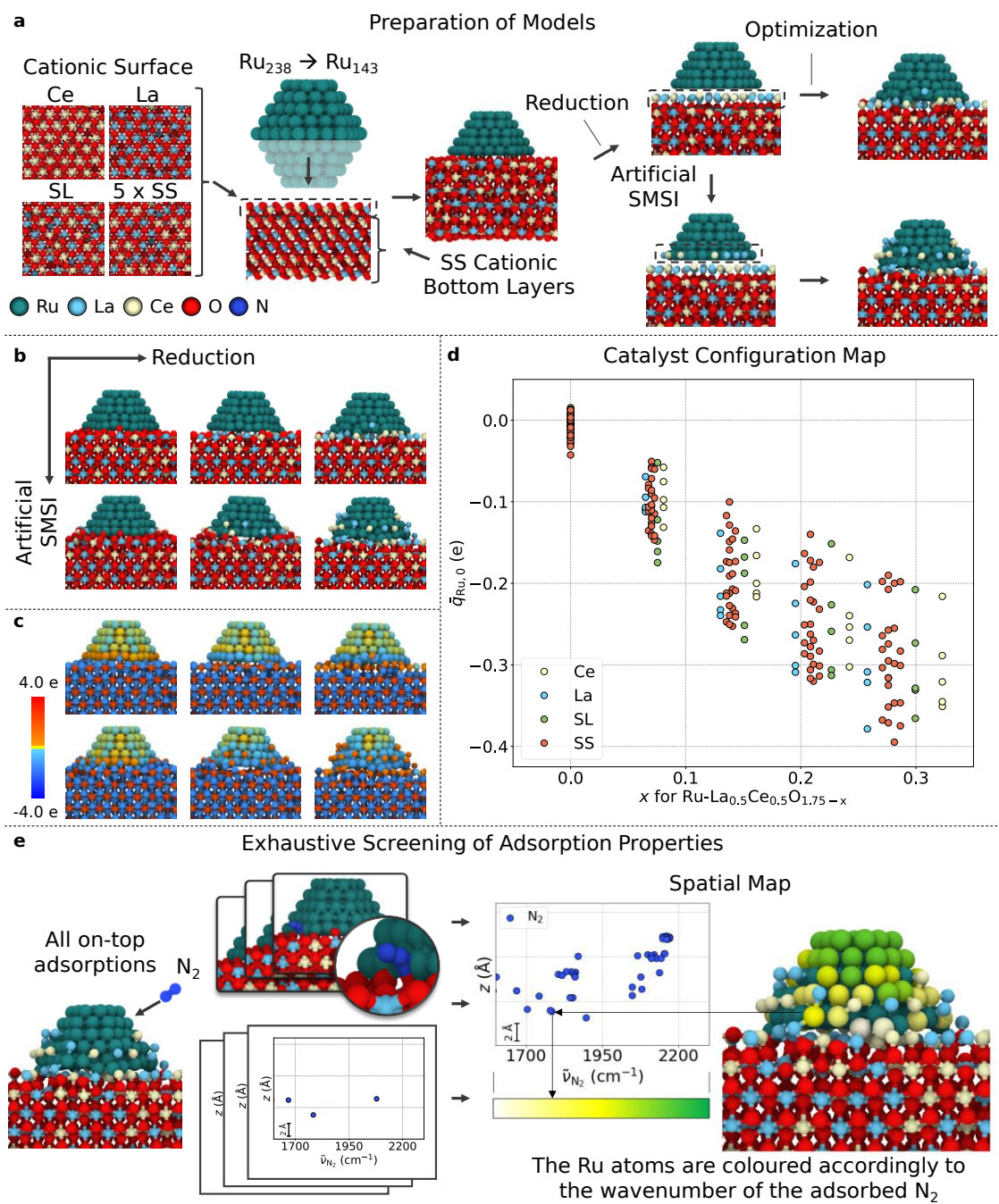

**Fig. 1. Strategy for modelling and simulating N$_2$ dissociation on Ru/La$_{0.5}$Ce$_{0.5}$O$_{1.75-x}$ catalysts. a,** Automated preparation of the models. **b,** Optimised catalyst configurations for three reduction degrees of the cationic surface and two artificial SMSIs. **c,** Charge distribution for the configurations in (**b**). **d**, Catalyst configuration map spread by the average charge $q_{Ru,0}$ of Ru atoms on the nanoparticle surface layer and reduction degree $x$ for all 200 catalyst configurations analysed in this study ($x$ corresponds to the composition of the 4 × 4 × 3 La$_{0.5}$Ce$_{0.5}$O$_{1.75-x}$ slab). **e**, Screening method used to calculate

the N$_2$ properties and create the spatial maps. The N$_2$ wavenumber $\tilde{\nu}_{N_2}$ for each N$_2$ adsorption state was plotted over the position of the Ru site for each configuration. The Ru sites were then coloured according to the value of the wavenumber. The coordinate z is given arbitrarily for the Ru sites.

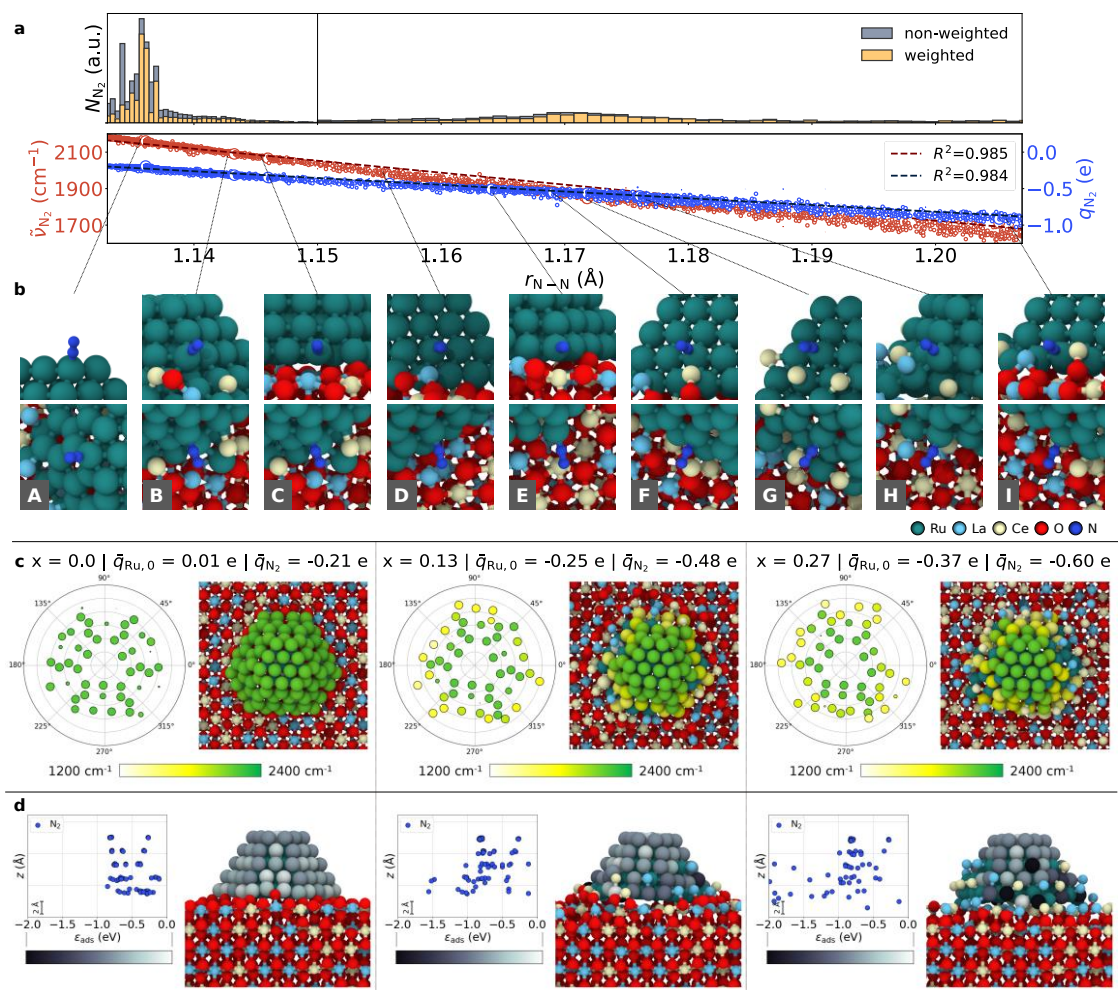

**Fig. 2. Simulated N$_2$ adsorption on different Ru/La$_{0.5}$Ce$_{0.5}$O$_{1.75-x}$ catalyst configurations. a**, **Top:** Weighted and non-weighted histograms of the N–N bond length $r_{N-N}$ distribution. A total of 150 bins for $r_{N-N}$ > 1.15 Å and 50 bins for $r_{N-N}$ < 1.15 Å were arbitrarily chosen. The N$_2$ coverage $\theta$ at 25 °C and 6 kPa was used as the weighting factor. **Bottom:** Simulated wavenumbers $\tilde{\nu}_{N_2}$ and charges $q_{N2}$ of single adsorbed N$_2$ molecules for all adsorption sites within the depicted range. **b**, Representative adsorption sites. **c**, Calculated spatial maps of the N$_2$ wavenumber, which were built as described in Fig. 1e. Here, the top view and polar coordinates of the catalyst were used. The diameter of the markers is given according to the N$_2$ coverage $\theta$ at 25 °C and 6 kPa. **d**, Spatial maps of the N$_2$ adsorption energy $\varepsilon_{ads}$, which were built as described in Fig. 1e.

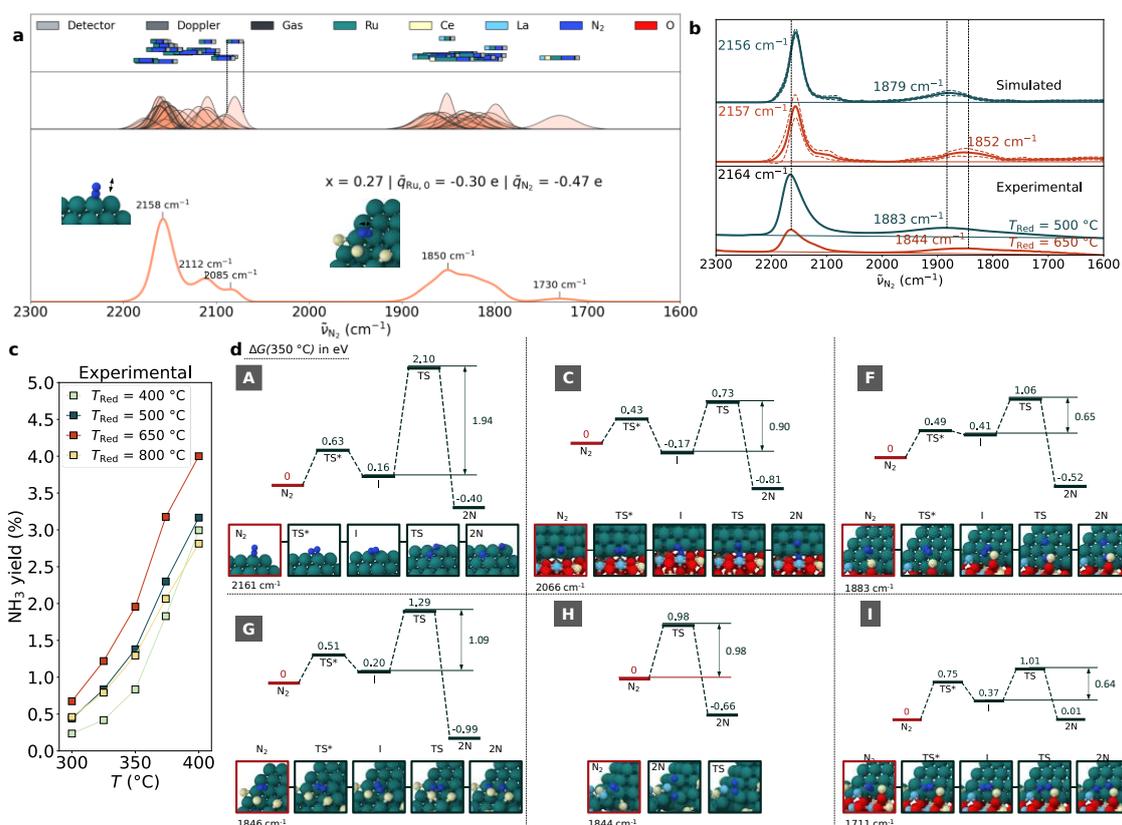

**Fig. 3. Simulated catalytic properties for N$_2$ dissociation on different Ru/La$_{0.5}$Ce$_{0.5}$O$_{1.75-x}$ catalyst configurations. a,** Example of a simulated N$_2$ wavenumber distribution. **Top:** Calculated spectral line widths coloured according to the calculated contributions: broadening due to the detector resolution, Doppler effect, and collisions with the N$_2$ gas atmosphere and neighbouring Ru/Ce/La/O and adsorbed N$_2$. **Middle:** Weighted Gaussians (weighting factor: N$_2$ coverage $\theta$). **Bottom:** Wavenumber distribution obtained from the sum of the Gaussians. The values were obtained at 25 °C and 6 kPa. **b,** Calculated average N$_2$ wavenumber distributions and measured IR spectra[59]. Averaging was performed over the distributions that are statistically similar to the experimental IR spectra.[59] The dashed lines indicate the standard error of averaging with a 95% confidence interval. The thin solid lines are the baselines of the spectra. **c,** Measured NH$_3$ yield (in %) as a function of the reduction temperature of the catalysts[46]. **d,** Calculated N$_2$ dissociation paths for the adsorption states A, C, F, G, H, and I shown in Fig. 2b. For each path, the label N$_2$ describes the initial state, TS* and TS indicate the transition states, I indicates the intermediate state, and 2N indicates the final dissociation state. The energy values indicated correspond to the Gibbs energy difference $\Delta G$(350 °C) relative to the initial state. The energies for the rate-determining step are also provided. All experimental data were taken from the literature[46].

**Table**

**Table 1. Average number of adsorption states for the simulated configurations assigned to real catalysts.**

| States | A | C | G | H | F | I |
|---|---|---|---|---|---|---|
| ΔG(350 °C) (eV) | 1.94 | 0.90 | 1.09 | 0.98 | 0.65 | 0.64 |
| $\overline{N}_{N_2}(T_{\text{red}} = 500\ °C)$ | 12.14 | 0.06 | 2.78 | | 2.15 | |
| $\overline{N}_{N_2}(T_{\text{red}} = 650\ °C)$ | 12.26 | 0.50 | 1.69 | | 2.56 | |

ΔG(350 °C) is the activation barrier at 350 °C. $\overline{N}_{N_2}(T_{\text{red}})$ is the average weighted number of adsorption states corresponding to the states A, C, G, H, F, and I (Fig. 1d) for the configurations assigned to the catalyst at a reduction temperature of $T_{\text{red}}$. The $N_2$ coverage θ(T = 350 °C and 1 MPa) was used as the weighting factor.

**Acknowledgments**

This work was supported by the Alexander von Humboldt Stiftung and by JSPS KAKENHI Grant Numbers JP21F30701 and JP20H05623. The authors are especially grateful to the participants of the Data-Driven AI Laboratory of the Research Initiative for Supra Materials at Shinshu University for their comments and suggestions during the whole project. All images of the atomic models within the article were rendered using OVITO.[64]

**Author contributions**

M.K., K.N., and K.S. proposed and supervised the project. G.V.H. wrote the article, performed the modelling work, designed the numerical framework, and conducted the calculations and evaluation based on the universal neural network potential. K.H. contributed to the statistical evaluation of the data. All authors contributed to the discussion of the results and article preparation.

**Competing interests**

The authors declare no competing interests.

# Supplementary Information

# Cyber Catalysis: N$_2$ Dissociation over Ruthenium Catalyst with Strong Metal-Support Interaction

Gerardo Valadez Huerta*,[1], Kaoru Hisama[1], Katsutoshi Sato[2], Katsutoshi Nagaoka[2], Michihisa Koyama*,[1,3]

[1]Research Initiative for Supra Materials, Shinshu University, Nagano 380-8553, Japan. [2]Department of Chemical Systems Engineering, Nagoya University, Nagoya 464-8601, Japan. [3]Open Innovation Institute, Kyoto University, Kyoto 606-8501, Japan

*Correspondence to: valadez@shinshu-u.ac.jp, koyama_michihisa@shinshu-u.ac.jp

**Table of Contents**

**Supplementary Notes**

*Supplementary Note 1. Catalyst Model Preparation*

Depictions of different views of all 25 initial configurations of the catalyst with a solid solution (SS) cationic surface, their optimised structures, and the resulting charge distribution can be found in the 'Supplementary Data' file (directory paths: Figures/Catalyst/Initial; Figures/Catalyst/Initial/Optimization). The depictions of other configurations may be obtained from the authors upon reasonable request. The optimised $Ru_{238}$ HCP nanoparticle structure and frames for the configuration labelled 165-La-1-1-xO25-1Cat75 are given in Supplementary Fig. 6. The directories containing the figures are labelled in the following manner. The number '165' refers to the rotational angle $\beta$ of the nanoparticle relative to the optimised structure and is followed by the cationic surface composition of the support (here, 'La'). The first number in '1-1' refers to the configuration number, and the second refers to the number of reduced cationic layers. 'xO25' refers to the percentage of oxygen remaining in the layers after reduction (e.g. '25' corresponds to a 75% reduction in oxygen). The numbers '1' and '75' in '1Cat75' refer to the number of nanoparticle layers wrapped by cations and percentage of occupied Ru atoms at the lowest nanoparticle layer, respectively. All values are relative to the initial configuration.

Next, we provide an explanation of the catalyst configuration map in Fig. 1d in the article. A non-reduced surface results in a slab composition of $La_{0.5}Ce_{0.5}O_{1.75}$ with $Ce^{4+}$ and $La^{3+}$ cations. Using our method, we can prepare slabs with a composition of $La_{0.5}Ce_{0.5}O_{1.5}$ (or slightly lower oxygen concentrations) with $Ce^{3+}$ and $La^{3+}$ cations. Beyond these limits, further reduction of the cationic layers in the slab may not have any practical significance.

*Supplementary Note 2. Simulated Spectra for Each Configuration*

The dipole moment is not calculated by the UNNP version used in the article and, thus, it is not considered by the wavenumber calculation. We repeated the calculations for the wavenumber spectra by accounting for the $^{15}$N isotope. Thus, we adjusted the N mass to 15 u and compared the results with the experimental data. The spectra depicted in Supplementary Fig. 7 correspond to that shown in Fig. 3a in the article. We can reproduce the red-shift of the peaks of $^{14}$N relative to the peaks of $^{15}$N by a factor of $(m(^{14}N)/m(^{15}N))^{1/2}$ as physically expected and experimentally confirmed. All spectra are provided in the 'Supplementary Data' file (directory paths: $^{14}$N: Figures/Catalytic/m14; $^{15}$N: Figures/Catalytic/m15). The values of $x$, $\bar{q}_{Ru,0}$, and $\bar{q}_{N_2}$ are summarised in the 'FreqHist_Legende_T25C-p6kPa.xlsx' file within each directory.

*Supplementary Note 3. Similarity Analysis of the Simulated and Measured Spectra*

We first show that the simulated and measured curves are statistically comparable. We conducted a Kolmogorov–Smirnov (KS) test for the two-sided null hypothesis $H_{01}$: $\rho_i$ is similar to $\rho_{Exp}$ vs. $H_{11}$: $\rho_i$ is not similar to $\rho_{Exp}$, where $\rho_i$ and $\rho_{Exp}$ are the calculated and measured vibrational frequency distributions, respectively (Fig. 3b in the article), at a significance level of $\alpha$ = 0.05. We calculated the wavenumber distributions using the *D* value, which is given by:

$$D_i = \max |F_i(x) - F_{Exp}(x)|,$$

where $F(x)$ is the normalised cumulative distribution function. We calculated $F(x)$ by numerically integrating $\rho_i$ and $\rho_{Exp}$ using the trapezoidal method. The hypothesis cannot be rejected when the values are lower than the critical value

$$D_{crit} \approx \sqrt{\frac{-0.5 \cdot \ln\left(\frac{\alpha}{2}\right)}{n}} = 0.1537,$$

with $n = 78$ as the number of calculated adsorption states. We found 87 and 59 distributions that were comparable with the IR spectra obtained for a catalyst reduced at 500 and 650 °C, respectively, for which the hypothesis cannot be rejected.

After statistically confirming the similarity between the spectra, we searched for a suitable similarity factor. Recent literature[1] indicates that the Pearson correlation coefficient

$$S_{PCC,i} = \frac{\sum_{j=1}^{n}\left(\rho(\tilde{\nu}_{N_2,j}) - \bar{\rho}_i\right) \cdot \left(\rho_{Exp}(x_j) - \bar{\rho}_{Exp}\right)}{\sum_{j=1}^{n}\left(\rho(\tilde{\nu}_{N_2,j}) - \bar{\rho}_i\right)^2 \sum_{j=1}^{n}\left(\rho_{Exp}(x_j) - \bar{\rho}_{Exp}\right)^2}$$

is the most suitable similarity factor for comparing the experimental and computed IR spectra. Scores of 0.684 or higher are considered to reflect high similarity. Note that a score of 1 is only given for an equal distribution. We first conducted a cross-check between the two experimental spectra depicted in Fig. 3b in the article and obtained a similarity factor of 0.98. Therefore, $S_{PCC}$ is not a suitable similarity factor for comparing spectra in our study because this factor suggests that both experimental distributions are almost the same. Moreover, we did not observe any trend between the simulated distributions with high $S_{PCC}$ factors and the experimental IR spectra, particularly in terms of wavenumber peaks. These observations agree with the literature[1], which states that established distance-based methods, such as the PCC, only weakly reflect any translation in the functions. Therefore, they may not be applicable for our case, where the distributions show a large broadening and relatively small differences between peaks. In addition, if we use a more traditional method of comparing and matching peaks[1], the limited number of adsorption sites may result in a distribution with different peaks but similar intensities (see Supplementary Fig. 8a) owing to the size of the catalyst model. Thus, simply comparing and matching peaks is insufficient. To overcome this problem, we divided each of the experimental spectra (depicted in Fig. 3b in the article) into two distributions A, ranging from 1600 to 2000 cm$^{-1}$, and B, ranging from 2000 to 2400 cm$^{-1}$. Considering the baselines, we calculated the $k$-th percentiles $P(k_{Exp}^{A,lo}) = \tilde{v}_{N_2,Exp}^{lo}$ and $P(k_{Exp}^{B,hi}) = \tilde{v}_{N_2,Exp}^{hi}$ corresponding to the peaks (see Supplementary Table 2). Next, we calculated the percentiles $\tilde{v}_{N_2}(k_{Exp}^{A,lo})$ and $\tilde{v}_{N_2}(k_{Exp}^{A,lo})$ for the simulated distributions and performed the comparison. Finally, we calculated the root-mean-square deviation and combined this result with the $D$ value obtained from the KS test to evaluate the similarity of the shapes of the distributions:

$$S = \left[\left(1 + \sqrt{\frac{\left|\tilde{v}_{N_2}(k_{Exp}^{A,lo}) - \tilde{v}_{N_2,Exp}^{lo}\right|^2 + \left|\tilde{v}_{N_2}(k_{Exp}^{B,hi}) - \tilde{v}_{N_2,Exp}^{hi}\right|^2}{2}}\right) \cdot D\right]^{-1}.$$

The higher the $S$, the higher the degree of similarity between the compared spectra. We cross-checked our results by comparing the experimental curves with each other and obtained a low $S$ of 0.19. The calculated distributions considered similar in the article showed $S$ values higher than 1.1, which means the so-defined similarity factor allows us to not only differentiate both experimental distributions but also find distributions with high similarity to each of the IR spectra considered. The wavenumber distribution with the highest similarity to the experimental IR spectrum of the catalyst reduced at 500 °C is depicted in Supplementary Fig. 8b. All results on the simulated distributions can be found in the 'FreqHist_T25C_p6kPa.xlsx' file within the 'Doc' directory of the 'Supplementary Data' file.

*Supplementary Note 4. Structural Analysis of Adsorption States*

To analyse the structures of the adsorption states (Fig. 3d in the article), we first searched for variables to determine the local structure of the adsorption state. We used the centre of mass of the $N_2$ molecule, defined a cut-off radius, and counted the number of cations $N_{Ce/La}$ and oxygen atoms $N_O$ for the adsorption states given in Fig. 3d. The results for different cut-off radii are summarised in Supplementary Table 3.

Using a cut-off radius of 5 Å or higher, we found the following conditions to count the number of adsorption states in each configuration:

$$A: \tilde{v}_{N_2} \geq 2100 \text{ cm}^{-1}, N_{Ce/La} = N_O = 0,$$

$$C: 2100 \text{ cm}^{-1} \geq \tilde{v}_{N_2} \geq 2000 \text{ cm}^{-1}, N_{Ce/La} > 0, N_O > 0,$$

$$G, H: \tilde{v}_{N_2} \leq 2000 \text{ cm}^{-1}, N_{Ce/La} > 0, N_O = 0,$$

$$F, I: \tilde{v}_{N_2} \leq 2000 \text{ cm}^{-1}, N_{Ce/La} \geq N_O, N_O > 0.$$

**Supplementary Methods**

*1. Energetic Analysis of the Ru$_{143}$ Half-Nanoparticle Orientation on the Support*

The nanoparticle is rotated by an angle of $\beta$ = 165° for the 200 models used in the study. The calculated values of the interfacial energy $\varepsilon_{\text{inter}}$ between the nanoparticle and support for 20 catalyst models and 5 nanoparticle orientations are summarised in Supplementary Table 4. The chosen angle range was sufficiently wide to achieve accurate analysis owing to the symmetry of the support. The interfacial energy was calculated as follows:

$$\varepsilon_{\text{inter}} = \varepsilon_{\text{Ru238}-\text{La}_{0.5}\text{Ce}_{0.5}\text{O}_{1.75}} - \frac{N_{\text{Ru}_{143}}}{N_{\text{Ru}_{238}}}\varepsilon_{\text{Ru238}} - \varepsilon_{\text{La}_{0.5}\text{Ce}_{0.5}\text{O}_{1.75}},$$

where $\varepsilon_{\text{Ru238}-\text{La}_{0.5}\text{Ce}_{0.5}\text{O}_{1.75}}$ is the potential energy of the optimised catalyst structure with a non-reduced support, $N$ is the number of atoms for the (half) Ru nanoparticle, and $\varepsilon_{\text{La}_{0.5}\text{Ce}_{0.5}\text{O}_{1.75}}$ is the potential energy of the non-reduced slab after the MD simulation. In general, the nanoparticle rotated by $\beta$ = 165° is the most energetically favourable. The orientation for this particular case can be taken from the top view of Supplementary Fig. 6.

*2. Validation of the UNNP to Describe the Ru Nanoparticle, La$_{0.5}$Ce$_{0.5}$O$_{1.75}$ Slab, and N$_2$ Adsorption on Ru Slabs*

We validated the UNNP to describe Ru nanoparticles in previous work[2]. Furthermore, we optimised the La$_{0.5}$Ce$_{0.5}$O$_{1.75}$ configuration with the fluorite structure provided in the Materials Project Database[3] using the UNNP. We calculated a lattice constant of 7.980 Å, which deviates from the DFT value of 7.939 Å by only 0.52%. The calculated bulk modulus (131 GPa) was also in good agreement with the value recorded in the Materials Project Database (130 GPa).

To validate the accuracy of the UNNP for calculating the adsorption of N$_2$ on Ru, we summarised all property values calculated using the UNNP, as well as DFT and experimental values obtained from previous studies, in Supplementary Table 5.
.

**Supplementary Figures**

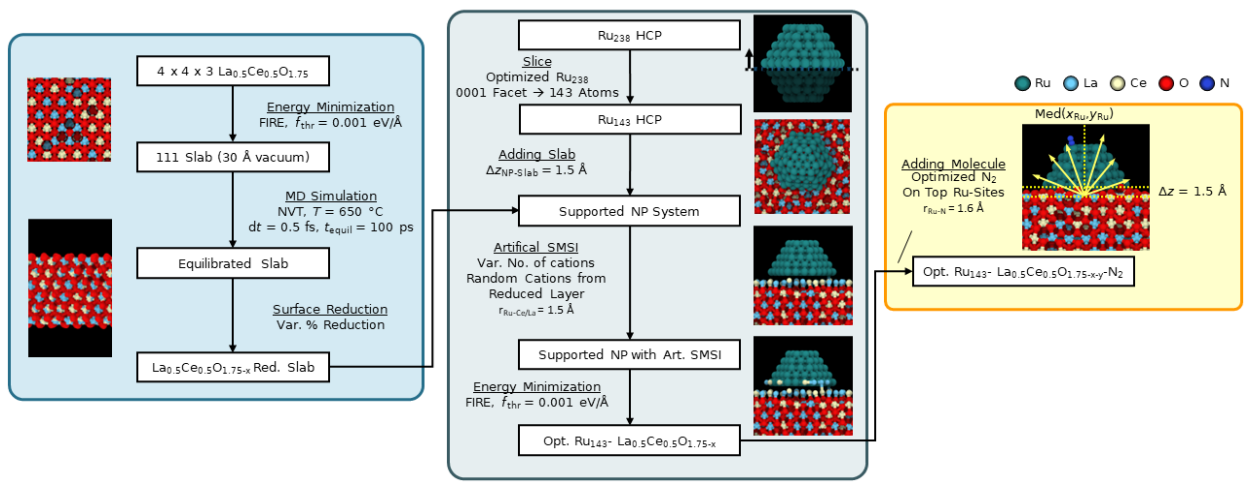

**Supplementary Fig. 1** Methods applied to model the catalyst systems.

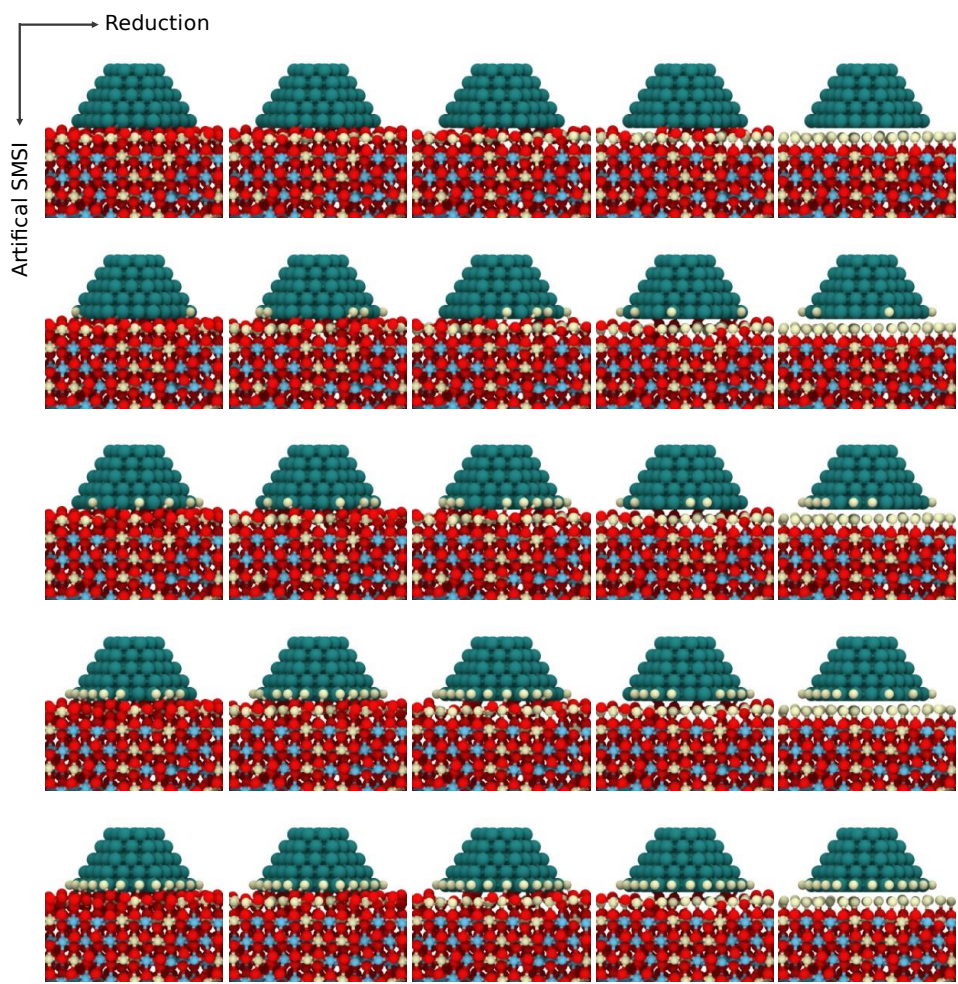

**Supplementary Fig. 2.** Example of 25 initial configurations obtained by varying the reduction of the Ce cationic surface and artificial SMSI. The cationic surface was reduced as follows from left to right: no removal of oxygen anions, removal of 1/4 of the oxygen anions, and so on, until the complete removal of oxygen anions on the cationic surface was achieved. The artificial SMSI follows a similar rule from top to bottom, and the number of cations attached to the lowest nanoparticle layer is varied.

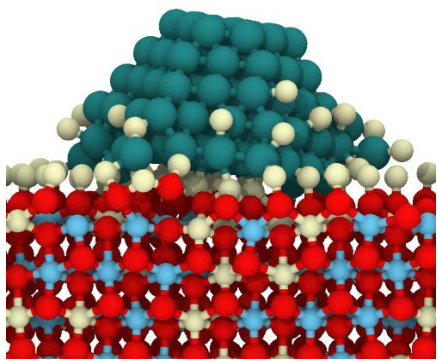

**Supplementary Fig. 3.** Optimised structure of the configuration with the highest SMSI and reduction degree.

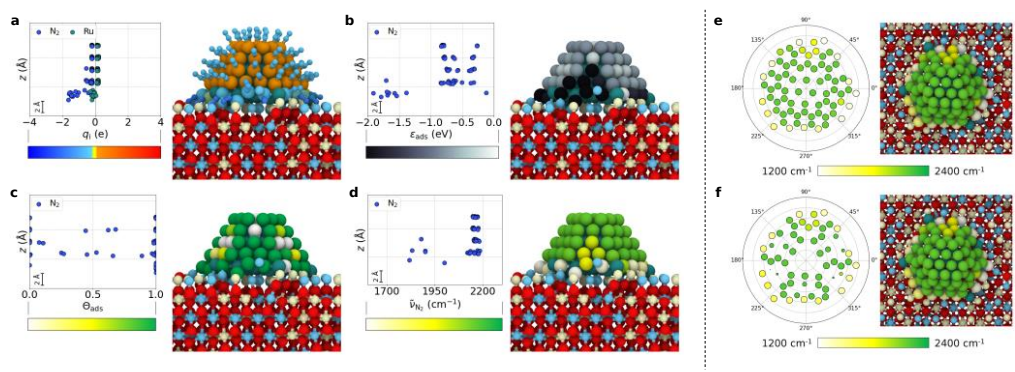

**Supplementary Fig. 4. a–d,** Spatial maps of the (**a**) charge $q_i$ for individual adsorbed $N_2$ molecules and the Ru site, (**b**) adsorption energy $\varepsilon_{ads}$, (**c**) $N_2$ coverage $\theta$ ($T = 25\,°C, p = 6\,kPa$), and (**d**) wavenumber $\nu_{N_2}$. **e, f,** (**e**) Weighted and (**f**) non-weighted spatial maps of the $N_2$ wavenumber in polar coordinates (weighting factor for the marker size: $\theta(T = 25\,°C, p = 6\,kPa)$). The 'Supplementary Data' file (Directory: Figures/Adsorption) includes all figures for the 25 catalyst configurations with an SS cationic surface. Images for other configurations may be obtained from the authors upon reasonable request.

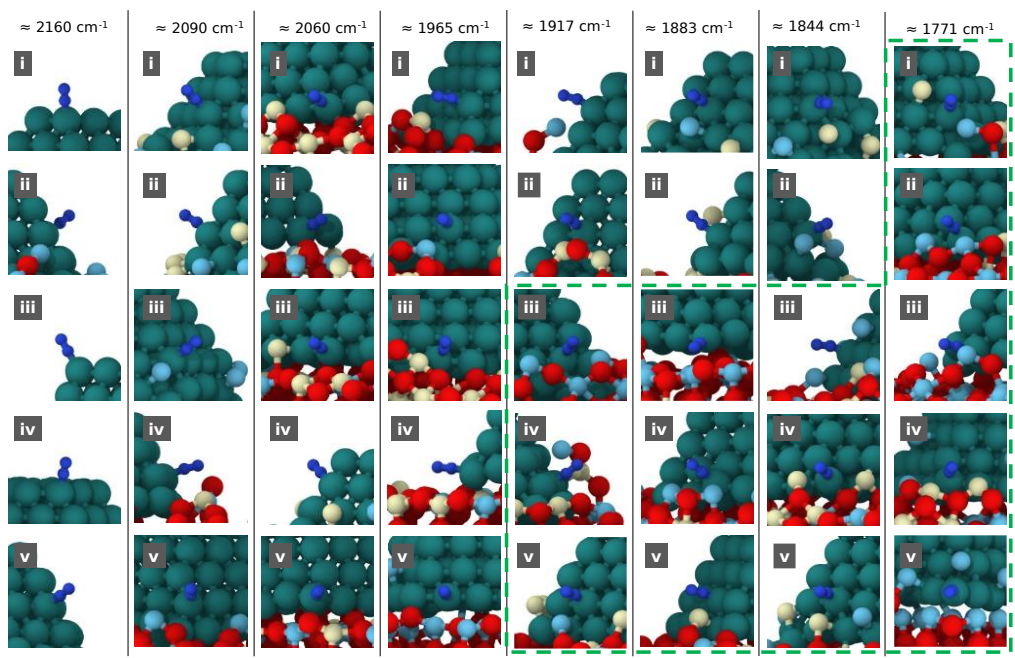

**Supplementary Fig. 5.** Additional adsorption states (compared with Fig. 2a in the article). Green regions mark adsorption states similar to F and I in Fig. 2 in the article that are expected to have a low activation barrier. All adsorption states have a high coverage of ≈1.0 at 25 °C and 6 kPa. The 'AdsorptionSites.xlsx' file in the 'Doc' directory of the 'Supplementary Data' file contains the properties of these and other adsorption states.

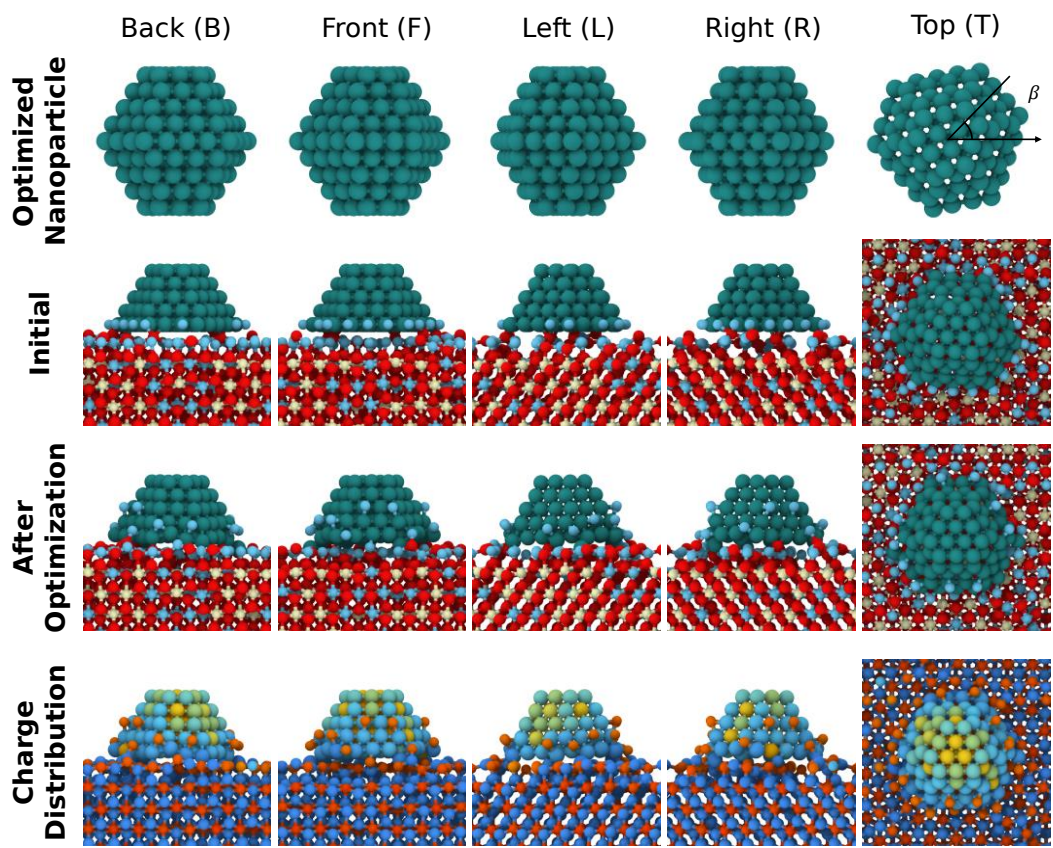

**Supplementary Fig. 6** Optimised $Ru_{238}$ HCP nanoparticle structure including the definition of the angle $β$ and frames for the configuration labelled 165-La-1-1-xO25-1Cat75 (see Supplementary Note 1 for details).

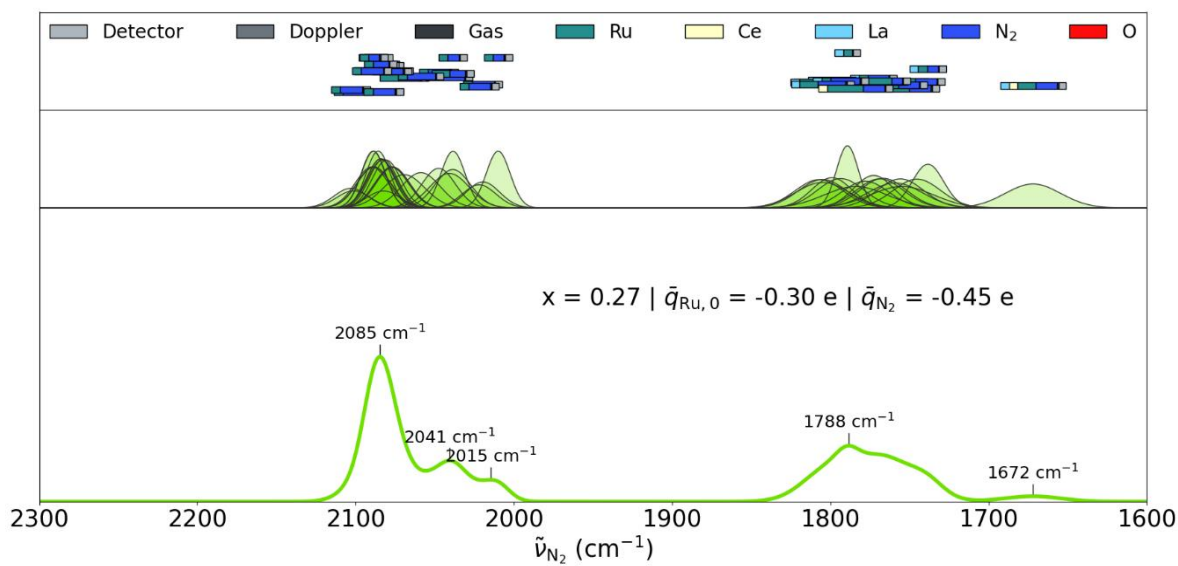

**Supplementary Fig. 7** Wavenumber distribution corresponding to Fig. 3a in the article, but for a $^{15}$N isotope.

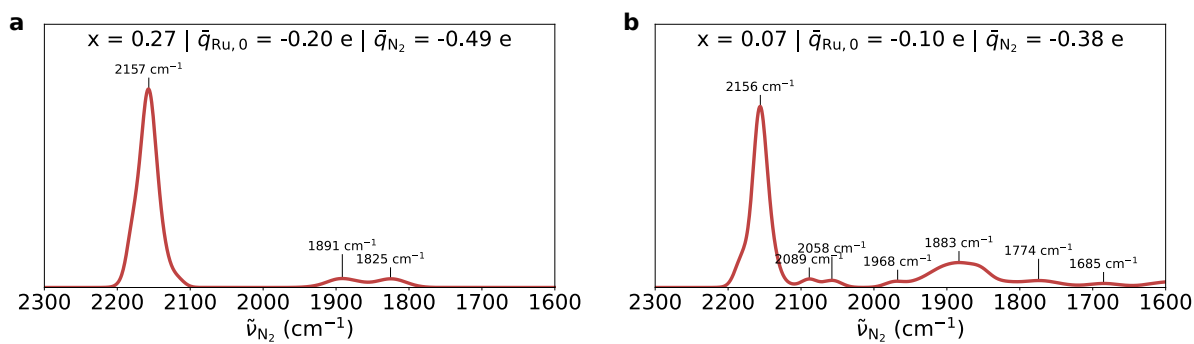

**Supplementary Fig. 8 a,** Example wavenumber distribution for a configuration with a fully reduced SS surface and no artificial SMSI. **b,** Simulated wavenumber distribution with the highest similarity to the experimental IR spectrum of the catalyst reduced at 500 °C.

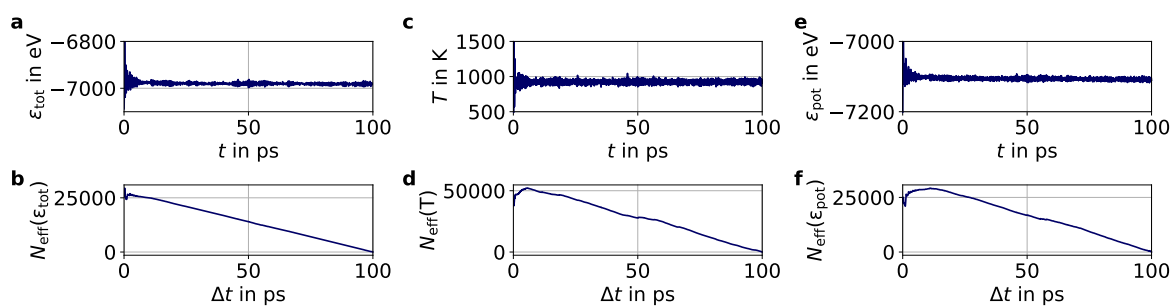

**Supplementary Fig. 9 a, c, e,** Changes in (**a**) total energy $\varepsilon_{\text{tot}}$, (**c**) temperature $T$, and (**e**) potential energy $\varepsilon_{\text{pot}}$ over the simulation time $t$. **b, d, f,** Corresponding number of uncorrelated samples $N_{\text{eff}}$ over the equilibration timespan $\Delta t$. All depictions are given as an example for a slab with a Ce surface and represent all other configurations. The system is considered to reach equilibrium over 100 ps because the number of uncorrelated samples is highest within this timespan and linearly approaches zero afterwards[6].

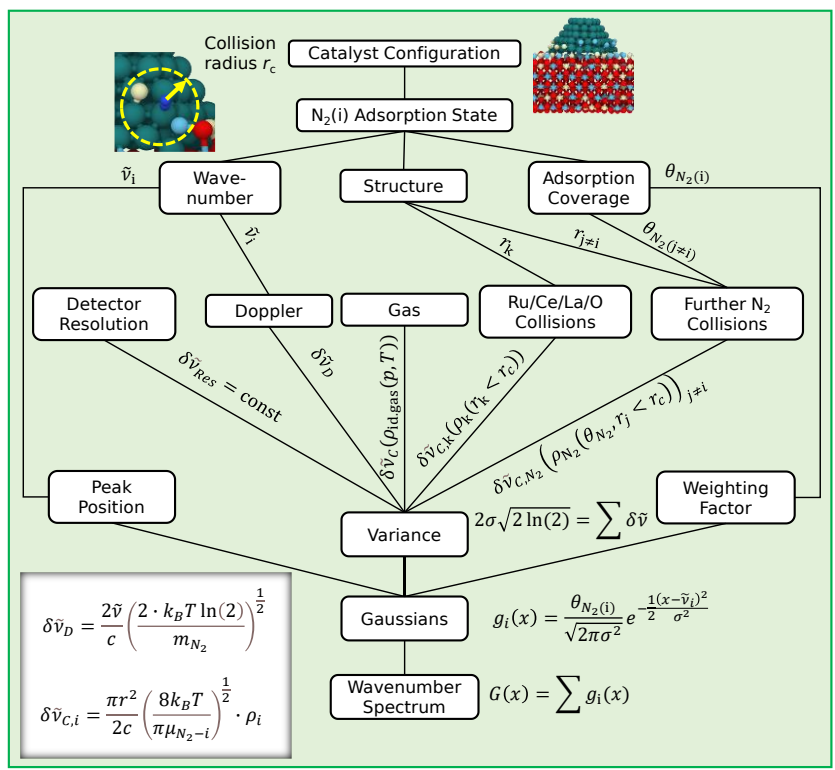

**Supplementary Fig. 10** Methods used to calculate the wavenumber distribution of adsorbed $N_2$ molecules on each catalyst configuration.

## Supplementary Tables

**Supplementary Table 1.** Comparison of the calculated Gibbs energy barriers for the dissociation of a $N_2$ molecule on a Ru slab (on-top) toward two hollow sites (2N) using the UNNP and DFT values. The Gibbs energy is defined as $\Delta G = G(TS)-G(N_2)$, similar to the definition in Fig. 3d in the article. The DFT values correspond to $\Delta G = \Delta G^{\ddagger}+\Delta G_r$ because the values are given for $\Delta G^{\ddagger} = G(TS)-G(2N)$ and $\Delta G_r = G(2N)-G(N_2)$ in the literature[7]. The UNNP calculations were performed for a 2 × 2 × 6 Ru slab (HCP) with a 0001 surface by fixing the four lowest slab layers. The DFT calculations were performed for a 2 × 2 × 2 Ru slab (HCP) by fixing the lowest layer[7].

| $T$ in K | $\Delta G_{UNNP}$ (eV) | $\Delta G_{DFT}$ (eV) |
|---|---|---|
| 100 | 1.56 | 1.66 |
| 300 | 1.60 | 1.70 |
| 700 | 1.73 | 1.80 |
| 1,000 | 1.83 | 1.88 |
| 1,400 | 1.97 | 1.99 |

**Supplementary Table 2** Peaks at low and high wavenumbers in the experimental IR spectra depicted in Fig. 3b in the article and their corresponding location $k$. The IR spectra were divided into two distributions: A (range: 1600–2000 cm$^{-1}$) and B (range: 2000–2400 cm$^{-1}$).

| $T_{Red}$ in °C | $\tilde{\nu}^{hi}_{N_2,Exp}$ (cm$^{-1}$) | $\tilde{\nu}^{lo}_{N_2,Exp}$ (cm$^{-1}$) | $k^{A,lo}_{Exp}$ | $k^{B,hi}_{Exp}$ |
|---|---|---|---|---|
| 500 | 2,164 | 1,883 | 60% | 65% |
| 650 | 2,164 | 1,846 | 60% | 60% |

**Supplementary Table 3** Number of Ce/La/O ions considered in the local environment of the adsorbed nitrogen molecule depending on the chosen cut-off radius.

| States | A | C | F | G | H | I |
|---|---|---|---|---|---|---|
| $\nu_{N_2}$ (cm$^{-1}$) | 2161 | 2066 | 1883 | 1846 | 1844 | 1711 |
| $z$-$z_C$ (Å) | 8.17 | 0.0 | 2.35 | 4.34 | 1.59 | 0.32 |
| $r_{Cut}$ = 2 Å: $N_{Ce/La}$ \| $N_O$ | 0\|0 | 0\|0 | 0\|0 | 0\|0 | 0\|0 | 0\|0 |
| $r_{Cut}$ = 3 Å: $N_{Ce/La}$ \| $N_O$ | 0\|0 | 0\|0 | 1\|0 | 1\|0 | 1\|0 | 2\|0 |
| $r_{Cut}$ = 4 Å: $N_{Ce/La}$ \| $N_O$ | 0\|0 | 1\|2 | 1\|0 | 1\|0 | 1\|0 | 3\|2 |
| $r_{Cut}$ = 5 Å: $N_{Ce/La}$ \| $N_O$ | 0\|0 | 3\|3 | 2\|1 | 2\|0 | 2\|0 | 4\|4 |
| $r_{Cut}$ = 6 Å: $N_{Ce/La}$ \| $N_O$ | 0\|0 | 5\|7 | 3\|2 | 2\|0 | 4\|0 | 6\|6 |

**Supplementary Table 4** Angular variations of the Ru nanoparticle on the La$_{0.5}$Ce$_{0.5}$O$_{1.75-x}$ slab (see Supplementary Fig. 6 for the definition of angle β) and corresponding interfacial energy $\varepsilon_{inter}$ for different catalyst configurations.

|  | Ce-1 | La-1 | SL-1 | SS-1 |
|---|---|---|---|---|
| β (°) | $\varepsilon_{inter}$ (eV) | $\varepsilon_{inter}$ (eV) | $\varepsilon_{inter}$ (eV) | $\varepsilon_{inter}$ (eV) |
| 75 | -133.33 | -132.33 | -133.07 | -126.42 |
| 105 | -116.79 | -135.11 | -137.93 | -131.94 |
| 135 | -133.79 | -129.65 | -132.12 | -127.40 |
| 165 | -140.58 | -137.59 | -137.53 | -132.51 |
| 195 | -133.33 | -132.33 | -133.07 | -126.42 |

**Supplementary Table 5** Various properties of N$_2$ on-top adsorption on $\sqrt{3} \times \sqrt{3} \times 6$ and $2 \times 2 \times 6$ Ru slabs calculated using the UNNP, as well as DFT and experimental values obtained in previous studies. Properties: $r_{i-j}$ = bond length between atoms $i$ and $j$; $b$ = buckling; $\varepsilon_{vib,N_2}$ = vibrational energy; $\nu_{N_2}$ = wavenumber; $\varepsilon_{ads}$ = adsorption energy; $\Delta\varepsilon_{act,top-b-top}$ = diffusion barrier for the on-top/bridge/on-top diffusion path; $\Delta\varepsilon_{act,N_2 \rightarrow 2N}$ = activation barrier dissociation energy for N$_2$ (on-top site) to 2N (closest hollow sites); and $\varepsilon_{def}$ = deformation energy of the slab.

| $\sqrt{3} \times \sqrt{3} \times 6$ ($\theta = 1/3$) | | | | | |
|---|---|---|---|---|---|
| Property | DFT[4] | DFT(PW91)[5] | DFT(RPBE)[5] | UNNP | Exp.[4,5] |
| $r_{N-N}$ (Å) | 1.11 | - | - | 1.13 | 1.10(4) |
| $r_{Ru-N}$ (Å) | 2.00 | - | - | 1.96 | 2.00(5) |
| $b$ (Å) | 0.15 | - | - | 0.00 | 0.00(5) |
| $\varepsilon_{vib,N_2}$ (meV) | - | - | - | 273 | 274 |
| $\nu_{N_2}$ (cm$^{-1}$) | 2239 | - | - | 2202 | 2195 |
| $\varepsilon_{ads}$ (eV) | -0.61 | - | - | -0.43 | -0.44 |
| $2 \times 2 \times 6$ ($\theta = 1/4$) | | | | | |
| Property | DFT[4] | DFT(PW91)[5] | DFT(RPBE)[5] | UNNP | Exp.[4,5] |
| $r_{N-N}$ (Å) |  | 1.14 |  | 1.13 | - |
| $r_{Ru-N}$ (Å) |  | 2.02 |  | 1.95 | - |
| $\nu_{N_2}$ (cm$^{-1}$) |  | 2229 |  | 2202 | - |
| $\varepsilon_{ads}$ (eV) | -0.47 | -0.74 | -0.40 | -0.44 | -0.31 |
| $\Delta\varepsilon_{act,top-b-top}$ (eV) |  | 0.45 | 0.51 | 0.56 | - |
| $\Delta\varepsilon_{act,N_2 \rightarrow 2N}$ (eV) | 1.92 | - | - | 1.63 | - |
| $\varepsilon_{def}$ (eV) |  | 0.08 | 0.07 | 0.07 | - |

**Supplementary References**

# Auto-Summary for Heterogeneous Catalysis

# Cyber Catalysis: N$_2$ Dissociation over Ruthenium Catalyst with Strong Metal-Support Interaction

Gerardo Valadez Huerta*,[1], Kaoru Hisama[1], Katsutoshi Sato[2], Katsutoshi Nagaoka[2], Michihisa Koyama*,[1,3]

[1]Research Initiative for Supra Materials, Shinshu University, Nagano 380-8553, Japan. [2]Department of Chemical Systems Engineering, Nagoya University, Nagoya 464-8601, Japan. [3]Open Innovation Institute, Kyoto University, Kyoto 606-8501, Japan

*Correspondence to: valadez@shinshu-u.ac.jp, koyama_michihisa@shinshu-u.ac.jp

Table of Contents

### Introduction & Explanation Notes

The following auto-summary was created using the Dimensions[1] Auto-Summarization 3.0 tool from Springer Nature. This tool is identical to that applied to create, e. g., the machine-generated review book 'Lithium-Ion Batteries'[2]. The datasets for the first, second, and fourth chapters were prepared by searching for articles from 2016 up to the present using the keywords [heterogeneous AND (catalysis OR catalysis)] within the "SN Insights" database from Springer Nature. The database was divided into five clusters, from which only three were kept because we could not identify any content pattern within the other clusters. The cluster used to build the chapter 'Synthesis & Characterization' contained 120 samples, the cluster for 'Towards Higher Catalytic Activity' contained 336, and the cluster for 'Recent Progress & Perspectives' contained 337; these samples accounted for 1,026 of the most downloaded documents in the database. The final summary is given in the next section and only considers the 30 samples with the highest metrics closest to the chapters' topic.

We applied the same methodology to create the cluster of the chapter 'Theoretical Studies'. Here, we use the keywords [heterogeneous AND (catalysis OR catalyst) and (DFT OR computational OR "machine learning" OR numerical OR simulation OR "neural network" OR ReaxFF OR "molecular dynamics" OR "monte carlo" OR "ab initio" OR "first principles" OR quantum OR "artificial intelligence" OR "neural network" OR descriptor OR "kinetic model" OR CFD OR "computational fluid dynamics")]. Finally, chapters 1–3 only include an (extended) abstract summary, whereas chapter 4 also includes a summary of the perspectives provided in the various summarised reviews. We thank Dr. Shin'ichi Koizumi from Springer for helping us prepare the datasets to be used with the auto-summarisation tool. After this section, the machine-generated section follows.

# Auto-Summary

## 1. Synthesis & Characterization

### Immobilized N-Heterocyclic Carbene-Palladium(II) Complex on Graphene Oxide as Efficient and Recyclable Catalyst for Suzuki–Miyaura Cross-Coupling and Reduction of Nitroarenes

DOI: 10.1007/s10562-019-03083-0

Publication outline: Introduction | Experimental | Results and Discussion | Conclusions

**Abstract-Summary**

This new air- and moisture-stable NHC-Pd@GO heterogeneous catalytic system was found to be a highly active catalyst in the Suzuki–Miyaura cross-coupling between phenylboronic acid and various aryl halides (bromides/chlorides/iodides) and in the reduction of nitroarenes. NHC-Pd@GO heterogeneous catalyst could be recovered easily and reused at least eleven times in Suzuki–Miyaura cross-coupling and nine times in reduction of nitroarenes without any considerable loss of its catalytic activity. The stability and good selectivity of the NHC-Pd@GO heterogeneous catalyst in recycling experiments signify that it could be useful for practical application in various organic transformations.

Extended:

This new air- and moisture-stable heterogeneous catalyst was found to be highly active in Suzuki–Miyaura cross-coupling and reduction of nitroarenes reactions in an aqueous ethanol and aqueous methanol as green solvents using an extremely small amount of palladium under mild conditions (room temperature and short reaction time). NHC-Pd@GO heterogeneous catalyst could be easily recovered from the reaction media by simple filtration and reused at least 12 and ten cycles in the Suzuki–Miyaura cross-coupling and reduction of nitroarenes respectively without significant loss of its activity.

### Pd Supported IRMOF-3: Heterogeneous, Efficient and Reusable Catalyst for Heck Reaction

DOI: 10.1007/s10562-019-02756-0

Publication outline: Introduction | Experimental | Results and Discussion | Conclusions

**Abstract-Summary**

The prepared porous catalyst was effectively used in the Heck coupling reaction in the presence of an organic base. The reaction parameters such as the type of base, amounts of catalyst and solvents, temperature were optimized. The catalyst was then easily separated, washed, and reused 4 times without significant losses of catalytic activity.

### Vanadium–Schiff base complex-functionalized SBA-15 as a heterogeneous catalyst: synthesis, characterization and application in pharmaceutical sulfoxidation of sulfids

DOI: 10.1007/s11164-016-2589-5

Publication outline: Introduction | Materials and methods | Results and discussion | Conclusion

**Abstract-Summary**

$VO_2$(picolinichydrazone) complex as a catalyst was stabilized on a SBA-15 mesoporous silica as a catalytic support by using (3-chloropropyl)triethoxysilane as a connector. The immobilization of a metal–Schiff base complex to the surface area of SBA-15 can improve its catalytic effects by increasing the catalytic surface area. A vanadium–Schiff base complex-functionalized SBA-15 was synthesized by covalency connected by a pre-synthesised $VO_2$(picolinichydrazone) complex to silanated SBA-15. The synthesized vanadium–Schiff base complex was characterized by proton nuclear magnetic resonance ($^1$H NMR) spectroscopy, carbon nuclear magnetic resonance ($^{13}$C NMR) spectroscopy and Fourier transform infrared spectroscopy (FT-IR), and the final V/SBA-15 was characterized by FT-IR, ultraviolet–visible spectrophotometry and X-ray powder diffraction. The catalytic effect was examined by using V/SBA-15 as a heterogeneous catalyst in sulfoxidation reactions. Of modafinil and modafinic acid by pharmaceutical sulfoxidation of solfides was carried out and the effects of different solvents, reaction times and also recoverability and reusability of the heterogeneous catalyst were investigated.

Extended:

Of mesoporous silicas was suggested for the first time in 1992 [9] and, since then, considerable progress has been made in the morphology and pore size regulation and application developments of various types of mesoporous silica [10–11].

### Green Synthesis of Pd Nanoparticles Supported on Magnetic Graphene Oxide by Origanum vulgare Leaf Plant Extract: Catalytic Activity in the Reduction of Organic Dyes and Suzuki–Miyaura Cross-Coupling Reaction

DOI: 10.1007/s10562-017-2220-4

Publication outline: Introduction | Experimental | Result and Discussions | Conclusions

**Abstract-Summary**

The prepared magnetically nanocatalyst was characterized by transmission electron microscopy, scanning electron microscopy, X-ray diffraction, vibrating sample magnetometer (VSM) and Fourier transform infrared spectroscopy. The heterogeneous catalytic system was investigated for the reduction of 4-nitrophenol, methylene blue and methyl orange in the presence of NaBH$_4$ as a reducing reagent and also for Suzuki–Miyaura cross-coupling reactions between phenylboronic acid and a range of aryl halides (X = I, Br, Cl).

### A New Type of Magnetically-Recoverable Heteropolyacid Nanocatalyst Supported on Zirconia-Encapsulated Fe$_3$O$_4$ Nanoparticles as a Stable and Strong Solid Acid for Multicomponent Reactions

DOI: 10.1007/s10562-017-2015-7

Publication outline: Introduction | Result and Discussion | Conclusion

**Abstract-Summary**

The prepared HPW supported on nano-Fe$_3$O$_4$@ZrO$_2$ heterogeneous acid catalyst (or n-Fe$_3$O$_4$@ZrO$_2$/HPW) was fully characterized by several physicochemical techniques such as: Fourier transform infrared spectroscopy (FT-IR), field emission scanning electron microscopy, transmission electron microscopy, energy-dispersive X-ray spectroscopy, X-ray diffraction, vibrating sample magnetometry and thermogravimetric analysis. It was surprising that this very strong solid acid catalyst exhibited an excellent acid strength which was as a result of possessing a higher number of surface active sites compared to its homogeneous analogues. The catalytic activity of the as-prepared novel nano-Fe$_3$O$_4$@ZrO$_2$/HPW was explored through the one-pot three-component synthesis of different 3,4-dihydropyrimidin-2(1H)-ones (i.e. Biginelli reaction) and 1,4-dihydropyridines (i.e. Hantzsh reaction) under solvent free condition. The sample of 40 wt% showed higher acidity and activity in the catalytic transformation. After the reaction, the catalyst/product isolation could be easily achieved with an external magnetic field and the catalyst could be easily recycled for at least five times without any decrease in its high catalytic activity.

### Magnetic Nanoparticle-Supported N-Heterocyclic Carbene-Palladium(II): A Convenient, Efficient and Recyclable Catalyst for Suzuki–Miyaura Cross-Coupling Reactions

DOI: 10.1007/s10562-017-1987-7

Publication outline: Introduction | Experimental | Results and Discussion | Conclusion

**Abstract-Summary**

The nanomagnetic catalyst was used as convenient and efficient catalyst for the Suzuki–Miyaura cross-coupling reaction of various aryl bromides/chlorides/iodide with phenylboronic acid. The notable advantages of this heterogeneous nanomagnetic catalyst are excellent yields, mild reaction conditions, short reaction times and easy work-up. The new nanomagnetic catalyst could be easily recovered with an external magnet and could be reused at least five times without loss of its catalytic activity.

Extended:

The nanomagnetic catalyst is air- and water-stable, and can be synthesized in high yield with high purity from inexpensive commercially available starting materials.

### Glycerol Cu(II) Complex Supported on Fe$_3$O$_4$ Magnetic Nanoparticles: A New and Highly Efficient Reusable Catalyst for the Formation of Aryl-Sulfur and Aryl-Oxygen Bonds

DOI: 10.1007/s10562-019-02973-7

Publication outline: Introduction | Experimental | Result and Discussion | Conclusion

**Abstract-Summary**

This magnetic nanocatalyst utilized as an effective catalyst for the C–S and C–O cross-coupling reactions of aryl halides with thiourea and phenol. A green and novel Fe$_3$O$_4$@SiO$_2$–Glycerole–Cu(II) catalyst successfully was prepared and characterized. This catalyst can be used for the C– and C–O cross coupling reactions.

Extended:

The results show that the leaching of Cu into reaction media is negligible.

### Magnetic cobalt ferrite composite as an efficient catalyst for photocatalytic oxidation of carbamazepine

DOI: 10.1007/s11356-016-7978-1

Publication outline: Introduction | Experimental | Results and discussion | Conclusions

**Abstract-Summary**

A magnetic spinel cobalt ferrite nanoparticle composite (CFO) was prepared via an ultrasonication-assisted co-precipitation method. The morphological structure and surface composition of CFO before and after reaction were investigated by using X-ray diffraction, scanning electron microscopy, transmission electron microscopy, energy dispersive X-ray, and Fourier transform infrared spectroscopy, indicating the consumption of iron oxide during photodegradation. The prepared CFO was then used for the photocatalytic degradation of carbamazepine (CBZ) as an example of pharmaceuticals and personal care products (PPCPs) from aqueous solution. The reactive species for CBZ degradation in the CFO/UV system was identified as hydroxyl radicals by the methanol scavenging method. Combined with the detection of leached iron ions during the process, the CBZ degradation mechanism can be presumed to be heterogeneous and homogeneous photocatalytic degradation in the CFO/UV system.

### Design and development of a novel magnetic camphor nanospheres core/shell nanostructure

DOI: 10.1007/s40097-017-0224-7

Publication outline: Introduction | Experimental | Results and discussion | Conclusions

**Abstract-Summary**

The first synthesis, characterization and catalytic performance of Fe$_3$O$_4$/camphor core/shell nanospheres as a magnetic composite nanocatalyst were reported. Further advantages of this new protocol are short reaction times, high yields, easy workup procedure, inexpensive and eco-friendly protocol and magnetically recoverability and several times reusability of the nanocatalyst without significant decrease in catalytic activity.

### Microspherical ReS$_2$ as a High-Performance Hydrodesulfurization Catalyst

DOI: 10.1007/s10562-017-2024-6

Publication outline: Introduction | Experimental | Results and Discussion | Conclusions

**Abstract-Summary**

An unsupported microspherical ReS$_2$ catalyst, consisting in self-assembled nano-layers, was evaluated in the hydrodesulfurization (HDS) of 3-methylthiophene showing an excellent catalytic activity. The presence of a defect-rich structure in the microspheres, with short and randomly-orientated ReS$_2$ slabs, results in the exposure of additional edge sites, which improve the catalytic performance of this material. This microspherical ReS$_2$ composite, with good HDS performance, is a promising catalyst for the desulfurization of fuel oils; the solvothermal reaction conditions are also useful to tune and create exotic morphologies for the design of new ReS$_2$ catalysts.

### Synthesis and Characterization of Immobilized Lipase on Fe$_3$O$_4$ Nanoparticles as Nano biocatalyst for the Synthesis of Benzothiazepine and Spirobenzothiazine Chroman Derivatives

DOI: 10.1007/s10562-016-1797-3

Publication outline: Background | Experimental Section | Characterization | Results and Discussion | Conclusion

**Abstract-Summary**

A retrievable nanostructure heterogeneous catalyst, based on lipase (lipase from the fungus Aspergillus niger), namely Fe$_3$O$_4$-bonded lipase (Fe$_3$O$_4$ NPs@ lipase) was prepared in nano size and fully characterized by several techniques such as Fourier transform infrared spectroscopy, X-ray diffraction, Scanning electron microscopy, Transmission electron microscopy and Vibrating sample magnetometer.

### Four-Component Synthesis of 2-Amino-3-Cyanopyridine Derivatives Catalyzed by Cu@imineZCMNPs as a Novel, Efficient and Simple Nanocatalyst Under Solvent-Free Conditions

DOI: 10.1007/s10562-018-2318-3

Publication outline: Introduction | Results and Discussion | Screening of Reaction Conditions | Conclusion | Experimental

**Abstract-Summary**

An efficient and convenient method was investigated for synthesis of 2-amino-3-cyanopyridine derivatives via a one-pot four-component reaction of various types of aldehydes, acetophenone, malononitrile, and ammonium acetate in the presence of 10 mg Cu@imineZCMNPs catalyst. Due to the magnetic nature of the catalyst, it can be easily recovered by an external magnetic field and comfortable reused.

### Synthesis and characterization of ceria-coated silica nanospheres: their application in heterogeneous catalysis of organic pollutants

DOI: 10.1007/s42452-019-1613-y

Publication outline: Introduction | Materials and methods | Results and discussion | Catalytic activity of ScP33%CeP and CeO$_2$ for 4NP reduction | Conclusions

**Abstract-Summary**

The present work investigates the parameters for the successful coating of silicon oxide nanoparticles surface with a homogeneous cerium oxide shell in an effort to develop core–shell nanostructures. Several processing conditions, such as the precipitation pH, the cerium precursor concentration and the treatment of silica cores (i.e., their use either as-received after their biomimetic formation, or after their calcination followed by surface modification), were examined and optimized. The analysis revealed the formation of spherical core–shell nanostructures bearing a uniform shell layer of crystalline cerium oxide around each silica core which after calcination at 600 °C was comprised of cubic CeO$_2$ nanocrystals with sizes ranging between 2 and 6 nm.

### Magnetic nanoparticles tris(hydrogensulfato) boron as an efficient heterogeneous acid catalyst for the synthesis of α,ά-benzylidene bis(4-hydroxycoumarin) derivatives under solvent-free condition

DOI: 10.1007/s11164-019-03881-6

Publication outline: Introduction | Experimental | Results and Discussion | Conclusions

**Abstract-Summary**

This effective and magnetically recoverable catalyst was employed for the synthesis of α,ά-benzylidene bis(4-hydroxycoumarin) derivatives through the reaction of an aromatic aldehyde and 4-hydroxycoumarin under solvent-free conditions.

### Gold nanoparticles immobilized on single-layer α-zirconium phosphate nanosheets as a highly effective heterogeneous catalyst

DOI: 10.1007/s42114-019-00091-x

Publication outline: Introduction | Experimental section | Results and discussion | Conclusions

**Abstract-Summary**

Gold nanoparticles (Au NPs) were immobilized on single-layered α-zirconium phosphate (ZrP) nanosheets as a highly efficient heterogeneous catalyst. The characterizations showed that the Au NPs with a size distribution of 2.0 ± 1.0 nm were uniformly dispersed and immobilized on single-layer ZrP nanosheets.

### Synthesis and Application of Hyperbranched Polyester-Grafted Polyethylene (HBPE-g-PE) Containing Palladium Nanoparticles as Efficient Nanocatalyst

DOI: 10.1007/s10562-016-1720-y

Publication outline: Introduction | Experimental | Results and Discussion | Conclusions

**Abstract-Summary**

In nowadays, palladium-catalyzed reactions are of utmost importance in synthetic organic chemistry. We report a facile approach to prepare hyperbranched polyester-grafted-polyethylene supported palladium nanoparticles as an efficient catalyst in Suzuki reaction and solvent-free aerobic oxidation of alcohols. Because of the amplification effect of hyperbranched polyester, high loading capacities (around 0.276 mmol g$^{-1}$ for Pd) was achieved. A hyperbranched polyester based on 2,2-bis (methylol) propionic acid (bis-MPA) was synthesized and grafted onto carboxylic acid functionalized polyethylene powder. The final nanocatalyst was prepared via solution loading of PdCl$_2$ and reduction to palladium nanoparticles on the HBPE-g-PE. The activity and stability of prepared polymer-supported heterogeneous catalyst was carefully examined for Suzuki reaction and solvent-free oxidation of alcohols.

Extended:

We chose a number of primary or secondary alcohols to perform oxidation reaction and prepare relevant aldehydes or ketones under solvent-free condition.

### Five-membered N-Heterocycles Synthesis Catalyzed by Nano-silica Supported Copper(II)–2-imino-1,2-diphenylethan-1-ol Complex

DOI: 10.1007/s10562-017-2173-7

Publication outline: Introduction | Experimental Section | Results and Discussion | Conclusion

**Abstract-Summary**

A new amino functionalized nano silica was synthesized using 2-hydroxy-1,2-diphenylethan-1-one. The new 2-hydroxy-1,2-diphenylethan-1-one@amino functionalized nano-silica was used as a support material for copper(II) catalyst. The nano heterogeneous copper catalyst is recoverable up to 12 times without any significant leaching which indicated that the heterogeneous catalyst is stable and very active under the applied conditions.

### In situ incorporation of well-dispersed Cu–Fe oxides in the mesochannels of AMS and their utilization as catalysts towards the Fenton-like degradation of methylene blue

DOI: 10.1007/s10853-016-0436-0

Publication outline: Introduction | Experimental section | Results and discussion | Conclusion

**Abstract-Summary**

Multi-components active metal oxide-supported catalysts are highly promising in heterogeneous catalysis due to some special promoting effects. The results revealed that bimetallic Cu–Fe oxides were directly formed and highly dispersed in the mesochannels during the calcinations and the introduction of Cu$^{2+}$ and Fe$^{2+}$ on the micelles has influence on the structure properties. As compared to the monometallic Fe-modified AMS, the presence of Cu promotes the effects between Fe species and silica wall, leading to the better dispersion of Fe in the mesochannels of AMS. A series of Cu–Fe-modified AMS were used as Fenton-like catalysts and exhibited good catalytic activity in the degradation of methylene blue (MB), which resulted from high dispersion of Fe species and synergetic effect between Cu and Fe active sites.

### Synthesis of Polyhydroquinoline, 2,3-Dihydroquinazolin-4(1H)-one, Sulfide and Sulfoxide Derivatives Catalyzed by New Copper Complex Supported on MCM-41

DOI: 10.1007/s10562-018-2311-x

Publication outline: Introduction | Experimental | Result and Discussion | Conclusions

**Abstract-Summary**

A simple, efficient and less expensive protocol for the synthesis of Cu(II) immobilized on MCM-41@Serine has been reported. The obtained nanostructured compound were also employed as a green, efficient, heterogeneous and reusable catalytic system for the synthesis of polyhydroquinoline, 2,3-dihydroquinazolin-4(1H)-one, sulfide and sulfoxide derivatives. After characterization of this catalyst, the catalytic activity of this nanostructure compound has been investigated for the synthesis of polyhydroquinoline, 2,3-dihydroquinazolin-4(1H)-one, sulfide and sulfoxide derivatives.

### Preparation, characterization and catalytic performance of polyoxometalate immobilized on the surface of halloysite

DOI: 10.1007/s10853-018-3142-2

Publication outline: Introduction | Experimental section | Results and discussion | Conclusions

**Abstract-Summary**

The catalytic performance of this catalyst was tested in green oxidation system of n-tetradecane using air as oxidant under the mild reaction condition with normal atmospheric pressure and low temperature and without adding any solvents and additives. In order to find the optimum catalytic reaction condition, we prepared five catalysts containing different amounts of SbWRu, 0.97%, 1.94%, 2.80%, 4.23% and 5.87%. The results of the controlled experiments confirmed that the catalyst containing SbWRu of 1.94% exhibited high activity with a conversion (53.30%) and turnover frequency (TOF: 52396 h$^{-1}$) at the optimal reaction condition. The final conversion of n-tetradecane runs up to 87.97% after five consecutive cycles without the separation of the catalyst HNTs/Apts/SbWRu (1.94%).

Extended:

The catalytic performance of HNTs/Apts/SbWRu was evaluated in the oxidation reaction of n-tetradecane under environmentally friendly condition. This unique heterogeneous catalyst has bright potential application in the oxidation of n-alkanes in the future.

### P$_4$VPy–CuO nanoparticles as a novel and reusable catalyst: application at the protection of alcohols, phenols and amines

DOI: 10.1007/s13738-016-0887-x

Publication outline: Introduction | Experiment | Results and discussion | Conclusions

**Abstract-Summary**

P$_4$VPy–CuO nanoparticles were synthesized using ultrasound irradiations. Relevant properties of the synthesized nanoparticles were investigated by X-ray diffraction, scanning electron microscopy, transmission electron microscopy and Fourier transform infrared spectroscopy.

Extended:

It bodes well for the future of P$_4$VPy reagents in which their properties can be fine-tuned in organic reactions [198].

### Supported Au/MIL-53(Al): a reusable green solid catalyst for the three-component coupling reaction of aldehyde, alkyne, and amine

DOI: 10.1007/s11144-016-1034-5

Publication outline: Introduction | Experimental | Results and discussion | Conclusion

**Abstract-Summary**

A metal–organic-framework-supported Au-based heterogeneous catalyst (Au/MIL-53(Al)) was prepared using the impregnation method under mild conditions with HAuCl$_4$·4H$_2$O as the Au precursors. MIL-53(Al) indicates excellent chemical stability without structure degradation during the loading and catalysis process. The XPS spectra indicate that the catalyst Au/MIL-53(Al) contains coexisting Au$^0$ and Au$^{3+}$ ions. The results showed that the catalyst Au/MIL-53(Al) displayed high activity without any additives or an inert atmosphere (the yield reached 97.9 % for 5 h at 120 °C). Au/MIL-53(Al) has proven to be applicable to a wide range of substrates. Various aromatic/aliphatic aldehydes, aromatic alkynes, and piperidine were able to undergo A$^3$ coupling on the catalyst Au/MIL-53(Al).

Extended:

We believe that this methodology will find widespread use in organic synthesis for propargylamine preparation.

### Mixed cobalt/nickel metal–organic framework, an efficient catalyst for one-pot synthesis of substituted imidazoles

DOI: 10.1007/s00706-016-1776-9

Publication outline: Introduction | Results and discussion | Conclusion | Experimental

**Abstract-Summary**

A three-dimensional bimetallic metal–organic framework, [CoNi($\mu_3$-tp)$_2$($\mu_2$-pyz)$_2$] (tp = terephthalic acid, pyz = pyrazine) was applied as a highly efficient and recoverable heterogeneous catalyst. The catalyzed reaction was the one-pot three-component condensation reaction between benzil or benzoin with various substituted aromatic aldehydes and ammonium acetate to synthesize the corresponding imidazoles in solvent-free conditions and high to quantitative yields. 2,4,5-trisubstituted 1H-imidazoles were not previously synthesized using MOFs as heterogeneous catalyst.

Extended:

There are a few reports on the application of bimetallic MOFs as a catalyst in MCRs.

### Magnetic Nanoparticle Decorated N-Heterocyclic Carbene–Nickel Complex with Pendant Ferrocenyl Group for C–H Arylation of Benzoxazole

DOI: 10.1007/s10562-018-2514-1

Publication outline: Introduction | Experimental | Results and Discussion | Conclusion

**Abstract-Summary**

The complex proved to be an efficient heterogeneous catalyst for C–H arylation of benzoxazole with aryl boronic acids. The recycling studies revealed that complex could be reused for six times without significant decrease in catalytic activity.

### Facile synthesis of monodispersed Pd nanocatalysts decorated on graphene oxide for reduction of nitroaromatics in aqueous solution

DOI: 10.1007/s11164-018-3621-8

Publication outline: Introduction | Experimental | Results and discussion | Conclusions

**Abstract-Summary**

We synthesized the reproducible heterogeneous catalyst of graphene oxide (GO)-supported palladium nanoparticles (NPs) via a simple and green process. The GO-supported Pd NPs (Pd/GO nanocatalyst) exhibited excellent catalytic activity for the reduction of nitroaromatics to aminoaromatics in aqueous sodium borohydride. These features ensured the high catalytic activity of the introduced graphene oxide supported Pd nanocatalysts.

### A facile green synthesis of silver nanoparticle-decorated hydroxyapatite for efficient catalytic activity towards 4-nitrophenol reduction

DOI: 10.1007/s11164-017-3161-7

Publication outline: Introduction | Experimental | Characterization techniques | Results and discussion | Conclusion

**Abstract-Summary**

Silver nanoparticles (AgNPs) were gradually grown on a hydroxyapatite surface via a green self-reducing approach. Hydroxyapatite (HAP) was functionalized by dopamine and subsequently self-polymerized onto the HAP surface. Surface-adhered polydopamine acted as a self-reducing agent for the $Ag^+$ ion; thus the process is considered to be green. The AgNP-decorated HAP was applied as a heterogeneous catalyst for the reduction of 4-nitrophenol (4-NP), which obeyed first-order reduction kinetics.

### Synthesis and characterization of oxo-vanadium complex anchored onto SBA-15 as a green, novel and reusable nanocatalyst for the oxidation of sulfides and oxidative coupling of thiols

DOI: 10.1007/s11164-018-3367-3

Publication outline: Introduction | Experimental | Results and discussion | Conclusion

**Abstract-Summary**

The results of the developed procedure bring several benefits such as the use of commercially available, ecologically benign, operational simplicity, and cheap and chemically inert reagents. It shows good reaction times, practicability and high efficiency, and is easily recovered from the reaction mixture by simple filtration and reused for several consecutive cycles without noticeable change in its catalytic activity. High efficiency, simple and an inexpensive procedure, commercially available materials, easy separation, and an eco-friendly procedure are the several advantages of the currently employed heterogeneous catalytic system.

### Ru(III) complex anchored onto amino-functionalized MIL-101(Cr) framework via post-synthetic modification: an efficient heterogeneous catalyst for ring opening of epoxides

DOI: 10.1007/s13738-017-1277-8

Publication outline: Introduction | Experimental | Results and discussion | Conclusion

**Abstract-Summary**

The metallo Schiff base-functionalized metal–organic framework was prepared by post-synthetic method and used as an electron-deficient catalyst for the alcoholysis of epoxides. This new catalyst, MIL-101–$NH_2$–PC–Ru, was characterized by Fourier transform infrared, UV–Vis spectroscopic techniques, X-ray diffraction, BET, inductively coupled plasma atomic emission spectroscopy and field-emission scanning electron microscopy. In the presence of this heterogeneous catalyst, ring opening of epoxides was performed under mild condition to show the significant ability and successful applications of Lewis acid containing catalysts in corporation with metal–organic frameworks.

### A novel catalyst of MIL-101(Fe) doped with Co and Cu as persulfate activator: synthesis, characterization, and catalytic performance

DOI: 10.1007/s11696-017-0276-7

Publication outline: Introduction | Experimental | Results and discussion | Conclusions

**Abstract-Summary**

Effective and novel heterogeneous catalysts of Metal–organic framework MIL-101(Fe) doped with cobalt ($Co^{2+}$) or copper ($Cu^{2+}$) have been synthesized by post-synthesis method, namely Co–MIL-101(Fe) and Cu–MIL-101(Fe). The initial and metal-doped samples were tested to activate persulfate (PS) for removal of Acid Orange 7 (AO7) in water. The addition of $Cu^{2+}$ and $Co^{2+}$ could alter structure characteristics of MIL-101(Fe) in crystal structure and morphology. The alteration was reflected in the catalytic capacity of PS activation. An interesting note was that, whether Co or Cu doping, metal-doped MIL-101(Fe) greatly improved the PS activation as compared to unmodified MIL-101(Fe). The removal rate of AO7 was about 66, 92, 98% in MIL-101(Fe)/PS, 6%wtCu–MIL-101(Fe)/PS and 6%wtCo–MIL-101(Fe)/PS system, respectively. Such an enhancement in activity may be attributed to the main reasons: the new active sites created by metal additives; an increase in number of active Fe sites produced by Co and Cu doping which results in alteration of morphology and structure of catalysts.

### Heterogeneous Ziegler–Natta catalysts with various sizes of $MgCl_2$ crystallites: synthesis and characterization

DOI: 10.1007/s13726-016-0424-x

Publication outline: Introduction | Experimental | Results and discussion | Conclusion

**Abstract-Summary**

A $MgCl_2$-based Ziegler–Natta catalyst was characterized using X-ray diffraction (XRD) patterns, scanning electron microscopy (SEM) and transmission electron microscopy (TEM) and IR spectra. We focused on the XRD reflection at $2\theta = 50°$ to determine the thickness of $MgCl_2$ crystals, and validated these results with TEM pictures. SEM pictures were taken in order to measure the size of the nanoparticles formed by the $MgCl_2$ crystals. The thickness of the catalyst crystals was calculated from the XRD reflection at $2\theta = 50°$. Changing the precipitation temperature within a range from 40 to 90 °C altered the thickness of the $MgCl_2$ crystal plates. We estimated that a typical catalyst particle with a diameter of 20 μm contained about one million nanoparticles, each of which consisted of about 25,000 $MgCl_2$ crystal units.

Extended:

We focused on the XRD at 50° 2θ to determine whether additional information can be gained from the crystallographic data. SEM pictures were used to find out whether a correlation exists between the size of the nanoparticles [267] in the produced material and the crystal thickness measured by X-ray at 50° 2θ. The thickness of the MgCl$_2$ crystal was calculated from the reflection peak at 2θ = 50°, on which our study concentrated. The thickness of the catalysts was measured by X-ray diffraction using the reflection at 2θ = 50°. As part of future work, we will evaluate the effect of MgCl$_2$ crystal thickness on polymerization activity, molecular weight capability, width of molecular weight distribution, and amount of soluble polymer material in co-polymers, especially when higher amounts of co-monomer are incorporated.

## 2. Towards Higher Catalytic Activity

### Highly Efficient and Convenient Supported Ionic Liquid TiCl$_5$-DMIL@SiO$_2$@Fe$_3$O$_4$-Catalyzed Cycloaddition of CO$_2$ and Epoxides to Cyclic Carbonates

DOI: 10.1007/s10562-017-2051-3

Publication outline: Introduction | Experimental | Results and Discussion | Conclusion

### Abstract-Summary

A series of silica coated magnetic nanoparticles supported ILs were prepared, characterized and their catalytic performances in cycloaddition of CO$_2$ to epoxides were investigated. Research shows that TiCl$_5$-DMIL@SiO$_2$@Fe$_3$O$_4$ could be used as the highest efficient catalyst in cycloaddition of CO$_2$ and epoxides to gives the corresponding cyclic carbonates. The heterogeneous supported catalyst is easily recoverable by filtration, and could be used for five consecutive runs without significant loss in catalytic activity.

Extended:

Research shows that TiCl$_5$-DMIL@SiO$_2$@Fe$_3$O$_4$ is the most effective catalyst in the titled reaction.

### Eco-friendly highly efficient solvent free synthesis of benzimidazole derivatives over sulfonic acid functionalized graphene oxide in ambient condition

DOI: 10.1007/s11164-016-2745-y

Publication outline: Introduction | Experimental | Results and discussion | Conclusions

### Abstract-Summary

Using prepared heterogeneous catalyst GO-HSO$_3$, benzimidazole synthesis was carried out by means of reacting diamine and aldehyde at room temperature in solvent free condition. The catalyst GO-HSO$_3$ showed tremendous catalytic activity in selective synthesis of benzimidazole, as a result 100 % conversion of reactants and up to 89.0 % yield of respective benzimidazole was achieved using 0.1 mg of catalyst in very short reaction duration. The present method is found eco-friendly, highly efficient, solvent free, high yielding, and clean method for the synthesis benzimidazole derivatives at room temperature.

Extended:

These results indicate that, the present methodology is highly eco-friendly and green approach for the preparation of benzimidazole derivatives.

### Ruthenium Nanoparticles Immobilized on Nano-silica Functionalized with Thiol-Based Dendrimer: A Nanocomposite Material for Oxidation of Alcohols and Epoxidation of Alkenes

DOI: 10.1007/s10562-018-2313-8

Publication outline: Introduction | Experimental Section | Results and Discussion | Conclusion

### Abstract-Summary

Ruthenium nanoparticles were immobilized on thiol-based dendrimer functionalized nano-silica and its catalytic activity was investigated in the oxidation reactions. The catalytic activity of this nanocomposite material as a heterogeneous catalyst was studied in the epoxidation of alkenes and oxidation of alcohols with tert-butyl hydroperoxide (tert-BuOOH) and the corresponding products were obtained in good to excellent yields. Ru$_{np}$–nSTDP provided a highly stable, active, reusable, and solid-phase catalyst for preparation of a series of epoxides and aldehydes.

Extended:

The average core diameter (D) is about 2 nm.

### Suzuki–Miyaura Cross-Coupling in Aqueous Medium Using Recyclable Palladium/Amide-Silica Catalyst

DOI: 10.1007/s10562-016-1796-4

Publication outline: Introduction | Results and Discussion | Conclusions | Experimental Sections | Preparation of Catalyst

**Abstract-Summary**

The complex exhibits excellent catalytic activity and stability for Suzuki–Miyaura cross-coupling reaction under mild aqueous reaction condition. Various aryl bromides were coupled with arylboronic acids in i-PrOH/$H_2O$, under open air at room temperature, in the presence of 0.22 mol% of the catalyst to afford corresponding cross-coupled products in high yields.

### Highly Active and Recyclable Mesoporous Molecular Sieves CaO(SrO,BaO)/SBA-15 with Base Sites as Heterogeneous Catalysts for Methanolysis of Polycarbonate

DOI: 10.1007/s10562-017-2193-3

Publication outline: Introduction | Experimental | Results and Discussion | Conclusion

**Abstract-Summary**

Mesoporous molecular sieves CaO(BaO, SrO)/SBA-15, which can produce basic sites, were synthesized by impregnation of a certain amount of CaO(BaO, SrO) on the synthesis of SBA-15 support, and used as heterogeneous catalysts for the methanolysis of polycarbonate (PC). The results showed that CaO/SBA-15 exhibited better catalytic activity than BaO/SBA-15 and SrO/SBA-15, and its catalytic activity was related to the loading amount of CaO. Moreover, the obtained product from the methanolysis of PC was mainly bisphenol A (BPA) with high purity. The influneces on reaction conditions of PC methanolysis catalyzed by 10%CaO/SBA-15 were also investigated. The stability of catalyst showed that 10%CaO/SBA-15 catalysts could be recycled and reused with negligible loss in activity over five cycles.

### Highly active recyclable SBA-15-EDTA-Pd catalyst for Mizoroki-Heck, Stille and Kumada C–C coupling reactions

DOI: 10.1007/s10934-016-0323-8

Publication outline: Introduction | Experimental | Results and discussion | Catalytic activity | Conclusions

**Abstract-Summary**

SBA-15-EDTA-Pd(11) catalyst was found to exhibit excellent catalytic activity in appreciable yield for Heck, Stille and Kumada cross-coupling reactions. Covalently anchored heterogeneous SBA-15-EDTA-Pd(11) catalyst can be recycled for more than five times without noticeable loss in activity and selectivity.

### Sustainable Oxidative Cleavage of Vegetable Oils into Diacids by Organo-Modified Molybdenum Oxide Heterogeneous Catalysts

DOI: 10.1007/s11746-017-3047-2

Publication outline: Introduction | Experimental | Results and Discussion | Catalytic Test Results | Conclusions

**Abstract-Summary**

A novel catalytic system based on organo-modified molybdenum trioxide was synthesized by a green hydrothermal method in one simple step, using Mo powder as precursor, hydrogen peroxide, and amphiphilic surfactants cetyltrimethylammonium bromide (CTAB) and tetramethylammonium bromide (TMAB) as capping agents. The synthesized catalysts were employed in oxidative cleavage of oleic acid, the most abundant unsaturated fatty acid, to produce azelaic and pelargonic acids with a benign oxidant, $H_2O_2$.

Excellent catalytic activities resulting in full conversion of initial oleic acid were obtained, particularly for CTAB-capped molybdenum oxide (CTAB/Mo molar ratio of 1:3) that gave 83 and 68% yields of production of azelaic and pelargonic acids, respectively. The CTAB-capped catalyst could be conveniently separated from the reaction mixture by simple centrifugation and reused without significant loss of activity up to at least four cycles.

Extended:

We have tried to push the proved potential of molybdenum oxide as an oxidizing solid catalyst [46, 47–48] to oxidative cleavage reaction of unsaturated fatty acids.

### Functionalized Polystyrene Supported Copper(I) Complex as an Effective and Reusable Catalyst for Propargylamines Synthesis in Aqueous Medium

DOI: 10.1007/s10562-016-1728-3

Publication outline: Introduction | Experimental | Results and Discussion | Test for Heterogeneity | Recycling of Catalyst | Comparative Study | Conclusions

**Abstract-Summary**

An efficient procedure for the one-pot synthesis of propargylamines is $A^3$ coupling in which an alkyne, an aldehyde, and an amine are coupled together. We report that polystyrene supported Cu(I) catalyst is excellent for $A^3$ coupling in water without using any additives or hazardous organic solvents. The polystyrene supported Cu(I) catalyst was synthesized and its catalytic activity was also evaluated in the synthesis of propargylamines in oxidative $A^3$-coupling reaction with benzyl alcohols instead of their aldehydes in water. Polymer supported copper catalyst, Cu-PS-ala, acts as a heterogeneous catalyst for the one pot synthesis of propargylamines via $A^3$ coupling reaction in water using aldehydes or alcohols, amines and alkynes.

Extended:

The simple protocol, broad scope of the substrates and recycling of the catalyst permitted us to look forward to a good future of this process in academic as well as in industry.

### Modified Montmorillonite Clay Stabilized Silver Nanoparticles: An Active Heterogeneous Catalytic System for the Synthesis of Propargylamines

DOI: 10.1007/s10562-015-1679-0

Publication outline: Introduction | Experimental | Results and Discussion | Conclusion

**Abstract-Summary**

We have synthesized size selective silver nanoparticles (Ag-NPs) in the nanopores of modified montmorillonite clay by an "in situ" approach. The modification of montmorillonite K-10 was carried out with HCl under controlled condition for generating a high surface area and porous matrix which acted as hosts for the "in situ" generation of silver nanoparticles. The Ag-NPs stabilized on modified montmorillonite serves as an efficient green heterogeneous catalyst for the three component one-pot synthesis of propargylamines with 83–95 % yield and 100 % selectivity under mild reaction condition. Ag-NPs@mmt is found to be an efficient and robust heterogeneous catalyst for the synthesis of propargylamines under mild reaction condition.

### Synthesis of Dowex functionalised half-sandwich Ru (II) ($\eta^6$-p-cymene) complex as effective and selective hydrogenation of furfural to levulinic acid

DOI: 10.1007/s10965-021-02454-9

Publication outline: Introduction | Experimental section | Results and discussion | Conclusions

**Abstract-Summary**

The ruthenium(II) complex [($\eta^6$-p-cymene)Ru(Cl)(8-amino quinoline)] and [$\eta^6$-p-cymene)Ru(OH$_2$)(8-amino quinoline)] were synthesised, charactrised, and used to prepare novel ruthenium(II) based Dowex resin polymer [($\eta^6$-p-cymene)Ru(Cl / or OH$_2$)(8-amino quinoline)]@Dowex. The ruthenium(II) comlex based Dowex catalyst is synthesisedby treatment of [($\eta^6$-p-cymene)Ru(Cl)(8-amino quinoline)] or [$\eta^6$-p-cymene)Ru(OH$_2$)(8-amino quinoline)] with Dowex@Na via ion exchange synthesis, in the presence of DCM as a solvent forming [($\eta^6$-p-cymene)Ru(Cl / OH$_2$)(8-amino quinoline)]@Dowex, respectively. The catalytic transfer hydrogenation of furfural over both Ru(II) comlex based Dowex catalyst were investigated with formic acid as the hydrogen donor and water as the solvent at 80 ºC after 19 h. The [($\eta^6$-p-cymene)RuCl (8-amino quinoline)]@Dowex and [($\eta^6$-p-cymene)RuOH$_2$(8-amino quinoline)]@Dowex catalysts were shown to exhibit high activity in the transfer hydrogenation of furfural to form a sole levulinic acid (Lv) giving good conversion and excellent selectivity under mild reaction conditions. Recyclability and leaching tests confirmed [$\eta^6$-p-cymene)Ru(OH$_2$)(8-amino quinoline)] complex to function as a heterogeneous catalyst which could be reused for at least four consecutive reaction runs without significant loss of catalytic activity.

### Synthesis of a Cellulosic Pd(salen)-Type Catalytic Complex as a Green and Recyclable Catalyst for Cross-Coupling Reactions

DOI: 10.1007/s10562-020-03172-5

Publication outline: Introduction | Materials and Methods | Results and Discussion | Conclusion

**Abstract-Summary**

A green recyclable cellulose-supported Pd(salen)-type catalyst was synthesized through sequential three steps: chlorination with thionyl chloride, modification by ethylenediamine, and the formation of Schiff base with salicylaldehyde to immobilize palladium chloride through multiple binding sites. This novel heterogeneous cellulosic Pd(salen)-type catalytic complex was fully characterized by FT-IR, SEM, TEM, XPS, ICP-AES and TG. Studies have shown that the synthesized catalyst shows high activity and efficiency for all types of transformations, providing the corresponding carbon–carbon or carbon–heteroatom coupling products in a general and mild manner.

### Heteropolyacid supported on amine-functionalized halloysite nano clay as an efficient catalyst for the synthesis of pyrazolopyranopyrimidines via four-component domino reaction

DOI: 10.1007/s11164-016-2756-8

Publication outline: Introduction | Experimental | Results and discussion | Conclusion

**Abstract-Summary**

An efficient heterogeneous hybrid catalyst was developed by functionalization of halloysite clay nanotubes by γ-aminopropyltriethoxysilane and then immobilization of a Keggin type heteropolyacid, phosphotungstic acid. This hybrid catalyst was characterized by SEM/EDX, FTIR and XRD and its catalytic activity for the synthesis of pyrazolopyranopyrimidine derivatives via four-component domino reaction of barbituric acid, hydrazine hydrate, ethyl acetoacetate and benzaldehyde was investigated. The results indicated that the hybrid system can promote the reaction to afford the desired products in high yields and short reaction times. It was re-used at least three times with negligible loss of activity.

Extended:

The results indicated the formation of the desired product in negligible yield after 1 h; this also can prove the slight leaching of HPA. The results indicated that microwave irradiation promoted the reaction in very short reaction time, 5 min.

### Synthesis, characterization and catalytic activity of melamine immobilized MCM-41 for condensation reactions

DOI: 10.1007/s10934-017-0481-3

Publication outline: Introduction | Experimental | Results and discussion | Conclusions

**Abstract-Summary**

Melamine immobilized MCM-41 is synthesized by grafting on modified MCM-41. The surface area, pore size and pore volume of MCM-41-Mela were found to be decreased after immobilization of melamine. FTIR and Raman spectroscopy results revealed the successful grafting of melamine on the surface of MCM-41. The catalytic activity of MCM-41-Mela was then successfully carried out for Knoevenagel condensation with different substrates, giving excellent yields of the corresponding products. Simple preparation methods, high efficiency and reusability of the heterogeneous MCM-41-Mela catalyst demonstrate a great potential for future catalysis application.

Extended:

Melamine immobilized MCM-41 catalyst is successfully synthesized by feasible low cost route and utilised in condensation of furfural and acetylacetone under solvent-free reaction conditions.

### Citric acid stabilized on the surface of magnetic nanoparticles as an efficient and recyclable catalyst for transamidation of carboxamides, phthalimide, urea and thiourea with amines under neat conditions

DOI: 10.1007/s13738-018-1523-8

Publication outline: Introduction | Results and discussion | Experimental procedures | Conclusion

#### Abstract-Summary

This magnetic nanocatalyst was employed as an efficient, recyclable, and environmentally benign heterogeneous catalyst for the transamidation of carboxamides, phthalimide, urea and thiourea with amines. The catalyst could be easily separated from the reaction mixture using an external magnet and can be reused six times without any significant loss in its catalytic activity.

### Cu(OAc)$_2$ entrapped on ethylene glycol-modified melamine–formaldehyde polymer as an efficient heterogeneous catalyst for Suzuki–Miyaura coupling reactions

DOI: 10.1007/s11164-019-03984-0

Publication outline: Introduction | Result and discussion | Conclusions | Experimental section

#### Abstract-Summary

Morphological, physicochemical characteristics and catalytic activity of the heterogeneous catalyst (Cu@MCOP) were analyzed by various instrumental methods including powder X-ray diffraction, FT-IR, UV-DRS, X-ray photoelectron spectroscopy, SEM and elemental mapping which have been used to authenticate the polymeric materials Cu@MCOP. The catalytic performance of Cu@MCOP as solid heterogeneous catalyst was evaluated in synthesis of various biphenyl derivatives through Suzuki–Miyaura cross-coupling reactions of various aryl halides with substituted organoboranes under normal reaction conditions.

### Ru–PPh$_3$@porous organic polymer: efficient and stable catalyst for the trickle bed regioselective hydrogenation of cinnamaldehyde

DOI: 10.1007/s11144-016-1130-6

Publication outline: Introduction | Experimental | Results and discussion | Conclusions

#### Abstract-Summary

Ru supported on the porous vinyl-functional triphenylphosphine organic polymer (Ru–PPh$_3$@POP) as a heterogeneous catalyst for the selective hydrogenation of cinnamaldehyde (CAL) was synthesized. This catalyst, which combines the advantages of heterogeneous and homogeneous catalyst, showed much higher catalytic activity of continuous trickle bed hydrogenation of CAL than traditional inorganic material supported heterogeneous catalysts. The high efficiency of the Ru–PPh$_3$@POP catalyst was attributed to the formation of the strong coordination bonds between the Ru species and the exposure phosphorus atoms on the surface of PPh$_3$@POP support according to the results of EDS-mapping, $^{31}$P NMR, XPS analysis and HRTEM. Ru species maintained their nanoparticles without further aggregation during the long-term reaction, and thus the catalyst exhibited a high conversion of CAL of 55% and selectivity towards cinnamyl alcohol (COL) of 63% even over 312 h of time on stream of hydrogenation of CAL.

### Pd Nanoparticles Immobilized on Supported Magnetic GO@PAMPS as an Auspicious Catalyst for Suzuki–Miyaura Coupling Reaction

DOI: 10.1007/s10562-017-2089-2

Publication outline: Introduction | Experimental | Results and Discussion | Conclusion

#### Abstract-Summary

A novel catalytic system based on palladium nanoparticles (Pd-NPs) immobilized onto the surface of graphene oxide (GO) modified by poly 2-acrylamido-2-methyl-1-propansulfonic acid decorated with magnetic Fe$_3$O$_4$ was designed, prepared and fully characterized. The results showed excellent catalytic activity for the cross coupling of aryl bromides, alkyl iodides as well as aryl chlorides as the challenging substrates. It was easily separated by an external magnet and reused without any pre-activation at least in seven consecutive runs without any loss in its catalytic activity as well as any detectable Pd leaching.

### Characterization and catalytic performance evaluation of a novel heterogeneous mesoporous catalyst for methanol–acetic acid esterification

DOI: 10.1007/s10934-019-00764-4

Publication outline: Introduction | Experimental | Results and discussion | Conclusions

#### Abstract-Summary

A novel heterogeneous catalyst based on silicotungstic acid (STA) has been synthesised using hydrothermal and wet-impregnation methods for methanol–acetic acid esterification. Characterizations of MCM-48 and STA/MCM-48 catalysts were performed using BET, XRD, FT-IR, DRIFT, TGA/DTA, SEM, and EDX mapping analysis. The catalytic activity was also investigated in methyl acetate esterification reaction and it was shown that temperature, reaction time and the molar ratio of the reactants had significant effect on conversion rate. The highest catalytic activity was obtained as 84% with 20% catalyst at feed ratio of 1:3 (metanol:acetic acid, 353 K, 27 h).

Extended:

A novel heterogeneous mesoporous catalyst for MeOH–AA ER was prepared using silicotungstic acid (STA, surface area < 1 m$^2$/g) as an active agent and MCM-48 as a mesoporous support material. Characterization and catalytic performance of a novel heterogeneous catalyst, STA/MCM-48, has been reported for MeOH–AA ER. The highest catalytic activity was obtained as 84% with 20% STA loadings at the feed ratio of 1:3 (MeOH:AA) and 353 K at 27 h. However, the reaction did not yet reach its equilibrium which suggests further optimizations needed for reactant molar ratios and experiment duration.

### Immobilization of new macrocyclic Schiff base copper complex on graphene oxide nanosheets and its catalytic activity for olefins epoxidation

DOI: 10.1007/s10853-018-3035-4

Publication outline: Introduction | Experimental procedures | Results and Discussion | Conclusion

**Abstract-Summary**

A new heterogeneous copper–salophen catalyst was fabricated through the nucleophilic attack and esterification reaction of salophen Schiff base with graphene oxide (GO). The Schiff base was covalently grafted to the activated carboxyl group on GO through esterification reaction and also through the nucleophilic attack to epoxy groups. The prepared heterogeneous catalyst was used in the oxidation of olefins with $H_2O_2$/$NaHCO_3$ under mild condition. The influence of various parameters such as catalyst dosage, reaction time as well as reaction temperature on the epoxidation of norbornene was evaluated. The catalyst exhibits high activity and selectivity to target product; cyclohexene showed 100% conversion, without any by-product, and norbornene gave 100% conversion and 93% selectivity. The catalyst reused up to four more consecutive times without significant sacrificing activity. The high activity of catalyst can be ascribed to strong π–π stacking interaction between GO sheet and aromatic rings of salophen complex.

Extended:

The influence of various parameters such as reaction time, the dose of catalyst, as well as the reaction temperature on the epoxidation of norbornene was investigated, and the reaction was followed by GC. The high activity and high selectivity of GO-Cu-L as a heterogeneous catalyst can be attributed to the active sites of the Schiff base complex as a privileged ligand and the synergistic catalytic interaction between the copper complex and GO as support [127, 128].

### Palladium Nanoparticles Anchored on Thiol Functionalized Xylose Hydrochar Microspheres: An Efficient Heterogeneous Catalyst for Suzuki Cross-Coupling Reactions

DOI: 10.1007/s10562-019-02984-4

Publication outline: Introduction | Experimental Section | Results and Discussions | Conclusions

**Abstract-Summary**

Novel thiol functionalized xylose hydrochar microspheres supported palladium nanoparticles (C–SH–Pd) were synthesized by gentle heating of palladium (II) acetate and thiol functionalized xylose hydrochar (C–SH) in ethanol. The as-prepared C–SH–Pd exhibited high catalytic activity towards Suzuki reactions with a yield of high up to 100%. Due to the superior catalytic performance and stability of the C–SH–Pd catalyst, it can be exploited in other cross coupling reactions in the long run.

### First Report About the Use of Micellar Keggin Heteropolyacids as Catalysts in the Green Multicomponent Synthesis of Nifedipine Derivatives

DOI: 10.1007/s10562-016-1784-8

Publication outline: Introduction | Experimental | Results and Discussion | Conclusion

**Abstract-Summary**

Micellar Keggin heteropolyacid catalysts were prepared using hexadecyltrimethylammonium bromide (cetyltrimethylammonium bromide—CTAB), 1-hexadecyl-pyridinium chloride, and Keggin heteropolyacids $H_3PMo_{12}O_{40}$ and $H_4PMo_{11}VO_{40}$ as precursors. This methodology requires a reaction time of 8 h, and a temperature of 78 °C to obtain good to excellent yields of 1,4-dihydropyridine derivatives. The micellar Keggin catalysts are insoluble in polar media, which allows easy removal of the reaction products without affecting their catalytic activity. The leaching test showed that they have an excellent stability and can be used five times as heterogeneous catalysts without appreciable loss of the catalytic activity.

### Nitrogen-doped reduced graphene oxide-supported $Mn_3O_4$: An efficient heterogeneous catalyst for the oxidation of vanillyl alcohol to vanillin

DOI: 10.1007/s10853-016-0318-5

Publication outline: Introduction | Experimental section | Results and discussion | Conclusion

**Abstract-Summary**

The as-made N-RGO/$Mn_3O_4$ catalyst was used for the oxidation of vanillyl alcohol to vanillin. N,N-Dimethylformamide was found to be the best solvent, affording both high conversion vanillin selectivity. 92.5 % conversion of vanillyl alcohol and 91.4 % selectivity of vanillin were achieved after 12 h at 120 °C under oxygen balloon by the use of 40 mg of the N-GO/$Mn_3O_4$ catalyst. Kinetic studies revealed that the active energy for the oxidation of vanillyl alcohol to vanillin over N-GO/$Mn_3O_4$ catalyst was 39.67 kJ.$mol^{-1}$.

### Harnessing $CO_2$ into Carbonates Using Heterogeneous Waste Derivative Cellulose-Based Poly(ionic liquids) as Catalysts

DOI: 10.1007/s10562-018-2637-4

Publication outline: Introduction | Experimental | Results and Discussion | Conclusion

**Abstract-Summary**

CPILs were used as heterogeneous catalysts for $CO_2$ chemical transformation into cyclic carbonates by cycloaddition of $CO_2$ with epoxides [propylene (PO) and styrene oxides (SO)]. The effect of the cation present in CPILs in catalytic performance, use of $ZnBr_2$ as a co-catalyst and catalytic reaction parameters (temperature, pressure and time) were investigated just as well. Results demonstrate that CPILs cation variation influence their catalytic activity. A higher $CO_2$ yield and selectivity of 81.9%/95.3% for propylene carbonate (PC) and 78.7%/100% for styrene carbonate (SC) was obtained by CPIL-TBP/$ZnBr_2$ at conditions of 40 bar, 110 °C and 6 h, being easily separated and recycled without significant loss of catalytic activity until the fourth cycle.

### Preparation of metallic Pd nanoparticles using supercritical $CO_2$ deposition: An efficient catalyst for Suzuki cross-coupling reaction

DOI: 10.1007/s11051-018-4252-0

Publication outline: Introduction | Experimental | Results and discussion | Conclusions

**Abstract-Summary**

Ligand-free palladium nanoparticles supported on multi-walled carbon nanotubes (Pd/MWCNT) were prepared by the supercritical carbon dioxide (sc$CO_2$) deposition method using a novel sc$CO_2$-soluble Pd organometallic complex as a precursor. Pd/MWCNT was utilized as a heterogeneous catalyst in Suzuki cross-coupling reaction. The nanocatalyst was found very effective in Suzuki reaction and it could also be recovered easily from the reaction media and reused over several cycles without significant loss of catalytic activity under mild conditions.

### Copper(II)-Substituted Polyoxotungstates Immobilized on Amine-Functionalized SBA-15: Efficient Heterogeneous Catalysts for Liquid Phase Oxidative Reaction

DOI: 10.1007/s10562-016-1874-7

Publication outline: Introduction | Experimental | Results and Discussion | Conclusions

**Abstract-Summary**

Results demonstrated that the heteropolyanions are highly dispersed on mesoporous materials. The resulting catalysts exhibited highly efficient catalytic activity for both the selective oxidation of benzyl alcohol to the benzaldehyde and the oxidation of thiophene with 30 % hydrogen peroxide as oxidant.

### Carbon Quantum Dots-$TiO_2$ Nanocomposites with Enhanced Catalytic Activities for Selective Liquid Phase Oxidation of Alcohols

DOI: 10.1007/s10562-017-2065-x

Publication outline: Introduction | Experimental | Results and Discussion | Conclusions

**Abstract-Summary**

The obtained catalyst showed enhanced catalytic activities compared to the pure CQDs catalyst. The enhanced catalytic activities were mainly attributed to the highly dispersive CQDs with more active centers which caused by the $TiO_2$ support. The CQDs-$TiO_2$ catalyst can also be reused without any significant loss in their catalytic activities after five times.

### A Novel Magnetic Immobilized Para-Aminobenzoic Acid-Cu(II) Complex: A Green, Efficient and Reusable Catalyst for Aldol Condensation Reactions in Green Media

DOI: 10.1007/s10562-020-03216-w

Publication outline: Introduction | Experimental | Results and Discussion | Conclusion

**Abstract-Summary**

The structure of the newly synthesized heterogeneous magnetic nanocatalyst with enhanced and improved catalytic efficiency were determined by various instrumental techniques, including SEM, VSM, TGA, XRD, UV–VIS FT-IR and EDXA. The catalytic activity and efficient performance of $Fe_3O_4$@PABA-Cu(II) MNPs were analyzed toward the synthesis of novel 5-arylidenthiazolidine-2,4-diones and 5-arylidene-2-imidazolidine-2,4-dione derivatives via aldol condensation reactions between a variety of (hetero) aromatic aldehydes and hydantoin or thiazolidine-2,4-dione multifunctional privileged scaffolds under reflux condensations in ethanol as a benign solvent. It is the first report of aldol synthesis of new 5-arylidenthiazolidine-2,4-dione, and 5-arylidene-imidazolidine-2,4-dione derivatives using a reusable copper-PABA complex supported on $Fe_3O_4$ MNPs ($Fe_3O_4$@PABA-Cu(II)) catalyst in Green media.

Extended:

These results show that $Fe_3O_4$@PABA-Cu(II) as an efficient catalyst increase the reaction yields and rate with easy conditions, simple workup and reduced unwanted side product.

**Direct oxidative esterification of alcohols catalyzed by a nitrogen-doped carbon black-supported PdBi bimetallic catalyst under ambient conditions**

DOI: 10.1007/s10853-020-05726-9

Publication outline: Introduction | Experimental section | Characterization | Results and discussion | Conclusions

**Abstract-Summary**

The optimal PdBi/NCB shows outstanding catalytic performance with broad substrate scope, good functional group tolerance towards direct oxidative esterification of alcohols under mild conditions in a heterogeneous catalytic system with air as the sole oxidant. Superior catalytic activity is mainly attributable to the unique structure of the catalyst, including synergetic electronic effect between Pd and Bi, as well as modulated surface character by acidification and N doping for better active components' anchoring and dispersion, as well as reactants' adsorption. This study provides a facial, practical, eco-friendly and efficient catalytic system for oxidative esterification of alcohols and shows promising prospect in industrial production.

**Novel and Efficient Heterogeneous 4-Methylbenzenesulfonic Acid-Based Ionic Liquid Supported on Silica Gel for Greener Fischer Indole Synthesis**

DOI: 10.1007/s10562-016-1721-x

Publication outline: Introduction | Experimental | Results and Discussion | Conclusion

**Abstract-Summary**

IL-SO$_3$H-SiO$_2$ was utilized as an efficient and heterogeneous catalyst for the synthesis of indoles via the one-pot Fischer reaction of phenyl hydrazines with ketones or aldehydes at room temperature.

**The Efficient Oxidation of Biomass-Derived 5-Hydroxymethyl Furfural to Produce 2,5-Diformylfuran Over Supported Cobalt Catalysts**

DOI: 10.1007/s12649-016-9724-9

Publication outline: Introduction | Experimental Section | Result and Discussions | Conclusions

**Abstract-Summary**

We report a novel cobalt-based catalyst (Co$_x$O$_y$-N@TiO$_2$) for the oxidation of HMF to DFF with 30 % aqueous tert-butyl hydroperoxide as the oxidant. Of catalytic test, it is found that a 91 % conversion of HMF and 40 % selectivity of DFF was obtained with the Co$_x$O$_y$-N@TiO$_2$ catalyst at 80 °C within 5 h. In addition, the Co$_x$O$_y$-N@TiO$_2$ can be recycled up to five times without significant loss of activity. The efficient oxidation of 5-hydroxymethyl furfural, a biomass-derived platform compound, to produce 2,5-diformylfuran is achieved using Co$_x$O$_y$-N@TiO$_2$ catalyst under mild conditions.

## 3. Theoretical Studies

### Carbon Permeation: The Prerequisite Elementary Step in Iron-Catalyzed Fischer–Tropsch Synthesis

DOI: 10.1007/s10562-018-02651-0

Publication outline: Introduction | Molecular Dynamic (MD) Simulation Methods | DFT Methods | Energy and Rate Constants | Model Surfaces | Carbon Permeation from MD Simulations | Dicarbon | Permeation Will Vary Substantially with Surface | Permeation vs Surface Migration | Rate Constant Estimates for Permeation | Conclusions

**Abstract-Summary**

The interaction of C atoms with five model surfaces (Fe (100), (110), (111), (211), (310)) was studied in six distinct ways. In the first, the random deposition of C atoms on the Fe surfaces was simulated by molecular dynamics, with C atoms released gradually. It shows that the early stages of carburization is a C permeation process, without much disturbance to the Fe surfaces. C atoms were approached to the surfaces sequentially. They bind readily (by 7–9 eV per C) to the surfaces, but to a different extent—strongest on Fe (100), and weakest on Fe (111). At a certain coverage, different on each surface, C atoms prefer in calculation to go subsurface. In a third approach, detailed transition paths of C permeation subsurface were calculated, with associated barriers in the order Fe (100) > (111) > (310) > (211) > (110). Comparing C permeation with surface migration on clean surfaces, the barrier of the former is smaller than that of the latter for most of the surfaces, except Fe (111). At intermediate C coverage, the (100) surface also prefers migration to permeation. In a fifth approach, the position in energy of the d-band centers of the Fe surfaces upon C permeation was studied. For all the surfaces, the d-band centers move away from the Fermi level with increasing C coverage, and start to resemble those of the bulk carbide phases at high C coverage. A general picture emerges of C permeation on Fe surfaces as a stepwise process with opposite thermodynamic and kinetic preferences.

Extended:

For all the surfaces studied, C permeation moves the d-band centers of the surfaces further away from the Fermi level.

**First-principles-based multiscale modelling of heterogeneous catalysis**

DOI: 10.1038/s41929-019-0298-3

Publication outline: Main | A generic multiscale modelling framework | Mechanistic complexity | Structural complexity | Environmental complexity | Conclusion and perspective

**Abstract-Summary**

They provide invaluable mechanistic insight and allow screening of vast materials spaces for promising new catalysts — in silico and at predictive quality. We briefly review methodological cornerstones of existing approaches and highlight successes and ongoing developments. On the road towards a higher structural, mechanistic and environmental complexity, it is, in particular, the fusion with machine learning methodology that promises rapid advances in the years to come.

**Achievements and Expectations in the Field of Computational Heterogeneous Catalysis in an Innovation Context**

DOI: 10.1007/s11244-021-01489-y

Publication outline: Introduction | Understanding the Nature of the Surface Active Sites | Quantification of Reaction Profiles | Prediction of the Reaction Rates and of Catalytic Performance | Prediction of New Active Phases | Current Challenges: Towards More Relevance, Accuracy, Efficiency

**Abstract-Summary**

The present short review discusses and illustrates these various stages of the catalyst understanding and performance prediction where computational catalysis has a crucial role to play. Selected achievements in the field are reviewed, with a focus on the simulation of complex metallic and zeolite catalysts of industrial relevance. Future directions are suggested, on the basis of the need for ever more exhaustive and accurate models of catalytic sites and catalytic reactions representative of industrial systems, and for speed up in catalyst understanding and discovery.

**The Role of Zeolite Framework in Zeolite Stability and Catalysis from Recent Atomic Simulation**

DOI: 10.1007/s11244-021-01473-6

Publication outline: Introduction | Theoretical Advances in Understanding Zeolite Stability | Atomic Simulations on the Zeolite–Molecule Interaction | Perspective

**Abstract-Summary**

Theoretical simulations, particularly those based on first principles calculations, have advanced significantly the understandings on zeolite, from structure to adsorption kinetics and to catalytic reactivity. This short review overviews the theoretical insights into the role of zeolite framework in zeolite stability and catalysis revealed from atomic simulation in recent years. We will mainly focus on two key aspects: (i) the theory on zeolite stability, including the templating effect of structure directing agents and the zeolite bonding pattern analysis; (ii) the confinement effect of zeolite pores that affects the catalytic conversion of molecules in zeolite.

**Catalysis at Metal/Oxide Interfaces: Density Functional Theory and Microkinetic Modeling of Water Gas Shift at Pt/MgO Boundaries**

DOI: 10.1007/s11244-020-01257-4

Publication outline: Introduction | Methods | Results and Discussion | Conclusions

**Abstract-Summary**

The impact of metal/oxide interfaces on the catalytic properties of oxide-supported metal nanoparticles is a topic of longstanding interest in the field of heterogeneous catalysis. Using a combination of periodic Density Functional Theory calculations and microkinetic modeling, we present a comprehensive analysis of the WGS mechanism at the interface between a quasi-one dimensional platinum nanowire and an irreducible MgO support. The nanowire is lattice matched to the MgO support to remove spurious strain at the metal/oxide interface, and reactions both on the nanowire and at the three-phase boundary itself are considered in the mechanistic analysis. These results are combined with detailed calculations of adsorbate entropies and dual-site microkinetic modeling to determine the kinetically significant features of the WGS reaction network which are subsequently, validated through experimental measurements of apparent reaction orders and activation barrier. The analysis demonstrates the important role that the metal/oxide interface plays in the reaction, with the water dissociation step being facile at the interface compared to the pure metal or oxide surfaces. Further, explicit consideration of CO interactions with other adsorbates at the metal/oxide interface is found to be central to correctly determining reaction mechanisms, rate determining steps, reaction orders, and effective activation barriers.

Extended:

The nanowire is lattice matched to the MgO surface to remove spurious strain effects that sometimes arise in models of metal/oxide interfaces, and the irreducibility of the MgO, in turn, leads to simplification of the space of possible reaction mechanisms, since lattice oxygen does not directly affect the WGS chemistry.

**DFT calculations on subnanometric metal catalysts: a short review on new supported materials**

DOI: 10.1007/s00214-018-2236-x

Publication outline: Introduction | Computational details | Discussion | Conclusions

**Abstract-Summary**

Metal clusters have been used in catalysis for a long time, even in industrial production protocols, and a large number of theoretical and experimental studies aimed at characterizing their structure and reactivity, either when supported or not, are already present in the literature. In the last years the advances made in the control of the synthesis and stabilization of subnanometric clusters promoted a renewed interest in the field. If supported, subnanometric clusters could be highly influenced by the interactions with the support that could affect geometric and electronic properties of the catalyst. The outstanding position of this corner of science will be highlighted through a selected number of examples present in the literature, concerning the growth and reactivity of subnanometric supported metal catalysts.

**Theory-guided design of catalytic materials using scaling relationships and reactivity descriptors**

DOI: 10.1038/s41578-019-0152-x

Publication outline: Introduction | [Section 2] | Descriptors on metal surfaces | Descriptors on metal-oxide surfaces | Universal and other descriptors | How to break scaling relationships | Outlook

**Abstract-Summary**

A physical or chemical property of the reaction system, termed as a reactivity descriptor, scales with another property often in a linear manner, which can describe and/or predict the catalytic performance. In this Review, we describe scaling relationships and reactivity descriptors for heterogeneous catalysis, including electronic descriptors represented by d-band theory, structural descriptors, which can be directly applied to catalyst design, and, ultimately, universal descriptors. The prediction of trends in catalytic performance using reactivity descriptors can enable the rational design of catalysts and the efficient screening of high-throughput catalysts.

Extended:

In this Review, we discuss reactivity descriptors in different catalytic systems, taking into consideration the limitations of scaling relationships and strategies to break these established scaling relationships. We aim to lay foundations for the future computational design of catalysts.

**Modelling heterogeneous interfaces for solar water splitting**

DOI: 10.1038/nmat4803

Publication outline: Main | Computational methods | Photoelectrode/water interfaces | Photoelectrode/catalyst/water interfaces | Outlook and conclusions

**Abstract-Summary**

The key of a successful solar-to-fuel technology is the design of efficient, long-lasting and low-cost photoelectrochemical cells, which are responsible for absorbing sunlight and driving water splitting reactions. We review recent progress and open challenges in predicting physicochemical properties of heterogeneous interfaces for solar water splitting applications using first-principles-based approaches, and highlights the key role of these calculations in interpreting increasingly complex experiments.

**Computational optimization of electric fields for better catalysis design**

DOI: 10.1038/s41929-018-0109-2

Publication outline: Main | Natural and synthetic biocatalysis | Porous supermolecular capsules and zeolite catalysts | Electric fields for heterogeneous and homogeneous catalysis | Summary and future directions

**Abstract-Summary**

We illustrate how electric fields have been used to computationally optimize biocatalytic performance of a synthetic enzyme, and how they could be used as a unifying descriptor for catalytic design across a range of homogeneous and heterogeneous catalysts.

Extended:

The data that supports the findings of this study are available from the corresponding author upon request.

**Properties of Isolated and TiO$_2$(110) Supported Pt$_{13}$ Clusters: A Theoretical Study**

DOI: 10.1007/s11244-019-01182-1

Publication outline: Introduction | Computational Details and Cluster Models | Results and Discussion | Conclusions

**Abstract-Summary**

We determine their equilibrium geometries, cohesive energies, magnetic moments, electronic and vibrational density of states. The analysis of the vibrational modes reveal that the Oh and Ih structures are dynamically unstable unlike the layered structures that have lower energies. We then examine the Pt$_{13}$-titania system to characterize the cluster/substrate interaction for both the stoichiometric and reduced surfaces. We characterize different aspects of the metal-oxide interaction by determining their equilibrium geometries, adsorption energies, charge transfer effects and electronic density of states. We find that the Pt$_{13}$ cluster suffers a strong restructuration when adsorbed on the surface, it deforms towards increasing the interaction of the platinum atoms with the surface, leading to a high value of the adsorption energy and getting oxidized. The calculated surface oxygen vacancy formation energy is prefered by the cluster deposition as a fact that favours the use of this system in the CO oxidation reactions from a surface oxygen, an important step in the water gas shift reaction.

Extended:

The analysis of the vibrational modes reveal that the Oh and Ih structures are dynamically unstable, while the layered configurations, named DT-1 and DT-2, are dynamically stable. We find that the $Pt_{13}$ cluster suffers a restructuration when adsorbed on the surface.

### Evaluating differences in the active-site electronics of supported Au nanoparticle catalysts using Hammett and DFT studies

DOI: 10.1038/nchem.2911

Publication outline: Main | Results | Additional information

**Abstract-Summary**

Supported metal catalysts, which are composed of metal nanoparticles dispersed on metal oxides or other high-surface-area materials, are ubiquitous in industrially catalysed reactions. Metal–support interactions have an enormous impact on the chemistry of the catalytic active site and can determine the optimum support for a reaction; however, few direct probes of these interactions are available. We show how benzyl alcohol oxidation Hammett studies can be used to characterize differences in the catalytic activity of Au nanoparticles hosted on various metal-oxide supports.

### Optimum Performance of Vanadyl Pyrophosphate Catalysts

DOI: 10.1007/s11244-016-0673-0

Publication outline: Introduction | Experimental | Results and Discussion | Summary

**Abstract-Summary**

A scheme is proposed for the dynamic, catalytically active vanadium-phosphorus-mixed oxide surface of industrially used catalysts for the selective oxidation of n-butane to maleic anhydride. This scheme is used as basis for a two-dimensional, heterogeneous reactor model describing the observed performance changes as function of the underlying phosphorus surface dynamics. The dynamic model comprises two reversible reactions: slow phosphorus adsorption, and water adsorption reaching its equilibrium faster. The kinetic model distinguishes explicitly between the intrinsic kinetics and phosphorus/water induced activity dynamics. In the presented study all phosphorus and water related processes appeared to be completely reversible, and the developed reactor model fully describes dynamic performance changes up to 400 h on stream.

### Oxygen evolution reaction over catalytic single-site Co in a well-defined brookite $TiO_2$ nanorod surface

DOI: 10.1038/s41929-020-00550-5

Publication outline: Main | Results | Conclusions | Methods

**Abstract-Summary**

The structural complexity of heterogeneous electrocatalysts makes it a great challenge to elucidate the surface catalytic sites and OER mechanisms. We report that catalytic single-site Co in a well-defined brookite $TiO_2$ nanorod (210) surface (Co-$TiO_2$) presents turnover frequencies that are among the highest for Co-based heterogeneous catalysts reported to date, reaching $6.6 \pm 1.2$ and $181.4 \pm 28\,s^{-1}$ at 300 and 400 mV overpotentials, respectively. Based on grand canonical quantum mechanics calculations and the single-site Co atomic structure validated by in situ and ex situ spectroscopic probes, we have established a full description of the catalytic reaction kinetics for Co-$TiO_2$ as a function of applied potential, revealing an adsorbate evolution mechanism for the OER.

### Dependency of solvation effects on metal identity in surface reactions

DOI: 10.1038/s42004-020-00428-4

Publication outline: Introduction | Results | Discussion | Methods

**Abstract-Summary**

Solvent interactions with adsorbed moieties involved in surface reactions are often believed to be similar for different metal surfaces. To reveal the importance of metal identity on aqueous solvent effects in heterogeneous catalysis, we studied solvent effects on the activation free energies of the O–H and C–H bond cleavages of ethylene glycol over the (111) facet of six transition metals (Ni, Pd, Pt, Cu, Ag, Au) using an explicit solvation approach based on a hybrid quantum mechanical/molecular mechanical (QM/MM) description of the potential energy surface. The main reason for this dependence could be traced back to a different amount of charge-transfer between the adsorbed moieties and metals in the reactant and transition states for the different metal surfaces.

### Using statistical learning to predict interactions between single metal atoms and modified MgO(100) supports

DOI: 10.1038/s41524-020-00371-x

Publication outline: Introduction | Results | Discussion | Methods

**Abstract-Summary**

We use density functional theory (DFT) and statistical learning (SL) to derive models for predicting how the adsorption strength of metal atoms on MgO(100) surfaces can be enhanced by modifications of the support. We use SL methods (i.e., LASSO, Horseshoe prior, and Dirichlet–Laplace prior) that are trained against DFT data to identify physical descriptors for predicting how the adsorption energy of metal atoms will change in response to support modification. These SL-derived feature selection tools are used to screen through more than one million candidate descriptors that are generated from simple chemical properties of the adsorbed metals, MgO, dopants, and adsorbates. Among the tested SL tools, we demonstrate that Dirichlet–Laplace prior predicts metal adsorption energies on MgO most accurately, while also identifying descriptors that are most transferable to chemically similar oxides, such as CaO, BaO, and ZnO.

Extended:

We will explore the performance of both LASSO-based FS methods and state-of-the-art Bayesian FS methods [164, 165] for finding descriptors that can predict changes in metal binding energy caused by MgO surface modification.

### Integration of theory and experiment in the modelling of heterogeneous electrocatalysis

DOI: 10.1038/s41560-021-00827-4

Publication outline: Main | Critical elements to model heterogeneous systems | Modelling reactions at metal electrodes | Modelling reactions at oxides | Outlook

**Abstract-Summary**

Theoretical and computational approaches are essential to interpret experimental data and provide the mechanistic understanding necessary to design more effective catalysts. Automated general procedures to build predictive theoretical and computational frameworks are not readily available; specific choices must be made in terms of the atomistic structural model and the level of theory, as well as the experimental data used to inform and validate these choices. The level of theory should be chosen for the specific system and properties of interest, and experimental validation is essential from the beginning to the end of the study.

### CO Oxidation Promoted by a $Pt_4$/$TiO_2$ Catalyst: Role of Lattice Oxygen at the Metal/Oxide Interface

DOI: 10.1007/s10562-018-2610-2

Publication outline: Introduction | Computational Details | Results and Discussion | Conclusions

**Abstract-Summary**

CO oxidation promoted by a subnano $Pt_4$ cluster deposited on the anatase a-$TiO_2$(101) surface has been investigated by means of DFT + U calculations. The focus of the study is on the role of supported $Pt_4$ in favoring the formation of an oxygen vacancy at interface sites between $Pt_4$ and the $TiO_2$ surface, a key step in CO oxidation reactions according to a Mars–van Krevelen mechanism. The motivation is to compare this reaction mechanism with other processes described in the literature for Pt clusters on anatase $TiO_2$ where the reaction involves $O_2$ dissociation at the surface of the metal particle or its activation at the metal/oxide interface. The processes is slightly endothermic, and occurs with barriers comparable, or even lower, than found for the case of Au nanoparticles supported on the same a-$TiO_2$ (101) surface.

Extended:

The motivation is to compare, using the same approach, the behavior of the Pt/$TiO_2$ catalyst with that of Au/$TiO_2$ [188], or Au/$ZrO_2$ [189] systems, where a MvK mechanism has been established [188] or suggested [189].

### Structural diversity of metallacycle intermediates for ethylene dimerization on heterogeneous NiMCM-41 catalyst: a quantum chemical perspective

DOI: 10.1007/s11224-018-1184-3

Publication outline: Introduction | Computational | Results and discussion | Conclusion

**Abstract-Summary**

Nanocluster models were investigated to explore the diversity of metallacycle intermediates for ethylene dimerization over NiMCM-41 at B3LYP/6-311+G* and M06/Def2-TZVP. The thermodynamic favorability of the formation of matallacycle with respect to the ring size of silica varied in the sequence of 6T < 3T < 2T < 5T < 4T in terms of Gibbs free energy (ranging from − 10.01 to 16.66 kcal/mol at B3LYP/6-311+G*). The formation of the intermediate and π complexation of 1-butene led to positive total charges on the hydrocarbon segment of the complex, being maximized on four-membered sites and minimized on two-membered ones.

Extended:

The formation of the catalytically active species from such pre-catalysts is an interesting less-addressed topic for which molecular simulation can be extremely useful. The formation of both types of clusters were found to be exothermic on all of the active sites at both levels of theory with the A4T and A5T complexes to be the most exothermic intermediates (with the enthalpies of − 43.03 and − 40.66 kcal/mol at L2, respectively).

### A review on the computational studies of the reaction mechanisms of $CO_2$ conversion on pure and bimetals of late 3d metals

DOI: 10.1007/s00894-021-04811-3

Publication outline: Introduction | $CO_2$ conversion studies on single metal crystals | Scope of review | Reaction mechanisms | Conclusions

**Abstract-Summary**

Despite series of experimental studies that reveal unique activities of late 3d transition metals and their role in microorganisms known for $CO_2$ conversion, these surfaces are not industrially viable yet. This review covers both experimental and theoretical studies into the mechanisms of $CO_2$ reduction into CO and methane, on single metals and bimetals of late 3d transition metals, i.e. Fe, Co, Ni, Cu and Zn. These mechanistic studies reveal $CO_2$ activation and reduction mechanisms are specific to both composition and surface facet. Surfaces with least $CO_2$ binding potential are seen to favour CO and O binding and provide higher barriers to dissociation. Hydrogen-assisted dissociation is seen to be generally favoured kinetically on Cu and Ni surfaces over direct dissociation except on the Ni (211) surface. Methane production on Cu and Ni surfaces is seen to occur via the non-formate pathway. Hydrogenation reactions have focused on Cu and Ni, and more needs to be done on other surfaces, i.e. Co, Fe and Zn.

### First principles investigation on the applicability of ruthenium as a potential ORR catalyst

DOI: 10.1007/s12039-019-1691-9

Publication outline: Introduction | Models and computational details | Results and Discussion | Conclusions

**Abstract-Summary**

Realizing the wide acceptance of ruthenium as a promising catalyst for various catalytic reactions, we have investigated the plausibility of Ru to perform in the bulk as well as nanoparticle forms as an efficient oxygen reduction reaction catalyst. We report here that Ru cannot be an alternative to the Pt-based catalysts owing to a high overpotential. The catalytic activity of ruthenium catalysts for ORR was investigated with reference to the recent advancements in the development of ruthenium catalysts for various electrochemical reactions. The results reveal that both the surface as well as nanocluster based ruthenium systems do not outperform the Pt(111) catalyst.

### Ability of density functional theory methods to accurately model the reaction energy pathways of the oxidation of CO on gold cluster: A benchmark study

DOI: 10.1007/s00214-016-1852-6

Publication outline: Introduction | Methods | Results and discussion | Summary

**Abstract-Summary**

Density functional theory (DFT) is the method of choice for the investigation of energy pathways of reactions assisted by metal nanoparticles due to their computational efficiency. The reliability of such theoretical studies depends to a large extent on the choice of the DFT functional used. The adsorption energies, barrier heights and reaction energies calculated using the DFT methods lie in a wide range with some methods showing high deviations from the CCSD(T) results. The percentage of Hartree–Fock exchange included in the DFT functional also plays a crucial role in predicting the correct pathway. Extensive benchmark study, it is suggested that the computationally less expensive hybrid density functionals, PBE0, B3PW91 and B3P86, are better suited for accurate modeling of this class of reactions.

### Effects of correlated parameters and uncertainty in electronic-structure-based chemical kinetic modelling

DOI: 10.1038/nchem.2454

Publication outline: Main | Results | Discussion | Conclusions | Methods

**Abstract-Summary**

Kinetic models based on first principles are becoming common place in heterogeneous catalysis because of their ability to interpret experimental data, identify the rate-controlling step, guide experiments and predict novel materials. To overcome the tremendous computational cost of estimating parameters of complex networks on metal catalysts, approximate quantum mechanical calculations are employed that render models potentially inaccurate. We rationalize why models often underpredict reaction rates and show that, despite the uncertainty being large, the method can, in conjunction with experimental data, identify influential missing reaction pathways and provide insights into the catalyst active site and the kinetic reliability of a model.

### Stability of heterogeneous single-atom catalysts: a scaling law mapping thermodynamics to kinetics

DOI: 10.1038/s41524-020-00411-6

Publication outline: Introduction | Results | Discussion | Methods

**Abstract-Summary**

We extend the current thermodynamic view of SAC stability in terms of the binding energy ($E_{bind}$) of single-metal atoms on a support to a kinetic (transport) one by considering the activation barrier for metal atom diffusion. A rapid computational screening approach allows predicting diffusion barriers for metal–support pairs based on $E_{bind}$ of a metal atom to the support and the cohesive energy of the bulk metal ($E_c$). This diffusion scaling-law provides a simple model for screening thermodynamics to kinetics of metal adatom on a support.

### Bulk and surface theoretical investigation of Nb-doped δ-FeOOH as a promising bifunctional catalyst

DOI: 10.1007/s00894-021-04864-4

Publication outline: Introduction | Theoretical methods | Results and discussion | Final remarks

**Abstract-Summary**

Iron oxyhydroxides have been identified as low-cost bifunctional catalysts. This work consists in the study, through DFT calculations, of the properties of the bulk and the surface of feroxyhyte (δ-FeOOH) doped with niobium, as a potential bifunctional catalyst. We identified the formation of stronger van der Waals interactions among the doped δ-FeOOH layers, which can increase the thermal stability of the catalyst. Evidence has been found that the insertion of Nb increases Brönsted acidity and gives rise to new Lewis acid sites on the catalyst surface.

### Methanol Oxidation on Pt(111) from First-Principles in Heterogeneous and Electrocatalysis

DOI: 10.1007/s12678-017-0370-1

Publication outline: Introduction | Computational Details | Chemical and Electrochemical Environments | Energetics of Reaction Intermediates | Hydrogen-Covered Pt(111) Electrode | Conclusion

**Abstract-Summary**

We find characteristic differences between the methanol oxidation paths in heterogeneous and electro-catalysis. The presence of the aqueous electrolyte stabilizes reaction intermediates containing hydrophilic groups thus also influencing the selectivity in the methanol oxidation. Adsorbed hydrogen on Pt(111) is shown to render the electro-oxidation of methanol less efficient.

Extended:

The presence of the aqueous electrolyte leads to the stabilization of reaction intermediates that contain hydrophilic groups such as hydroxyl when isomers are compared. The presence of adsorbed hydrogen atoms on Pt(111) is shown to be detrimental for methanol oxidation.

### Graph theory approach to determine configurations of multidentate and high coverage adsorbates for heterogeneous catalysis

DOI: 10.1038/s41524-020-0345-2

Publication outline: Introduction | Results and discussion | Methods

**Abstract-Summary**

Growing computational power has permitted the extension of such studies to complex reaction networks involving either high adsorbate coverages or multidentate adsorbates, which bind to the surface through multiple atoms. Describing all possible adsorbate configurations for such systems, however, is often not possible based on chemical intuition alone. To systematically treat such complexities, we present a generalized Python-based graph theory approach to convert atomic scale models into undirected graph representations. These representations, when combined with workflows such as evolutionary algorithms, can systematically generate high coverage adsorbate models and classify unique minimum energy multidentate adsorbate configurations for surfaces of low symmetry, including multi-elemental alloy surfaces, steps, and kinks. Two case studies are presented which demonstrate these capabilities; first, an analysis of a coverage-dependent phase diagram of absorbate NO on the $Pt_3Sn(111)$ terrace surface, and second, an investigation of adsorption energies, together with identifying unique minimum energy configurations, for the reaction intermediate propyne ($CHCCH_3$*) adsorbed on a PdIn(021) step surface.

### A Bayesian framework for adsorption energy prediction on bimetallic alloy catalysts

DOI: 10.1038/s41524-020-00447-8

Publication outline: Introduction | Results and discussion | Methods

**Abstract-Summary**

We extend the single descriptor linear scaling relation to a multi-descriptor linear regression models to leverage the correlation between adsorption energy of any two pair of adsorbates. With a large dataset, we use Bayesian Information Criteria (BIC) as the model evidence to select the best linear regression model. Gaussian Process Regression (GPR) based on the meaningful convolution of physical properties of the metal-adsorbate complex can be used to predict the baseline residual of the selected model. This integrated Bayesian model selection and Gaussian process regression, dubbed as residual learning, can achieve performance comparable to standard DFT error (0.1 eV) for most adsorbate system. For sparse and small datasets, we propose an ad hoc Bayesian Model Averaging (BMA) approach to make a robust prediction.

Extended:

We demonstrate a Bayesian framework for the model selection and model averaging for efficient and robust prediction of chemisorption energies of a few important multi-atom mono-dentate adsorbates on a bimetallic alloy dataset with the goal of novel catalytic materials discovery for hydrocarbon or nitrogen containing chemical reaction processes, such as, methanation, Fischer–Tropsch synthesis, and ammonia synthesis.

### The Effect of Hartree-Fock Exchange on Scaling Relations and Reaction Energetics for C–H Activation Catalysts

DOI: 10.1007/s11244-021-01482-5

Publication outline: Introduction | Reaction Mechanism | Computational Details | Results and Discussion | Conclusions

**Abstract-Summary**

In open shell transition metal complexes (TMCs) that are promising for challenging reactions (e.g., C–H activation), the predictive power of DFT has been challenged, and properties are known to be strongly dependent on the admixture of Hartree-Fock (HF) exchange. We carry out a large-scale study of the effect of HF exchange on the predicted catalytic properties of over 1200 mid-row (i.e., Cr, Mn, Fe, Co) 3d TMCs for direct methane-to-methanol conversion. Narrower metal/oxidation/spin-state specific LFERs perform better and are less sensitive to HF exchange than absolute reaction energetics, except in the case of some intermediate/high-spin states. The interplay between spin-state dependent reaction energetics and exchange effects on spin-state ordering means that the choice of DFT functional strongly influences whether the minimum energy pathway is spin-conserved. Despite these caveats, LFERs involving catalysts that can be expected to have closed shell intermediates and low-spin ground states retain significant predictive power.

### Microscopic understanding of electrocatalytic reduction of $CO_2$ on Pd-polyaniline composite: an ab initio study

DOI: 10.1007/s00894-018-3762-0

Publication outline: Introduction | Methods | Results and discussion | Conclusions

**Abstract-Summary**

It is found that both PANI and Pd/PANI show high selectivity for the formation of formic acid (HCOOH) over the methanol ($CH_3OH$) production. The electroreduction of $CO_2$ towards formic acid (HCOOH) follows two different pathways, depending on the catalyst: on PANI the formation of HCOOH occurs through the *COOH intermediate, whereas for the case of Pd/PANI, the same reaction proceeds through the formation of formate (*OCHO). While the formation of $CH_3OH$ from $CO_2$ on PANI is not feasible, electroreduction of $CO_2$ towards $CH_3OH$ on Pd/PANI occurs through the formation of CO.

Extended:

It is also observed that the smaller-sized nanoparticles are more efficient for this reduction reaction [257]. It is found that the formation of *OCHO is 0.34 eV more favorable than that of *COOH.

**Molecular Insights on the Role of (CTA$^+$)(SiO$^-$) Ion Pair into the Catalytic Activity of [CTA$^+$]–Si–MCM-41**

DOI: 10.1007/s11244-019-01181-2

Publication outline: Introduction | Method and Calculation Details | Results and Discussion | Conclusions

**Abstract-Summary**

A theoretical study in conjunction with a spectroscopic analysis by FTIR were carried out in order to obtain molecular insights on the role of (CTA$^+$)($\equiv$SiO$^-$) ion pair into the catalytic activity of [CTA$^+$]–Si–MCM-41, the non-calcined form of Si–MCM-41. The reaction of transesterification of ethyl acetate with methanol was used as a model of transesterification of vegetal oils using the mesoporous heterogeneous catalyst. It is found that the most favorable reaction mechanism is the dual site mechanism and that the presence of the (CTA$^+$)($\equiv$SiO$^-$) ionic pair is fundamental for the catalytic activity.

Extended:

The reaction of transesterification of ethyl acetate (which is the simplest known ester) with methanol was used as a model of transesterification of vegetal oils. Further studies on the nature of interactions on these systems are under investigation in our laboratory, and results will be published in the near future.

## 4. Recent Progress & Perspectives

### Non-Colloidal Nanocatalysts Fabricated Using Arc Plasma Deposition and Their Application in Heterogenous Catalysis and Photocatalysis

DOI: 10.1007/s11244-017-0746-8

**Abstract-Summary**

To understand technologically complex catalytic systems and to tailor both activity and selectivity in heterogeneous catalysis and photocatalysis, there is the challenge of bridging the materials gap that exists between two- and three-dimensional model catalytic systems and real catalysts, which are comprised of highly dispersed, oxide-supported metal nanoparticles. Coaxial vacuum arc plasma deposition (APD) is a method to fabricate non-colloidal nanocatalysts that has the potential for large-scale synthesis of nanocatalysts and, at the same time, allows for systematic investigation of intrinsic factors that affect catalytic activity. Direct vaporization of metallic materials to deposit active materials on two-dimensional or three-dimensional oxide supports has drawn considerable interest due to its simplicity, high reproducibility, and the possibility for large-scale production. In the case of photocatalysis, APD was used to fabricate metal nanoparticles on a hierarchically porous oxide; enhanced hydrogen evolution by doping and porosity was also demonstrated. The materials gap in catalysis can be bridged by the potential for large-scale synthesis of a variety of catalysts using APD.

Extended:

In the case of powder-type substrates, the powder is constantly stirred for effective dispersion of the deposited nanoparticles on the powder.

**Conclusion**

We have discussed the most recent advances in the preparation of catalyst nanoparticles on catalyst supports fabricated using APD. With the advent of pulsed APD, new applications for preparing metallic nanoparticles in a direct and dry process are emerging. It is also easy to control the size of the generated nanoparticles at nanometer scale by controlling the APD parameters, e.g. the number of arc plasma pulse shots, the arc discharge voltage, and the arc discharge condenser capacitance. As for the substrates for deposition, the APD method provides a simple and easy method for direct and dry deposition of metallic nanoparticles on a variety of substrates, e.g. 2-dimensional thin films and 3-dimensional powders.

**Graphitized nanocarbon-supported metal catalysts: synthesis, properties, and applications in heterogeneous catalysis**

DOI: 10.1007/s40843-017-9160-7

**Abstract-Summary**

Graphitized nanocarbon materials can be an ideal catalyst support for heterogeneous catalytic systems. This review summarizes recent relevant applications in supported catalysts that use graphitized nanocarbon as supports for catalytic oxidation, hydrogenation, dehydrogenation, and C–C coupling reactions in liquid-phase and gas-solid phase-reaction systems. The latest developments in specific features derived from the morphology and characteristics of graphitized nanocarbon-supported metal catalysts are highlighted, as well as the differences in the catalytic behavior of graphitized nanocarbon-supported metal catalysts versus other related catalysts.

**Recent Advances on the Preparation and Catalytic Applications of Metal Complexes Supported-Mesoporous Silica MCM-41 (Review)**

DOI: 10.1007/s10904-020-01689-1

**Abstract-Summary**

Mesoporous silica, in particular MCM-41 was widely used in several fields, including catalysis and catalytic support, due to its remarkable textural properties. The immobilization of metal complexes on the surface of mesoporous silica is one of the most popular methods. The metal complexes immobilized on the mesoporous silica MCM-41 exhibit significant catalytic activities due to the increased dispersion of the active sites and the faster diffusion of large organic molecules. This review mainly discusses the immobilization strategies of different metal complexes supported on the mesoporous silica MCM-41. A recent study was presented and well detailed based on the different preparation methods, catalytic applications and the stability of metal complexes supported on mesoporous silica MCM-41.

Extended:

Several reviews were published on metal complexes supported on different mesoporous silicas and their applications as heterogeneous catalysts for a wide variety in organic synthesis [14, 15, 16–17]. The metal complexes supported on MCM-41 have known wide application in organic synthesis and were used in several types of reaction due to their diversity. The metal complexes supported on MCM-41 have attracted attention because of their wide applications in heterogeneous catalysis.

**Conclusions**

The mesoporous silica MCM-41 supported by metal complexes was presented. The metal complexes supported on MCM-41 have attracted attention because of their wide applications in heterogeneous catalysis. These kinds of catalysts have been widely tested in several types of reactions such as oxidation, ozonation, reduction and photocatalysis. It is very interesting to widen the application field of metals complexes via this kind of reaction. Polymers have also been known for their effectiveness to stabilize metallic nanoparticles, so it is very interesting to prepare and test the complexes polymer-metal supported on MCM-41 in water treatment.

**Graphene-supported metal single-atom catalysts: a concise review**

DOI: 10.1007/s40843-019-1286-1

**Abstract-Summary**

SACs could greatly increase the availabilities of the active metal atoms in many catalytic reactions by reducing the size to single atom scale. Graphene-supported metal SACs have also drawn considerable attention due to the unique lattice structure and physicochemical properties of graphene, resulting in superior activity and selectivity for several chemical reactions. We review recent progress in the fabrications, advanced characterization tools and advantages of graphene-supported metal SACs, focusing on their applications in catalytic reactions such as CO oxidation, the oxidation of benzene to phenol, hydrogen evolution reaction, methanol oxidation reaction, oxygen reduction reaction, hydrogenation and photoelectrocatalysis.

### N-Heterocyclic carbenes as tunable ligands for catalytic metal surfaces

DOI: 10.1038/s41929-021-00607-z

**Abstract-Summary**

The active sites of these heterogeneous catalysts are often metal surfaces, which consist of a large ensemble of metal atoms, and their properties differ significantly from metal complexes bearing highly developed ligands typically used in homogeneous catalysis. A highly interesting approach is the combination of these sophisticated ligand systems from homogeneous catalysis with metal surfaces to improve parameters such as stability, reactivity and selectivity. This review summarizes developments in this area, covering fundamental properties, preparation methods, important analytical techniques and reactions of NHC-modified metal surfaces.

**Conclusions and outlook**

NHCs have been used to control a variety of metal-surface properties, such as stability and solubility, but also important catalytic features like selectivity and activity. Crucial for this development has been the fundamental understanding of the binding of NHCs to surfaces. A plethora of analytical techniques for the analysis of surfaces modified with NHCs have been established, and further development is essential. Design strategies of tailor-made NHCs for specific applications in catalysis are highlighted and we expect that the growing understanding of NHC-modified surfaces will lead to further groundbreaking developments. Future developments should aim at expanding the repertoire of reactions with these catalytically active metal surfaces, making use of the full structural diversity of NHCs to broaden existing approaches and identify new design strategies. The key to success should be the establishment of principles to determine which subclass of NHCs is most promising for the modification of a given surface, which is so far still challenging.

### Heterogeneous single-atom catalysis

DOI: 10.1038/s41570-018-0010-1

**Abstract-Summary**

Aided by recent advances in practical synthetic methodologies, characterization techniques and computational modelling, we now have a large number of single-atom catalysts (SACs) that exhibit distinctive performances for a wide variety of chemical reactions. SACs have well-defined active centres, such that unique opportunities exist for the rational design of new catalysts with high activities, selectivities and stabilities. Given a certain practical application, we can often design a suitable SAC; thus, the field has developed very rapidly and afforded promising catalyst leads. The control we have over certain SAC structures paves the way for designing base metal catalysts with the activities of noble metal catalysts.

**Conclusions and future directions**

To supported metal NP or nanocluster catalysts, in which metal–metal bonding between like atoms dominates the chemistry, the present catalysts feature atomically dispersed metal centres — no homoatomic metal–metal bonds are present, and the metal atoms are attached to the support surface through either heteroatoms, in the case of SACs, or other metals, in the case of SAAs. The single metal sites in SACs more or less resemble the peripheral atoms at the NP–support interface of supported NP catalysts. The non-uniformity of support materials means that not all of the single metal centres in SACs are equally accessible or equally active. The rapid development of single-atom catalysis in recent years suggests that it is not unreasonable to expect speedy access to robust SACs with high stability, selectivity and activity for industrially important reactions.

### Ultra-stable metal nano-catalyst synthesis strategy: a perspective

DOI: 10.1007/s12598-019-01350-y

**Abstract-Summary**

The small particles exhibit superior catalytic activity in comparison with the larger particles because of an increase in low-coordinated metal atoms on the particle surface that work as active sites, such as edges and corner atoms. These small NPs are typically unstable and tend to migrate and coalescence to reduce their surface free energy during the real catalytic processes, particularly in high-temperature reactions. A means to fabricate stable small metal NP catalysts with excellent sinter-resistant performance is necessary for maintaining their high catalytic activity. At the end of this review, we highlight our recent work on the preparation of high-stability metal catalysts via a unique interfacial plasma electrolytic oxidation technology, that is, metal NPs are well embedded in a porous MgO layer that has both high thermal stability and excellent catalytic activity.

Extended:

We provide our viewpoints about the future development of ultra-stable supported metal nano-catalyst synthesis.

**Introduction**

Catalysts based on metal nanoparticles (NPs) have attracted great attention because of their excellent catalytic activities. Enormous efforts have been devoted to improving the performance of supported metal catalysts by downsizing NPs [76, 77]. Noble metal NPs, as a type of dominant-supported catalyst, show excellent catalytic activity in various reactions due to their special electron structures and small particle sizes that generate a high fraction of surface active atoms. Rationally and effectively stabilizing ultra-small metal NPs on supports against sintering and aggregation are of great significance for heterogeneous catalysis in practical applications. The ideal sinter-resistant catalyst typically has the following characteristics [78]: (1) its integrated components are thermally stable and without structural collapse under high-temperature conditions; (2) its active sites are fully accessible by reactants without substantial mass transfer resistance; (3) its catalytic activity remains unchanged even after treatment at elevated temperatures. Recent theoretical and experimental studies demonstrate that either a thermodynamic approach or a kinetic approach [78, 79, 80–81] can effectively stabilize noble metal NPs on supports.

**Thermodynamic approach**

The thermodynamic approach aims to reduce the difference in chemical potential and surface energy between the NPs by which the Ostwald ripening can be effectively suppressed. They found that particle size and distribution are essential factors for determining their stability and a narrow size and homogeneous distribution suppresses the Ostwald ripening rate due to the NPs similar chemical potential and surface energy. A support with high surface activation energy can generate the same effect for restraining the Ostwald ripening action.

**Kinetic approach**

The dynamical approach mainly takes advantage of an external medium to restrain the growth of NPs in the harsh condition during the catalytic reaction, e.g., high temperature, high pressure, etc In this section, three strategies including space confinement, coating structure and strong metal–support interaction are proposed to stabilize the NPs. There are also some other important approaches focusing on the stabilization of tiny metal NPs within zeolite matrix to improve their sintering resistance, for instance, incorporating Pt clusters into high-silica zeolites MCM-22 via the transformation of two-dimensional (2D) zeolite into a three-dimensional (3D) [82] structure, stabilizing CoO clusters within silicalite-1 crystals using a steam-assisted method [83], etc Even so, simpler and more universal strategies should be put forward to enrich the types of catalysts to meet the increasing demand of industrial applications. This unique structure makes the Au NPs partially encapsulated and partially exposed to the air, which generates a strong interaction between the Au NPs and the mixed supports that facilitate the stability of the Au particles, and also enhances their catalytic activity due to the partially exposed surface.

**Interfacial plasma electrolytic oxidation technology**

Unlike the methods above, a new preparation technology, termed interfacial plasma electrolytic oxidation technology (PEO), has been successfully applied to fabricate a high-quality Au/MgO catalyst with superior stability and activity for catalytic reactions. During the PEO process, noble metal precursors are in situ decomposed and transformed into metal NPs along with their encapsulation into the MgO porous layer. Owing to this unique embedded structure, the obtained Au/MgO catalysts possess excellent thermal stability and sintering resistance. Because the oxide layer is firmly attached to metal substrate, the concentration of oxygen vacancies in MgO cannot be annihilated by high-temperature annealing. The foremost characteristic of this approach is the simultaneous in situ formation of metal NPs and an oxide support that generates a unique embedded structure between them and endows the catalyst outstanding thermal stability and recyclability.

**Summary and outlook**

Two main paths including thermodynamic and kinetic approaches were performed during the entire developed solution. Most solutions involve the kinetic approaches including channel confinement, coating and SMSI. Consideration SMSI is only suitable for some special metal–support pairs that have strictly limited the application in the field of catalysis, our group developed an SMSI based on universal strategy– interfacial plasma electrolytic oxidation technology for the preparation of ultra-stable supported catalysts, which is appropriate for various noble or transition metal catalysts and hence largely extends the scope of SMSI.

## Whither Goest Thou, Catalysis

**Abstract-Summary**

While the history of catalysis goes back to antiquity, catalysis as we know it today began in the nineteenth century with catalytic hydrogenation having its beginning with the report by Sabatier in 1897 on the hydrogenation of ethylene over a reduced NiO catalyst. About the middle of the last century, though, catalysis, particularly catalytic hydrogenations and oxidations, began to be used more extensively in synthetic applications. The last quarter of the past century saw an increased interest in understanding the details of the chemistry taking place on the catalyst surface. EHMO calculations provided information on the nature of the interaction between the substrate and metal catalyst surface. This approach was recently replaced by the use of DFT calculations to obtain information on the energetics of the interaction of a metal surface and the substrate or presumed reaction intermediates. Another twenty-first century innovation was the introduction of "nano-technology" with nano-particles of metals being used as catalysts. Has the research leading to a more detailed understanding of the catalyst surface led to the preparation of more active and selective catalysts or are the better catalysts still being prepared by the old trial and error approach? More efficient catalysts could be designed if there were a better understanding of what was taking place on the surface of 'real world' catalysts rather than on idealized substitutes. The presentation will cover a brief review of catalysis with emphasis on catalytic hydrogenation including proposals about the nature of catalytic active sites which were made through the years.

**Where Do We Go From Here?**

A high percentage of these DFT calculations have been concerned with the adsorption energies of various species on the (111) face of a catalytically active metal even though it is generally accepted that few reactions take place on these face atoms. Another factor is that in studies of hydrogenation reactions these calculations seldom include the simultaneous adsorption of hydrogen even though it has been shown that covering the catalyst surface with hydrogen inhibits adsorption on (111) face atoms [112]. Could a mixed metal composition of a catalyst to be used for a specific, highly selective reaction be determined in advance of any reaction studies? We have a number of new techniques which have been developed to obtain information about the substrate-catalyst interaction, the detailed composition of many catalysts, the nature of the active sites on catalysts and other features of catalytic reactions.

## Oxygen Reduction Reactions of Fe-N-C Catalysts: Current Status and the Way Forward

DOI: 10.1007/s41918-019-00030-w

**Abstract-Summary**

Fe-N-C materials are considered to be among the most important oxygen reduction reaction (ORR) catalysts, because they are potential substitutes for Pt-based catalysts and are therefore promising in the development of non-noble metal-based catalysts. From a chemical stand point, improvements can be made through the better understanding of mechanisms in Fe-N-C-based ORR catalysis along with a deeper understanding of the chemical origin of active sites on Fe-N-C catalyst surfaces. Based on these, this comprehensive review will focus on the energy conversion, transformation kinetics and electron transfer of the ORR process as catalyzed by Fe-N-C catalysts. This review focuses on the profound understanding of heterogeneous oxygen reduction reaction on Fe-N-C materials from the following aspects: (1) thermodynamics of energy conversion in ORR processes, (2) kinetics of ORR processes based on Fe-N-C catalysts, (3) the textural features of Fe-N-C and analytic results known as far, (4) fundamental principle for Fe-N-C materials synthesis and (5) practical application for fuel cell and metal–air batteries.

**Introduction**

One may note that from the stand point of a larger impact, Fe-N-C ORR materials are also relevant to other areas, such as selective oxidation reaction [113, 114, 115–116], $CO_2$ reduction [117, 118] and hydrogenation [119, 120]. A timely comprehensive review on this rapidly growing category of Fe-N-C for ORR catalysis is of great significance to a vast body of readers. Several review articles [121, 122, 123, 124, 125, 126–127], related or referred to the preparation and electrochemical performance of Fe-N-C materials for ORR catalysis, have been published. We engage to the profound understanding of heterogeneous oxygen reduction reaction on Fe-N-C materials from the following aspects: (1) the thermodynamics of energy conversion in the ORR process, (2) the kinetics of the ORR process based on Fe-N-C catalysts, (3) the textural features of Fe-N-C and analytic results known as far, (4) the fundamental principles for Fe-N-C materials synthesis and (5) the practical applications in fuel cells and metal-air batteries.

**Thermodynamic Description of the ORR Process**

In fuel cells, the high efficiency of energy conversion (i.e., from chemical to electrical) in terms of the theoretical thermodynamics makes it be a promising technique for an energy-sufficient in future. The thermal efficiency of ideal fuel cells operating reversibly on pure hydrogen and oxygen at standard conditions is: Then, an actual cell thermal efficiency is: At standard state, in which $E_{ideal}$ = 1.23 V, and hence: In a fuel cell, oxygen ($O_2$) is electrochemically reduced at the cathode through the mediation of ORR. Taking the overpotential ($E_{op}$) into consideration for a real fuel cell, the actual cell efficiency is found to be: Therefore, a high $E_{ORR}$ with low $E_{op}$ is anticipated, where the $E_{ORR}$ depends on the process of electron transfer pathway and the $E_{op}$ is up to the cathodic electrode materials and the electrolyte.

**Kinetic Mechanisms of ORR Processes**

The kinetic mechanism of reaction can be changed by a catalyst, and doing so would not affect the thermodynamics at all. As a promising non-precious metal catalysis, studying the kinetics of ORR on Fe-N-C is very useful in understanding the mechanism of ORR and for further improving the catalytic performance. A suitably chosen cationic surfactant can modulate the ORR selectivity on a Fe-N-C catalyst by kinetic promotion through Coulombic interaction with the peroxo-Fe complexes [128]. As the first step during ORR, the adsorption and activation of $O_2$ on catalysts involves associative or dissociative pathways, or a combination of these. It may be noted that the kinetics of ORR is determined to a large extent by the choice of electrode materials and the electrolyte chosen. A clear understanding of the kinetic mechanisms involved in ORR on Fe-N-C would facilitate further research to design, synthesize and investigate ORR catalysts.

**[Section 4]**

The Fe atom in Fe-N-C materials could be coordinated by the pyridine/pyrrole nitrogen, carbon or a combination of C and N [129]. For the sake of the effective catalyst design, it is hence reasonable to identify and ideally, quantify the Fe-N-C structure–ORR catalytic activity correlations [130, 131]. To unravel the nature of Fe-N-C structure, the techniques of X-ray photoelectron spectroscopic (XPS), electron microscopy, Mössbauer spectroscopic and X-ray absorption fine structure (XAFS) are mostly available. XPS is effective to elucidate the chemical composition and nitrogen bonding configuration for Fe-N-C materials. This peak differentiation principle is widely used in the research on structure of the Fe-N-C catalysts and their correlations with the electrochemical performance for ORR [132]. Pure FeO, $Fe_2O_3$ or iron phthalocyanine are used as references to clarify that Fe in Fe-N-C materials is coordinated with nitrogen but not oxygen [132, 133, 134].

**[Section 5]**

For ORR on Fe-N-C materials, to improve the activity (i.e., for facilitating substance transformation and to enhance electron transfer rate), similar to the traditional heterogeneous catalyst, there are two strategies: (1) increasing the intrinsic activity of Fe-N-C or (2) increasing the number of active site on a given electrode (e.g., through increased loading or tuned catalyst structure to expose more active sites per gram) [135]. Hence, a catalyst consisting of abundant Fe-N-C site with high intrinsic activity is desirable; in particular there would be advantages to have sufficient $Fe-N_2$ species on the edge of carbon support. A high content of Fe-N-C active sites is essential to deliver high ORR catalytic performance. It is generally recognized that the avoidance of carbide formation during pyrolysis represents a promising way to enhance the density of ORR active sites on Fe-N-C catalysts [136].

**Application for Energy Conversion Systems**

Currently, the dominant limiting factor in PEMFCs, unlike APEFCs, which is the PEM, is the electrocatalyst [137], and therefore, the focus of this review will be on the application of Fe-N-C catalysts for PEMFCs. In comparison with Pt/C, the nominally high power density of the reported Fe-N-C catalyst was related to higher mass loadings. This can be explained by the facts that Fe-N-C catalysts are carbonized at high temperatures (> 700 °C) and are more hydrophobic, and that although the porous structure of Fe-N-C catalysts can contribute to higher surface areas for active site dispersion and mass transport enhancement, it also lightens bulk density and cause these catalysts to be much thicker than Pt/C at same mass loading. The Fe-N-C materials show promising applications for the booming activities in Zn-air battery (ZABs).

**Challenges and the Way Forward: a Perspective**

Significant progress has been made in developing Fe-N-C materials for oxygen reduction reactions in both alkaline and acidic media. Most of the work relating to Fe-N-C for ORR being reported recently is focused on novel methods for material preparation; however, the activity seems to be stagnate [138, 139]. In acidic media, the activity of Fe-N-C materials is much less than Pt/C with ~ 50 mV lower of $E_{1/2}$. Is it the extreme value for actual Fe-N-C materials, although the activity of Fe-N-C materials can be further theoretically enhanced by the structural optimization? Despite the many challenges, there is reason to be hopeful about Fe-N-C for ORR since advancements in situ and ex situ characterization techniques will continue to give insights of relevance to progressively move toward champion catalysts.

**Molecular enhancement of heterogeneous $CO_2$ reduction**

DOI: 10.1038/s41563-020-0610-2

**Abstract-Summary**

The electrocatalytic carbon dioxide reduction reaction ($CO_2RR$) addresses the need for storage of renewable energy in valuable carbon-based fuels and feedstocks, yet challenges remain in the improvement of electrosynthesis pathways for highly selective hydrocarbon production. Organic molecules or metal complexes adjacent to heterogeneous active sites provide additional binding interactions that may tune the stability of intermediates, improving catalytic performance by increasing Faradaic efficiency (product selectivity), as well as decreasing overpotential. We introduce present-day challenges in molecular strategies and describe a vision for $CO_2RR$ electrocatalysis towards multi-carbon products. These strategies provide potential avenues to address the challenges of catalyst activity, selectivity and stability in the further development of $CO_2RR$.

**Conclusion and outlook**

This Perspective has highlighted the potential of applying molecular strategies to enhance heterogeneous electrochemical $CO_2RR$. A distinct advantage of molecular enhancement approaches is that structure–function relationships can be more precisely defined. Molecularly enhanced heterogeneous $CO_2RR$ catalysts offer the potential to overcome major challenges in tunability and stability as a frontier opportunity for $CO_2RR$ electrocatalysis.

**Modeling Ceria-Based Nanomaterials for Catalysis and Related Applications**

DOI: 10.1007/s10562-016-1799-1

**Abstract-Summary**

Atomistic and electronic structure details of the functioning of ceria in catalysis can nowadays be successfully uncovered with the help of computational modeling based on the density functional theory (DFT). The majority of such computational studies undertaken so far relied on extended models of surfaces, which are adequate for the description of surface science processes and phenomena, but neglect the nanostructured nature of ceria in many catalysts. This Perspective focuses on discussing DFT calculations of various nanostructured models of ceria and its composites relevant for catalysis. Pivotal consequences of ceria nanostructuring for its role in catalysts derived from the computational studies are documented and supported by experimental results. How prone are metal particles deposited on ceria to sintering or dispersion and how is this interplay controlled by the nanostructuring of the support? Under what conditions will the transfer of lattice oxygen atoms from ceria support to the metal particles deposited thereon become energetically favorable? The discussed examples show that accounting for ceria nanostructuring in catalysts is essential for performing trustworthy computational modeling.

**Introduction**

It means that common surface-science models in the form of ordered extended single-crystal ceria surfaces often inadequately represent the structure and reactivity of active sites on the nanostructured ceria. In other articles reviewing DFT studies of metal-oxide nanostructures relevant for catalysis [150, 151] results for ceria-based nanomaterials were not presented in due details. We consider timely to discuss in this Perspective recent research efforts to theoretically model the structure and reactivity of ceria-related catalytic nanomaterials and to overview the achieved advances and remaining issues. The authors have been actively involved in the theoretical modeling of nanostructured ceria and its nanocomposites with transition metals from the opening of this research area 10 years ago [152, 153], over the past decade (e.g. [154, 155, 156]), and up to very recently (e.g. [157–158]). That whenever possible, we refrain from discussing in this article technical details and problems of the DFT-based description of metal-oxide materials such as ceria and of their reactivity and refer for more information to two recent Perspective articles [159, 160].

**Outlook and Concluding Remarks**

Despite of the successful progress, computational studies involving explicit ceria NP models are still scarce, mainly due to the substantial cost of DFT calculations involving size-representative models of large nanostructures combined with multitude of surface sites exposed by the latter. Our understanding of the catalytic properties of nanostructured ceria-based catalysts and, especially, of their particularities with respect to extended surfaces, should benefit from more systematically addressing reaction mechanisms of different reactions on such models. This should subsequently allow evaluating the catalytic properties of sites found on different ceria-based nanostructured systems without the need of explicit reaction mechanism calculations, which ought to be especially attractive for studying large nanostructured models. Despite of its rather recent commencement, such modeling has been successfully used in combination with experimental studies for characterizing systems where the properties of ceria-based systems are strongly affected by their nanostructure.

## Recent progress and prospect of carbon-free single-site catalysts for the hydrogen and oxygen evolution reactions

DOI: 10.1007/s12274-021-3680-9

### Abstract-Summary

The key challenge for scalable production of hydrogen from water lies in the rational design and preparation of high-performance and earth-abundant electrocatalysts to replace precious metal Pt and $IrO_2$ for hydrogen evolution reaction (HER) and oxygen evolution reaction (OER). Developing highly active and stable OER electrocatalysts is the key for electrochemical water splitting. This review presents feasible design strategies for fabricating carbon-free single-site catalysts and their applications in HER/OER and overall water splitting.

## Catalytic conversion of solar to chemical energy on plasmonic metal nanostructures

DOI: 10.1038/s41929-018-0138-x

### Abstract-Summary

The demonstrations of visible-light-driven chemical transformations on plasmonic metal nanostructures have led to the emergence of a new field in heterogeneous catalysis known as plasmonic catalysis. In this Review, we provide a fundamental overview of plasmonic catalysis with emphasis on recent advancements in the field. It is our objective to stress the importance of the underlying physical mechanisms at play in plasmonic catalysis and discuss possibilities and limitations in the field guided by these physical insights.

### Current trends and outlook

These multi-electron processes have been observed mainly when charge scavengers have been employed, so to fully understand these process it is critical to: (1) establish that the products are actually derived from the reactants ($N_2$ or $CO_2$) rather than from the scavengers or impurities, (2) evaluate the quantum efficiencies in terms of the reaction rates obtained and the light intensity used in the experiments — very low rates or large intensity requirements would be discouraging in this context — and (3) critically assess the overall thermodynamic efficiencies of these processes as these processes often employ high-energy scavenger molecules. Engineering selective chemistry will require engineering of not only optical properties of the plasmonic metal, but also the electronic structure of the reactants adsorbed on the active centres as well as the ground and excited potential energy surfaces.

## Bridging homogeneous and heterogeneous catalysis by heterogeneous single-metal-site catalysts

DOI: 10.1038/s41929-018-0090-9

### Abstract-Summary

In heterogeneous single-metal-site catalysts (HSMSCs) the active metal centres are located individually on a support and are stabilized by neighbouring surface atoms such as nitrogen, oxygen or sulfur. Modern characterization techniques allow the identification of these individual metal atoms on a given support, and the resulting materials are often referred as single-atom catalysts. Their electronic properties and catalytic activity are tuned by the interaction between the central metal and the neighbouring surface atoms, and their atomically dispersed nature allows for metal utilization of up to 100%.

### Challenges and perspectives

Heterogeneous single-metal-site catalysts combine features of homogeneous and heterogeneous catalysis. In metal-based catalysts, any metal atom constitutes a single low-coordinated active site. Researchers have already sensed the importance of using specific ligands for the fabrication of HSMSCs; for example, a well-defined active site containing single cobalt atoms was reported [191]. Structural studies at the atomic level play an important role for understanding the essential factors of single-metal-site catalysts. New characterization tools on an atomic level are required for the detailed understanding and coherent synthesis of HSMSCs; for example, in homogeneous catalysis the steric and electronic nature of the active centre are in part described as a result of the so-called cone angle or bite angle of the metal with certain ligands.

## Single-Atom Catalysts: From Design to Application

DOI: 10.1007/s41918-019-00050-6

### Abstract-Summary

Single-atom catalysis is a powerful and attractive technique with exceptional performance, drastic cost reduction and notable catalytic activity and selectivity. In single-atom catalysis, supported single-atom catalysts contain isolated individual atoms dispersed on, and/or coordinated with, surface atoms of appropriate supports, which not only maximize the atomic efficiency of metals, but also provide an alternative strategy to tune the activity and selectivity of catalytic reactions.

### Introduction

To address these issues, the downsizing of noble metals from nanoclusters to isolated single atoms is the most effective method to provide optimal active sites in corresponding catalysts to maximize metal atom efficiency and maintain necessary catalytic performances [192, 193–194]. Single-atom catalysts (SACs), a class of catalysts in which catalytically active individual and isolated metal atoms are anchored to supports, have emerged as a novel class of catalysts that can exhibit optimal metal utilization, with all metal atoms being exposed to reactants and available for catalytic reactions [194]. Despite the observed performances for various reactions however, significant challenges associated with the synthesis and stabilization of SACs exist as well in which a key challenge in the application of SACs is the stabilization of isolated metals on supports without the compromise of catalytic activity, especially at high temperatures or under harsh reaction conditions [195, 196, 197–198].

**Conclusion and Perspectives**

The choice of supports and corresponding properties are important for the design of novel SACs because different supports can provide different anchoring sites to stabilize single metal atoms and provide different activities [215, 216, 217, 218–219]. To achieve high loading of single metal atoms on supports, high surface area and robust supports with large numbers of anchor sites are needed in which anchor sites on support materials can determine the active and stability of metal SACs. A deep understanding of both sintering mechanisms and active mechanisms for single metal atoms under working conditions can provide insights into the design of high-performance catalytic systems and provide useful criteria for the design of industrial catalysts.4.Characterization methods advanced characterization methods such as HAADF-STEM, STM, XAS spectroscopy and IR spectroscopy are essential to develop highly active and stable SACs.

**Single-Atom Catalysts: Advances and Challenges in Metal-Support Interactions for Enhanced Electrocatalysis**

DOI: 10.1007/s41918-021-00124-4

**Abstract-Summary**

Since metal-support interactions (MSIs) in SACs exert a substantial influence on the catalytic properties, gaining a profound understanding and recognition of catalytic reactions depends greatly on investigating MSIs both experimentally and computationally. The engineering and modulation of MSIs are regarded as one of the most efficient methods to rationally design SACs with disruptively enhanced catalytic properties. We then discuss the existing MSIs in SACs and elucidate the significant role of strong MSIs in catalytic properties and mechanisms. The correlation between strong MSIs and electrocatalytic activities in SACs, including an outlook to increase our understanding of MSIs, is discussed. The present review provides some perspectives and an in-depth understanding of strong MSIs to advance high-performing SACs.

Extended:

The essential role of strong MSIs is not limited to improving electron transfer in SACs [220]. The strategies mentioned above, the ball-milling method [221], selected soft-landing method [222], plasma sputtering [223], photochemical reduction [224, 225], etc , aim to develop SACs by regulating MSIs. The investigation of MSIs in SACs is expected to play an essential role in the future development of heterogeneous electrocatalysis to develop renewable and environmentally friendly energy resources. The large-scale commercial application of SACs will be realized in the near future, and the development of economically green energy conversion and storage technologies as well as catalytic sciences will be improved by SACs with strong MSIs.

**Conclusion and Perspectives**

Depending on the energy conversional process, an ideal SAC can be prepared by effectively balancing its own activity and stability through MSIs, where neither an MSI that is too strong, which keeps SAs centered away from activity inhibition, nor an MSI that is too weak, which results in a decrease in stability, is favored. By precisely regulating the MSIs, the location, content, and distribution of the SAs as well as the architecture of the support, can be well controlled, and the electronic and geometric environments of the SACs can be further modified; therefore, a group of high-performance SACs can be generated to advance electrocatalytic applications. Efforts should be directed toward further dynamic and in situ characterizations of SACs and reaction intermediates and studying the electronic and geometric environment of the catalytic centers to gain a deeper understanding of MSIs.

**Beyond Nanoparticles: The Role of Sub-nanosized Metal Species in Heterogeneous Catalysis**

DOI: 10.1007/s10562-019-02734-6

**Abstract-Summary**

Titanate nanostructures are of great interest for catalytic applications because their high surface area and cation exchange capacity create the possibility to achieve high metal dispersion. Due to the large amount of defects, titanate nanotubes (TNT) can stabilize sub-nanosized gold clusters, presumably in $Au_{25}$ form. This perspective summarizes the previous results obtained in the photocatalytic transformation of methane in which the size of gold nanoclusters plays an important role. Photocatalytic measurements revealed that methane is active towards photo-oxidation. Based on recent additional results, we stress here that gold clusters ($Au_{25}$) may be directly involved in the photo-induced reactions, namely in the direct activation of the methane/$Au_{25}^{\delta+}$ complex during irradiation. Another new finding is that gold nanoparticles supported on TNT exhibit high catalytic activity in $CO_2$ hydrogenation. Our results revealed fundamental differences in the reaction schemes as the products of the two routes are CO (thermal process) and $CH_4$ (photocatalytic route), indicating the importance of photogenerated electron–hole pairs in the reaction. Gold in nano and sub-nano sizes promotes the adsorption and scission of reactants, important for both types of reactions. Gold ions ($Au^+$), in the cationic sites of the titanate lattice promote the photocatalytic transformation of formate (which is one of the intermediates), thus advancing the reaction further towards the fully reduced product.

**Concluding Remarks**

Recent literature data indicates that sub-nanosized gold clusters ($Au_8$) on MgO and $Au_{25}$ on $CeO_2$ rods can be prepared and characterized. Ion exchange allows titanate nanostructures to incorporate metal adatoms in their framework, which may create another type of active center besides metal clusters. Similar to $CeO_2$ rods, TNT can also stabilize sub-nanosized gold clusters in Au₂₅ form. Au/TNT decorated with sub-nanosized gold clusters exhibited the highest photocatalytic activity in methane transformation. Small metal clusters (Au$_{25}$, probably with a partial positive charge) can bind strongly to defect sites in TNT. These clusters may be directly involved in photo-induced reactions, namely in the direct activation of the methane-Au$_{25}$ complex during irradiation. We should mention that small, even sub-nanosized, clusters can be prepared on non-oxide type supports as well. Very small Au clusters can also be prepared on hexagonal boron nitride [258, 259].

### Advances in transition metal oxide catalysts for carbon monoxide oxidation: a review

DOI: 10.1007/s42114-019-00126-3

**Abstract-Summary**

The performances of transition metal catalysts are highly dependent on the crystallite size, surface area, and pore volume of the catalysts. The chemisorptions of CO over transitional metal and supported catalysts were studied in this review. The transition metal catalysts have been represented an excellent catalyst from lower cost, thermally, activity, and selectivity point of view. This investigation will show scientific basis for potential design of transition metal oxide catalysts for CO oxidation.

Extended:

The performances of most active transition metal oxides, such as Co$_3$O$_4$, CuOx, Ni$_2$O$_4$, ZnOx, or MnOx, are highly dependent on the reaction conditions [260]. The performances of silver catalysts are mostly dependent upon their surface structure and composition [261]. The transition metal oxides (TMOs) constructions, buildings, designing, and structures have been represented as one of the most important useful elements in the area of catalysis, fuel cells, energy storage, air pollution control, and so on. The transition metal oxides have superior potential for reduction of CO in the atmosphere so that it is mostly used in a catalytic converter [262, 263]. The transition metal catalysts are highly prominent metal oxide catalysts for ambient temperature CO oxidation.

**Conclusions**

The transition metal catalysts are highly prominent metal oxide catalysts for ambient temperature CO oxidation. The CO oxidation over transition metal catalysts to be associated in the presence of several forms of chemisorbs reactants, products, and intermediates on the surface. The accumulation of suitable promoters, supports, pretreatment, and superior preparation methods would result to improvement in the performances of transition metal catalyst toward CO oxidation. The relative ease of preparation, good thermal and chemical stability, and high activity of transition metal catalysts offer better activity for automobile vehicle pollution control applications.

### Recent advances in gold-metal oxide core-shell nanoparticles: Synthesis, characterization, and their application for heterogeneous catalysis

DOI: 10.1007/s11705-015-1551-1

**Abstract-Summary**

This paper reviews the most recent work and research in coreshell catalysts utilizing noble metals, specifically gold, as the core within a metal oxide shell. The advantage of the core-shell structure lies in its capacity to retain catalytic activity under thermal and mechanical stress, which is a pivotal consideration when synthesizing any catalyst. The selective oxidation of carbon monoxide and reduction of nitrogen containing compounds, such as nitrophenol and nitrostyrene, have been studied over the past few years to evaluate the functionality and stability of the core-shell catalysts. Different factors that could influence the catalyst's performance are the size, structure, choice of metal oxide shell and noble metal core and thereby the interfacial synergy and lattice mismatch between the core and shell. This review covers the synthesis and characterization of gold-metal oxide core-shell structures, as well as how they are utilized as catalysts for carbon monoxide (CO) oxidation and selective reduction of nitrogen-containing compounds.

### Chapter model systems in heterogeneous catalysis at the atomic level: a personal view

DOI: 10.1007/s11426-019-9671-0

**Abstract-Summary**

Novel instruments and technique developments, as well as their applications are reported, in an attempt to cover studies on model systems of increasing complexity, including some of the key ingredients of an industrially applied heterogeneous catalyst and its fabrication. The review is intended to demonstrate the power of model studies in understanding heterogeneous catalysis at the atomic level.

### Catalytic CO Oxidation on Nanocatalysts

DOI: 10.1007/s11244-018-0920-7

**Abstract-Summary**

Since CO can be easily adsorbed on active metals like Au, Pt, Pd, Rh, and Ru, many researchers have attempted to observe surface phenomena, including adsorption, surface restructuring, and oxidation kinetics on diverse metal surfaces ranging from single crystals to nanoparticles (NPs). The rapid advances in nanochemistry have further stimulated studies of catalytic reactions, and in-depth understanding of CO oxidation is possible based on analyses of how the size, shape, and composition affect the activity and determine the most effective active site of metal NPs. Recent research has demonstrated that the CO oxidation activity varies with the size of metallic NPs. The oxidation state of the NPs varies as a function of the size, and the surface oxide layers formed on NPs have been found to be important in enhancing or suppressing CO oxidation. The NP surfaces undergo significant adsorbate-induced structural changes, as confirmed by various in situ characterization techniques under catalytically relevant CO oxidation conditions. Using in situ characterization techniques combined with well-defined NPs, it is possible to study CO oxidation on the molecular level in order to gain vital mechanistic insights under catalytic working conditions.

## Summary and Outlook

Classical studies of CO adsorption and dissociation studies on single crystals offer valuable information about the properties of metal surfaces and their dynamic behaviors, such as adsorbate-induced surface restructuring and changes in the oxidation states. As single crystal studies shift to the use of NPs, CO oxidation remains a valuable model catalytic reaction. The current trends in heterogeneous catalysis involve molecular level investigation of nanostructured catalysts under catalytically relevant environments to study surface phenomena using in situ characterization tools. By maintaining catalytically relevant reaction environments in specially designed cells, the dynamic behavior on NP surfaces can be investigated to monitor the real motion of catalysts during reaction by employing surface instruments.

### Integrating the Fields of Catalysis: Active Site Engineering in Metal Cluster, Metal Organic Framework and Metal Single Site

DOI: 10.1007/s11244-020-01248-5

**Abstract-Summary**

Research evolved using nanoparticles synthesized and characterized under reaction conditions opened the door to study all three fields of catalysis: heterogeneous, homogenous, and enzyme. We envision that the combination of active site engineering and unifying fields of catalysis could be applied to solve practical issues of science-based technology and develop new fields useful in energy research.

**Introduction**

We successfully synthesized a heterogeneous-enzyme hybrid catalyst by incorporating the enzyme-like active sites into metal organic frameworks (MOFs) for selective oxidation methane to methanol. These supported DEMCs could be readily modulated to combine the advantages of both heterogeneous and homogeneous catalysts: (1) DEMCs could catalyze homogenous reactions while typical heterogeneous catalyst could not realize; (2) tunable selectivity and activity by changing the dendrimer generation and terminal functional groups; (3) easy recyclability compared to homogeneous catalysts. Although the catalysts deactivate after the first cycle as a result of water molecules strongly bonded to the active sites, MOF-808-Bzz-Cu shows the highest reported selective methanol productivity of 71.8 μmol $g^{-1}$ at this temperature. Zr-MOFs could also function as catalyst support to promote the activity and selectivity of Cu catalyst via the strong metal support interaction (SMSI) between the guest copper nanocrystals (NCs) and the zirconium oxide secondary building units (SBUs) [300].

**Outlook and Summary**

As summarized in our previous perspective, several approaches have been designed to construct enzyme hybrid catalysts, including support-bound, entrapment, cross-linking, cooperation with homogenous metal complexes, and dispersion in reverse micelles. MOF could serve as an ideal platform to incorporate artificial enzyme active sites. The design of enzyme mimicking catalysts with a high level of control could be achieved by MOF support with simplified enzyme active sites. Selective methane oxidation to methanol could be realized by incorporating only enzyme active site onto MOF catalysts. The research will focus on enhancing the reaction selectivity and use single site catalyst as a new example to bridge homogenous and heterogeneous catalysis.

### The kinetics of elementary thermal reactions in heterogeneous catalysis

DOI: 10.1038/s41570-019-0138-7

**Abstract-Summary**

The kinetics of elementary reactions is fundamental to our understanding of catalysis. We present a short Perspective on recent experimental advances in the measurement of rates of elementary reactions at surfaces that rely on a stroboscopic pump–probe concept for neutral matter. The topic is discussed within the context of a specific but highly typical surface reaction, CO oxidation on Pt, which, despite more than 40 years of study, was only clarified after experiments with velocity-resolved kinetics became possible. This deceptively simple reaction illustrates fundamental lessons concerning the coverage dependence of activation energies, the nature of reaction mechanisms involving multiple reaction sites, the validity of transition-state theory to describe reaction rates at surfaces and the dramatic changes in reaction mechanism that are possible when studying reactions at low temperatures.

**Outlook**

Modern synchrotron XPS and STM methods operating at high or near-ambient pressure have revealed significant structural changes to the surface, suggesting that it is challenging to extrapolate UHV experiments on model catalysts to 'real-world' systems [314]. The first combined experimental results obtained in UHV experiments on single-crystal metals with DFT calculations [315, 316, 317, 318, 319–320] to devise a kinetic model for a high-pressure reactor comprised of nanoparticles. In that study, basic knowledge derived from the study of model systems could be used to predict product selectivity for oxygen-assisted esterification of methanol over gold-nanoparticle catalysts under high-pressure catalytic conditions [321]. Had the leaders of that same company had greater vision, they might have invested a small fraction of these losses in the basic research of heterogeneous catalysis to improve our ability to design new catalysts by principles based on physical understanding.

### Design concept for electrocatalysts

DOI: 10.1007/s12274-021-3794-0

**Abstract-Summary**

Metal-based electrocatalysts with different sizes (single atoms, nanoclusters, and nanoparticles) show different catalytic behaviors for various electrocatalytic reactions. Regulating the coordination environment of active sites with precision to rationally design an efficient electrocatalyst is of great significance for boosting electrocatalytic reactions. This review summarizes the recent process of heterogeneous supported single atoms, nanoclusters, and nanoparticles catalysts in electrocatalytic reactions, respectively, and figures out the construct strategies and design concepts based on their strengths and weaknesses.

### Recent progress on functional mesoporous materials as catalysts in organic synthesis

DOI: 10.1007/s42247-020-00086-1

**Abstract-Summary**

Mesoporous materials have been used as catalysts or their supports for homogeneous catalysis in the past decades. This review concerns the very recent development in this area, including the catalyst design, the types of active species, and the reaction catalyst by this kind of catalysts. Mesoporous catalysts can be classified as supported catalyst and bulk one. The supported catalyst includes active site in the pore wall framework, the active site on the pore wall surface, and the active site encapsulated in the mesopores. The bulk one can be directly used as catalyst, which includes metal oxides and carbon nitrides. Four different types of active sites, nanoparticles, highly distributed metals, anchored organic catalysts, and immobilized enzymes, are summarized in this review.

**Summary and outlook**

Tremendous advances have been made for the design and synthesis of mesoporous material-based catalysts for heterogenization of homogeneous organic reactions. With the development of the mesoporous materials science, an increasing number of organic reactions are developed for various fields of daily lives of the human beings. This review has tried to summarize the very recent progress in the design and synthesis of mesoporous material-based catalysts and their applications in a range of organic reactions. In spite of the great development of the design and synthesis of mesoporous material-based catalysts, a simple and universal method for the preparation of the stable metal centers with ultrasmall size and adjusted mesostructures is still lacking.

### Catalysis using metal–organic framework-derived nanocarbons: Recent trends

DOI: 10.1557/jmr.2020.166

**Abstract-Summary**

Recent trends in the area of catalytic applications of metal–organic framework (MOF)-derived nanocarbons are covered. Differences between conventional porous carbons and MOF-derived carbons are in pore volumes, surface area, and presence of ad-atoms. The morphology of the MOF-derived nanocarbons can be adjustable with uniform dopant distribution. Good catalytic performance is related with highly dispersed heteroatoms, density of catalytic active sites, controllable porosity, and high surface area.

Extended:

Differences between conventional porous carbons and MOF-derived carbons are in pore volumes, surface area, presence of ad-atoms (doping atoms), all is in favor of the MOF-derived nanocarbons.

**Conclusions and further outlook**

Photoredox properties of MOFs and their derivatives are related with inherent high porosity, so the nanocarbons can act as hosts for photoredox species as metal nanoparticles. Structures of MOF-derived materials include MOF-derived porous carbon, MOF-derived hollow structures, compositions of MOF-derived materials (metal-free nanocarbon, transition metal/metal compound-decorated nanocarbon, and micro/nanostructured MOF-derived composites), and MOF-supported noble metal NPs. Differences between conventional porous carbons and MOF-derived carbons are in pore volumes, surface area, presence of ad-atoms (doping atoms), all is in favor of the MOF-derived nanocarbons. Transition metals and carbon are the basis of the MOF-derived electrocatalysts, together with N-dopation. Among many reported MOFs, a relatively small number of their derived nanocarbons are currently used for photocatalysis purposes (in particular, for photocatalytical $CO_2$ reduction), so, undoubtedly, this is also a niche for intensive further investigations, as well as for meeting requirements for practical applications.

### Recent progress and perspective of electrochemical $CO_2$ reduction towards $C_2$-$C_5$ products over non-precious metal heterogeneous electrocatalysts

DOI: 10.1007/s12274-021-3335-x

**Abstract-Summary**

Compared with traditional $C_1$ products, high-value multicarbon products converted from atmospheric $CO_2$ via $CO_2$ER have attracted dramatic interest due to their significant economic efficiency, however desired catalytic selectivity of multicarbon products is difficult to achieve because of the high thermodynamic barriers and complex reaction pathways. Although certain progress has been made, there are still few systematic reports on the non-precious metal heterogeneous (NPMH) $CO_2$ER electrocatalysts for efficient conversion of $CO_2$ to multicarbon products. We summarize the latest research advances in recent developments of NPMH electrocatalysts, including nanostructured Cu, Cu-based bimetallic catalysts, Cu-based complexes, and carbon-based Cu-free catalysts for electroreduction of $CO_2$ into high-value multicarbon products. Key strategies and characterization techniques for catalytic mechanism insights, and unsolved issues and future trends for enhancing the $CO_2$ER performance of NPMH electrocatalysts are highlighted, which provides a constructive guidance on the development of $CO_2$ER electrocatalysts with high activity and selectivity for multicarbon products.

### Metallic Nanoparticles in Heterogeneous Catalysis

DOI: 10.1007/s10562-020-03477-5

**Abstract-Summary**

Many factors including the particle size, shape and metal-support interfaces can have significant influences on the catalytic properties of metal catalysts. In this Review, the size and shape dependent catalytic chemistry of metal nanoparticles and their electronic properties will be discussed. The unique catalytic chemistry at the metal-support interfaces will be discussed in details.

**Conclusions**

Size and shape dependent catalytic chemistry is of great interest and importance for designing highly active and selective catalysts. Although particle size has a major effect on catalytic activity, other factors such as oxidation state of the metal NPs and presence of low coordinated step and corner sites are also playing significant role. Unusual high catalytic activity can be obtained by tuning the metal-support interfaces.

**Selective Oxidation: From a Still Immature Technology to the Roots of Catalysis Science**

DOI: 10.1007/s11244-016-0684-x

**Abstract-Summary**

The design of heterogeneous selective oxidation catalysts based upon complex metal oxides is governed at present by a set of empirical rules known as "pillars of oxidation catalysis". They serve as practical guidelines for catalyst development and guide the reasoning about the catalyst role in the process. The present work extends the ideas of the pillar rules and develops the concept of considering a selective oxidation catalyst as enabler for the execution of a reaction network. The enabling function is controlled by mutual interactions between catalyst and reactants. This concept is based upon physical observables and could allow for a design strategy based upon a kinetic description that combines the processes between reactants with the processes between catalyst and reactants.

Extended:

This concept is based upon rigid reconstructed active surface sites that are covered and uncovered by adsorption and desorption. It is the purpose of this paper to show that the worlds of local static active sites and sites on oxidation catalysts are not different from one another and that the unifying concept of dynamical active sites can lead to a functional understanding of selective oxidation connected with physical concepts.

**Theory-guided design of catalytic materials using scaling relationships and reactivity descriptors**

DOI: 10.1038/s41578-019-0152-x

**Abstract-Summary**

A physical or chemical property of the reaction system, termed as a reactivity descriptor, scales with another property often in a linear manner, which can describe and/or predict the catalytic performance. In this Review, we describe scaling relationships and reactivity descriptors for heterogeneous catalysis, including electronic descriptors represented by d-band theory, structural descriptors, which can be directly applied to catalyst design, and, ultimately, universal descriptors. The prediction of trends in catalytic performance using reactivity descriptors can enable the rational design of catalysts and the efficient screening of high-throughput catalysts.

Extended:

In this Review, we discuss reactivity descriptors in different catalytic systems, taking into consideration the limitations of scaling relationships and strategies to break these established scaling relationships. We aim to lay foundations for the future computational design of catalysts.

**Outlook**

The ability of scaling relationships to simplify our understanding of catalysis inspires us to extract specific reactivity descriptors to predict the reaction trends and to screen for new catalysts. The complexity of catalytic systems leads to numerous reactivity descriptors, which are specifically explored to describe certain sections of the catalysis. As our insight into catalysis deepens, a universal reactivity descriptor for multi-catalytic systems is gaining attention. The goal is to develop reactivity descriptors that can be used simultaneously in various types of catalyst and reaction systems, namely, universal descriptors. In the short term, the discovery of a universal reactivity descriptor for all materials and reactions is not achievable, but it is possible to extract regionally universal descriptors, which are accurate for catalytic reactions within a subfield, such as the ORR, the OER or dehydrogenation.